\documentclass[twocolumn,showpacs,preprintnumbers,amsmath,amssymb,nofootinbib]{revtex4-2}
\usepackage{epsfig}

\usepackage{comment}
\usepackage[table]{xcolor}
\usepackage{graphicx}
\usepackage{dcolumn}
\usepackage{bm}

\usepackage{hyperref}
\hypersetup{colorlinks=true,linkcolor=magenta,anchorcolor=green,citecolor=cyan,filecolor=black,menucolor=black,urlcolor=brown}
\usepackage{slashed}
\newcommand{\ba}{\begin{eqnarray}}
\newcommand{\ea}{\end{eqnarray}}
\newcommand{\be}{\begin{equation}}
\newcommand{\ee}{\end{equation}}

\begin{document}

\title{Vortices and rotating solitons in ultralight dark matter}

\author{Philippe Brax}
\affiliation{Universit\'{e} Paris-Saclay, CNRS, CEA, Institut de physique th\'{e}orique, 91191, Gif-sur-Yvette, France}
\author{Patrick Valageas}
\affiliation{Universit\'{e} Paris-Saclay, CNRS, CEA, Institut de physique th\'{e}orique, 91191, Gif-sur-Yvette, France}

\begin{abstract}

The dynamics of ultralight dark matter with non-negligible self-interactions are determined by
a nonlinear Schr\"odinger equation
rather than by the Vlasov equation of collisionless particles.
This leads to wave-like effects, such as interferences, the formation of solitons, and a velocity field
that is locally curl-free, implying that vorticity is carried by singularities associated
with vortices.
Using analytical derivations and numerical simulations in 2D, we study the evolution of such
a system from stochastic initial conditions with nonzero angular momentum.
Focusing on the Thomas-Fermi regime, where the de Broglie wavelength of the system is smaller than its
size, we show that a rotating soliton forms in a few dynamical times.
The rotation is not associated with a large orbital quantum number of the wave function.
Instead, it is generated by a regular lattice of vortices that gives rise to a solid-body rotation in the
continuum limit. Such rotating solitons have a maximal radius and rotation rate for a given central
density, while the vortices follow the matter flow on circular orbits.
We show that this configuration is a stable minimum of the energy at fixed angular momentum and
we check that the numerical results agree with the analytical derivations.
We expect most of these properties to extend to the 3D case where point vortices would be replaced
by vortex rings.

\end{abstract}

\date{\today}

\maketitle

\section{Introduction}
\label{sec:Introduction}

Although weakly interacting massive particles (WIMPs) remain a popular scenario for Cold Dark Matter
(CDM) \citep{Jungman:1995df,Drees:2004jm,Steigman:1984ac},
alternative models such as axions or more generally axion-like-particles (ALPs) have
generated a renewed interest in recent years.
A classic example is the QCD axion
\cite{Peccei:1977hh,Weinberg:1977ma,Wilczek:1977pj,Preskill:1982cy,Abbott:1982af}
but string theory can also lead to many ALPs with a wide range of masses
from $10^{-22}$ eV to $1$ eV
\cite{Svrcek:2006yi,Arvanitaki:2009fg,Halverson:2017deq,Bachlechner:2018gew}.
In addition to their interest for particle physics or string theory, these models
might alleviate the small-scale tensions of the standard CDM scenario
\cite{Weinberg:2013aya,DelPopolo:2016emo,Nakama:2017ohe,Salucci:2018hqu,DiLuzio:2020wdo}.
More generally, their distinct dynamics on small scales could allow us to discriminate between
these various dark matter candidates.

These models are generically described by scalar or pseudo-scalar fields
and for $m<1$ eV the very large occupation number means that they can be treated as classical fields
\cite{Sikivie:2009qn,Hui:2016ltb,Fan:2016rda},
despite obeying a nonlinear Schr\"odinger equation corresponding to the non-relativistic limit of the Klein-Gordon equation, also known as the
Gross-Pitaevskii equation when self-interactions are non-negligible.
For very low masses, $m \sim 10^{-22} - 10^{-20}$ eV, the de Broglie wavelength
can reach the kpc size and lead to flat DM density cores at the center of galaxies.
This model, where self-interactions are negligible and the departure from standard CDM
is due to the large size of the de Broglie wavelength, is often called ``Fuzzy Dark Matter''
(FDM) \cite{Hu:2000ke,Schive:2014dra,Hui:2016ltb,Nori:2018hud}.
However, because they suppress small-scale density fluctuations such low masses
are ruled out by Lyman-$\alpha$ forest observations
\cite{Irsic:2017yje,Armengaud:2017nkf}.
Therefore, ALPs should have higher masses (unless they only constitute a small fraction of the
dark matter or non-standard kinetic terms lead to a strong redshift dependence of the speed of sound
\cite{Lu:2024uyv}) and they may also have non-negligible self-interactions.
Such dark matter scenarios, with a dark matter field that shows fast oscillations at the frequency $m$,
can also lead to characteristic signatures on gravitational waves 
\cite{Khmelnitsky:2013lxt,Brax:2024yqh,Cai:2023ykr,Chowdhury:2023xvy,Blas:2024duy}.

As compared with the standard CDM scenario, two new scales appear, the de Broglie wavelength
$\lambda_{\rm dB}=2\pi/(mv)$, where $v$ the typical velocity (e.g., the virial velocity of a collapsed
dark matter halo), and the self-interaction Jeans scale $r_a \sim \sqrt{\lambda_4/{\cal G}}/m^2$,
where $\lambda_4$ is the coupling constant of repulsive quartic self-interactions
\cite{Rindler-Daller:2011afd,Chavanis:2011zi,Zhang:2024bjo}.
Then, whereas the evolution remains similar to that of CDM on larger scales, on smaller scales
new effects appear such as wave-like features or the formation of solitons
\cite{Goodman:2000tg,Peebles_2000,Schive:2014dra,Mocz:2017wlg,Veltmaat:2018dfz,Mirasola:2024pmw}.
In practice, the transition scale must be below $1 \, \rm kpc$ to be consistent with observations
of the cosmic web and galaxy profiles, as recalled above.

In this paper, we focus on scenarios where $r_a \gg \lambda_{\rm dB}$, that is,
the repulsive self-interactions dominate over the so-called quantum pressure.
This typically corresponds to the Thomas-Fermi regime \cite{Chavanis:2011zi}
(which however only applies inside smooth configurations like the soliton and not in the outer halo).
We also focus on scales of the order of $r_a$, which  corresponds to the size of the solitons.
This is a regime where the dynamics can depart from the standard CDM scenarios.
A characteristic feature that has already been studied in detail is the formation of solitons,
that is, hydrostatic equilibria at the center of virialized halos
\cite{Chavanis:2011zi,Schive:2014dra,Mocz:2017wlg,Veltmaat:2018dfz,Dawoodbhoy:2021beb,Garcia:2023abs,Chen:2024pyr}.
They correspond to the ground state of the Schr\"odinger equation, in a spherically
symmetric gravitational potential, with a vanishing velocity.
However, if the system has a nonzero initial angular momentum and because of the conservation
of angular momentum by the dynamics, we can expect the soliton to display some rotation
(unless all the angular momentum is expelled into the outer halo).
This question has been much less studied and it is the focus of this paper.
This is motivated by the fact that cosmological simulations show that dark matter halos
have a nonzero angular momentum, with the dimensionless spin parameter
\be
\lambda_L = L |E|^{1/2} /( {\cal G} M^{5/2})
\label{eq:lambda-L}
\ee
ranging from 0.01 to 0.1 \cite{Barnes:1987hu,Bullock:2000ry,Maccio-2007}.
With tidal torque theory, this can be understood from the growth of the angular momentum
in the quasi-linear regime, because of the tidal field due to neighbouring structures
\cite{Peebles:1969jm,Doroshkevich-1970,White-1984}.

In the nonrelativistic regime the scalar field associated with these ultralight bosonic dark matter particles
obeys a Gross-Pitaevskii equation.
This equation also describes the Bose-Einstein Condensates (BEC) studied in laboratory experiments
(e.g., cold atom gas).
These correspond to the so-called solitons in our context, except that in a dark matter halo distribution
the external confining potential is replaced by the halo's self-gravity.
It is well-known that BECs in the laboratory, placed within a rotating container, exhibit vortices
\cite{Abo-Shaeer-2001,Fetter_2001,pitaevskii2003bose,Pethick_Smith_2008}.
In our case, we have an isolated system (a single halo) and the rotation is not due to the
external apparatus but simply to the initial angular momentum. Then, our goal is to find out
whether these ultralight dark matter systems, governed by their self-gravity, also form vortices,
what are their properties and what is their impact on the solitons.

Numerical simulations of Fuzzy Dark Matter scenarios, in a cosmological
setting or for collapsing halos \cite{Mocz:2017wlg,Hui:2020hbq,Liu:2022rss}
have shown that vortices naturally appear outside  the soliton, because of the interferences
between uncorrelated excited modes that generate granules of size $\lambda_{\rm dB}$
and many vortex lines associated with the zeros of the wave function.
In models without self-interactions, it has been shown that rotating boson stars are unstable
\cite{Dmitriev:2021utv}, although a central vortex can be stabilized by the external gravitational
potential of a central black hole of a sufficiently large mass
\cite{Glennon:2023oqa}.
However, our focus in this paper is rather on the vortices that can appear inside the solitons,
in relation with the macroscopic angular momentum, for scenarios with large self-interactions.
A few studies have already considered angular momentum and vortices in such ultralight dark matter
cases \cite{Silverman:2002qx,Rindler-Daller:2011afd}.
In particular, Refs.\cite{Kain:2010rb,Rindler-Daller:2011afd} obtained
the critical rotation rate above which the energy is lowered by creating a vortex.
They find that this corresponds to a total angular momentum that is greater than the
quantum of angular momentum of a vortex, when self-interactions are non-negligible
and $r_a$ is below the size of the system. This is precisely the regime considered in this paper. 
In a similar and phenomenological vein, 
it has also been suggested that ultralight dark matter vortices could explain observations
of the spin of cosmic filaments on Mpc scales \cite{Alexander:2021zhx}, or that rotating solitons
could provide a good model in order to reproduce the galaxy rotation curves \cite{Boehmer:2007um}.

In contrast with most of these earlier works, in this paper we do not assume homogeneous halos
or solid-body rotation and we derive the radial profile of the rotating soliton;
we also consider stochastic initial conditions, without an initial soliton.
This allows us to check how rotating solitons naturally form within such virialized halos
that better resemble the configurations found in cosmology.

As we focus on the Thomas-Fermi regime, where the width and vorticity quantum of the vortices
are small, this necessitates a good numerical resolution to handle scales ranging from the vortex radius
$\xi$ to the size $L_\star$ of the system, which is greater than the self-interaction Jeans
scale $r_a$.
Thus, we typically have the hierarchy $\xi \ll d \ll r_a \lesssim L_\star$, where $d$ is the
distance between vortices.
This leads us to focus on isolated halos (rather than full cosmological simulations) in 2D.
The other advantage of working in 2D is that it is easier to identify the vortices and compare
the numerical results with analytical derivations.

In this paper, we tackle a situation where the initial halo has angular momentum.
What we observe numerically is that after a few dynamical times, a network
of quantized vortices appears inside a rotating soliton. Moreover, they follow circular trajectories
with a velocity generated by the other vortices of the network. When the Thomas-Fermi limit is taken,
i.e. when the de Broglie wavelength becomes very small compared to the halo size,
we find that the number of vortices grows while keeping the total angular momentum of the vortices
at a fraction of the initial halo angular momentum.
In this limit, the gas of vortices is dilute within the rotating soliton.
These results can be confirmed analytically. In particular, we find that stable rotating solitons
with a solid-body rotation correspond to the continuum limit of an infinite number of vortices,
with a homogeneous distribution within the axisymmetric soliton.
We also confirm analytically that rotating vortices are wider than their static counterparts,
with a maximal radius and a maximal rotation rate set by the central density.

This paper is organized as follows.
In Sec.~\ref{sec:GP} we recall the equations of motion associated with scalar field dark matter
with quartic self-interactions.
In the nonrelativistic limit this leads to a Gross-Pitaevskii equation, which can also be mapped
to hydrodynamical equations. We also recall the profile of static solitons, which correspond
to hydrostatic equilibria.
In Sec.~\ref{sec:vortices} we describe solutions that include vortices, associated with singularities
of the velocity field (but the wave function remains regular).
We first study the profile of a single vortex and next generalize to wave functions that contain
many vortices, deriving the equations of motion of the flow and of the vortices.
We take the continuum limit in Sec.~\ref{sec:continuum} and we show that a nonzero angular
momentum leads to a rotating soliton that displays a solid-body rotation and a uniform vorticity.
This corresponds to a uniform distribution of vortices.
We compare our analytical results with numerical simulations in Sec.~\ref{sec:simulations}
and we study the dependences on the de Broglie wavelength and on the initial angular momentum.
We conclude in Sec.~\ref{sec:Conclusion}.

We also provide three appendices, to derive the asymptotic profiles of the static soliton
in App.~\ref{app:static-soliton} and of the vortex in App.~\ref{app:vortex-profile}, and the
excess energy associated with the vortex in App.~\ref{app:vortex-energy}.

\section{Equations of motion}
\label{sec:GP}

\subsection{Nonrelativistic equation of motion}

We consider the following Lagrangian to describe the scalar field dark matter,
\be
\mathcal{L}_{\phi} = -\frac{1}{2}g^{\mu\nu}\partial_{\mu}\phi\partial_{\nu}\phi - V(\phi),
\label{eq:lagrangian-sfdm}
\ee
where $g^{\mu\nu}$ is the inverse metric, the first term is the standard kinetic term and $V(\phi)$
is the scalar potential given by
\be
V(\phi) = \frac{m^2}{2}\phi^2 + V_I(\phi) , \;\;\; V_I(\phi) = \frac{\lambda_4}{4} \phi^4 ,
\label{eq:potential-v-sfdm}
\ee
and we work in natural units, $c=\hbar=1$.
Typically, dark matter fields in the form of scalars could be pseudo-Goldstone bosons and
have a periodic potential.
Close to one of the minima of these potentials, the potential can be expanded in Taylor series
where the quadratic term corresponds to the mass term while the leading correction is quartic
for models with a parity symmetry. For axions, the sign of the quartic coupling is negative.
In this paper we focus on the opposite situation with positive quartic couplings, which can be associated
with axion monodromy models for instance.

In the weak gravity regime and neglecting the Hubble expansion, which is appropriate for galactic
and subgalactic scales, the scalar field obeys a nonlinear Klein-Gordon equation.
At leading order the field oscillates at the very high frequency $m$ and
to average over these fast oscillations it is convenient to introduce a complex scalar field
$\psi$ by \citep{Hu:2000ke,Hui:2016ltb},
\be
\phi= \frac{1}{\sqrt{2m}} ( \psi e^{-imt} +  \psi^* e^{imt}) .
\label{eq:phi-psi}
\ee
Substituting into the action or the equation of motion and averaging over these fast oscillations
gives the nonrelativistic equation of motion \citep{Brax:2019fzb}
\be
i \frac{\partial\psi}{\partial t} =
- \frac{\Delta\psi}{2m} + m ( \Phi_N + \Phi_I ) \psi , \;\;\;
\Phi_I = \frac{3 \lambda_4}{4 m^3} |\psi|^2 ,
\label{eq:Schrod}
\ee
which has the form of a Gross-Pitaevskii equation, except that the Newtonian potential $\Phi_N$
is not external but given by the self-gravity of the scalar field.

\subsection{Dimensionless quantities}

As usual, it is convenient to work with dimensionless quantities, which we define by
\be
\psi = \psi_\star \tilde\psi, \;\;\; t = T_\star \tilde t, \;\;\;
\vec r = L_\star \tilde{\vec r} , \;\;\;
\Phi = V_\star^2 \tilde\Phi,
\label{eq:dimesionless-def}
\ee
where $T_\star$, $L_\star$ and $V_\star = L_\star/T_\star$ are the characteristic time, length and
velocity scales of the system, and $T_\star = 1/\sqrt{{\cal G} m \psi_\star^2}$ where
${\cal G}$ is Newton's constant.
Under this rescaling, we obtain the dimensionless Schr\"odinger--Poisson
equations that govern the dynamics of small-scale structures,
\be
i \epsilon \frac{\partial\tilde\psi}{\partial\tilde t} =
- \frac{\epsilon^2}{2} \tilde\Delta \tilde\psi + (\tilde\Phi_{N} + \tilde\Phi_I ) \tilde\psi ,
\label{eq:Schrod-eps}
\ee
\be
\tilde\Delta \tilde\Phi_{N} = 4 \pi \tilde\rho , \;\;\; \tilde\Phi_I = \lambda \tilde\rho , \;\;\;
\mbox{and}  \;\;\; \tilde\rho = | \tilde\psi|^2 ,
\label{eq:Poisson-eps}
\ee
with the coupling constant $\lambda = 3\lambda_4/(4 {\cal G} m^4 L_\star^2)$.
The coefficient $\epsilon$ is given by
\be
\epsilon = \frac{T_\star}{m L_\star^2} .
\label{eq:scaling-eps}
\ee
If we compare this quantity with the typical de Broglie wavelength $\lambda_{\rm dB} = 2\pi/(m V_\star)$,
we have
\be
\epsilon \sim \frac{\lambda_{\rm dB}}{L_\star} .
\label{eq:epsilon-de-Broglie}
\ee
Therefore, the parameter $\epsilon$, which appears in the dimensionless Schr\"odinger equation
(\ref{eq:Schrod-eps}) plays the role of $\hbar$ in quantum mechanics. This parameter measures the
relevance of wave effects in the system, such as interferences, or the importance of the quantum pressure.
More precisely, we have for the rescaled de Broglie wave length
\be
\tilde\lambda_{\rm dB} = 2\pi \epsilon / \tilde v ,
\label{eq:de-Broglie-rescaled}
\ee
where $\tilde v=v/V_\star$ is the rescaled velocity.
In the following we omit the tilde to simplify the notations.

\subsection{Action}

The Gross-Pitaevskii equation (\ref{eq:Schrod-eps}) derives from the non-relativistic action
\ba
{\cal S}[\psi,\psi^\star] & = & \int d\vec r dt \biggl[ \frac{i\epsilon}{2}
\left( \psi^\star \frac{\partial \psi}{\partial t} - \psi \frac{\partial \psi^\star}{\partial t} \right)
- \frac{\epsilon^2}{2} | {\vec\nabla} \psi |^2 \nonumber \\
&& - \Phi_N | \psi |^2 - {\cal V}_I(|\psi|^2) \biggl] ,
\label{eq:action-psi}
\ea
where $\cal V_I(\rho)$ is related to $\Phi_I(\rho)$ by
\be
\Phi_I = \frac{d {\cal V}_I}{d\rho} .
\label{eq:V-I-rho}
\ee

The energy of a given configuration reads
\be
E[\psi] = \int d {\vec r} \left[ \frac{\epsilon^2}{2} | \vec\nabla \psi |^2
+ \frac{1}{2} \rho \Phi_N + {\cal V}_I \right] ,
\label{eq:E-psi}
\ee
where the factor $1/2$ in the gravitational term arises from the need to avoid double-counting as
the self-gravitational potential is sourced by the system itself.
It is conserved by the Gross-Pitaevskii equation (\ref{eq:Schrod-eps}), as well as the total mass,
linear momentum and angular momentum.

\subsection{Hydrodynamical picture}
\label{sec:hydro}

As in Eq.(\ref{eq:Poisson-eps}), the matter density is the square of the amplitude of the
wave function $\psi$. Defining a velocity field $\vec v$ from the phase $S$ by \citep{Madelung:1927ksh}
\be
\psi = \sqrt{\rho} \, e^{i S} , \;\;\; \vec v = \epsilon \vec\nabla S ,
\label{eq:Madelung}
\ee
and substituting into the equation of motion (\ref{eq:Schrod-eps}), the real and imaginary parts
give the continuity and Euler equations,
\be
\frac{\partial\rho}{\partial t} + \nabla\cdot(\rho \vec v) = 0 ,
\label{eq:continuity}
\ee
\be
\frac{\partial\vec v}{\partial t} + (\vec v \cdot \vec\nabla) \vec v =
- \vec\nabla( \Phi_Q + \Phi_N + \Phi_I ) ,
\label{eq:Euler}
\ee
where we introduced the so-called quantum pressure defined by
\be
\Phi_Q = - \frac{\epsilon^2}{2} \frac{\Delta \sqrt{\rho}}{\sqrt{\rho}} .
\label{eq:PhiQ-def}
\ee
In terms of the hydrodynamic variables, the action (\ref{eq:action-psi}) reads
\ba
{\cal S}[\rho,S] & = & \int d\vec r dt \biggl [ - \epsilon \rho \frac{\partial S}{\partial t}
- \frac{\epsilon^2}{8\rho} ( \vec\nabla \rho)^2 - \frac{\epsilon^2}{2} \rho (\vec\nabla S)^2
\nonumber \\
&& - \rho \Phi_N - {\cal V}_I \biggl ]  .
\label{eq:action-hydro}
\ea
Of course, the equations of motion (\ref{eq:continuity})-(\ref{eq:Euler}) can also be derived from
this action, by taking variations with respect to $\rho$ and $S$.
The energy (\ref{eq:E-psi}) now reads
\be
E[\rho,\vec v] = \int d {\vec r} \left[ \frac{\epsilon^2}{2} (\vec\nabla \sqrt\rho)^2
+ \frac{1}{2} \rho \vec v^{\,2} + \frac{1}{2} \rho \Phi_N + {\cal V}_I \right] .
\label{eq:E-rho-v}
\ee

\subsection{Static soliton}
\label{sec:static-soliton}

The Schr\"odinger equation (\ref{eq:Schrod-eps}) admits hydrostatic equilibria, also called solitons
\citep{Chavanis:2011zi,Chavanis:2011zm,Harko:2011jy,Brax:2019fzb},
which correspond to the ground state of the potential $\Phi_N+\Phi_I$.
These configurations, of the form $e^{-i \mu t/\epsilon} \hat\psi(\vec r)$, are
solutions of the time-independent Schr\"odinger equation
\be
\Phi_Q + \Phi_N + \Phi_I = \mu , \;\;\; \psi_{\rm sol}(\vec r,t) = e^{-i \mu t/\epsilon}
\hat\psi_{\rm sol}(r) ,
\label{eq:hydrostatic-full}
\ee
where we considered spherically symmetric solutions and the quantum pressure $\Phi_Q$ was
already defined in Eq.(\ref{eq:PhiQ-def}).
They also correspond to the hydrostatic equilibria of the hydrodynamical equations
(\ref{eq:continuity})-(\ref{eq:Euler}) and to minima of the energies (\ref{eq:E-psi}) and
(\ref{eq:E-rho-v}) at fixed mass.

In the Thomas-Fermi regime where we can neglect the quantum pressure because $\epsilon \ll 1$
and gravity is balanced by the repulsive self-interactions, the hydrostatic equilibrium is given by
\be
\mbox{TF regime} : \;\;\; \Phi_N + \Phi_I = \mu  .
\label{eq:TF-static}
\ee
In 2D this gives the density profile
\be
R_0 - r \gg \epsilon^{2/3} : \;\;\; \rho_{\rm TF,0}(r) = \rho_{0} J_0( z_0 r/R_0) ,
\label{eq:rho-TF-0}
\ee
with the radius
\be
R_0 = z_0 \sqrt{\frac{\lambda}{4\pi}} ,
\label{eq:R0-def}
\ee
where $\rho_0$ is the central density, $z_0 \simeq 2.405$ is the first zero of the Bessel function
of the first order, $J_0(z_0)=0$, and the subscript ``TF,0'' stands for the Thomas-Fermi regime
with zero rotation.
The soliton radius $R_0$ does not depend on its mass.

In (\ref{eq:rho-TF-0}) we noted that the Thomas-Fermi approximation breaks down within a boundary
layer of width $\epsilon^{2/3}$ to the left of $R_0$.
We present in App.~\ref{app:static-soliton} a derivation of the full soliton profile, over $0 \leq r < \infty$,
in the asymptotic limit $\epsilon \to 0$.
We distinguish three regions, the inner Thomas-Fermi region $0 \leq r \leq R_0 - \delta_-$
with the profile (\ref{eq:TF-hat-psi}) and $\epsilon^{2/3} \ll \delta_- \ll 1$,
the outer WKB region $R_0 + \delta_+ \leq r < \infty$ with the profile (\ref{eq:WKB-det})
and $\delta_+ \gg \epsilon^{2/3}$,
and the intermediate region $R_0 - \Delta_- \leq r \leq R_0 + \Delta_+$ with the profile
(\ref{eq:P-II}) and $\Delta_- \ll 1$ and $\Delta_+ \ll \epsilon^{2/5}$.
The hierarchies $\Delta_- \gg \delta_-$ and $\Delta_+ \gg \delta_+$ mean that there is a large
overlap region between the boundary layer solution (\ref{eq:P-II}) and the Thomas-Fermi and
WKB solutions (\ref{eq:TF-hat-psi}) and (\ref{eq:WKB-det}).
This allows us to perform a well-defined asymptotic matching between these three regions and
to obtain a global approximation in the limit $\epsilon \to 0$.
In particular, at $R_0$ the density is nonzero but of order $\epsilon^{2/3}$ from Eq.(\ref{eq:P-II}),
while at large distance it decays slightly faster than an exponential from Eq.(\ref{eq:psi-decay}).

\section{Vortices}
\label{sec:vortices}

\subsection{Single vortex}
\label{sec:single-vortex}

\subsubsection{Homogeneous background}
\label{sec:vortex-homogeneous-background}

We consider a single vortex at the center of the system, $\vec r=0$, of width much smaller than
the soliton radius. Therefore, in this section we neglect the slow spatial variation of the soliton density
and write $\Phi_N \simeq \Phi_{N0}$, $\Phi_I \simeq \Phi_{I0} = \lambda \rho_0$ and
$\mu=\Phi_{N0}+\Phi_{I0}$.
(See Refs.\cite{Brook:2009ku,Kain:2010rb} for studies of a large vortex taking into account
its self-gravity.)
Then, single vortices of spin $\sigma$ correspond to solutions of the Gross-Pitaevskii equation of the form
\be
\psi(\vec r,t) = e^{-i \mu t/\epsilon} \sqrt{\rho_0} f(r) e^{i \sigma \theta} , \;\;\;
\rho(r) = \rho_0 f^2(r) , \;\;\; \sigma \in \mathbb{Z} ,
\label{eq:single-vortex-psi}
\ee
with the boundary condition
\be
r \to \infty: \;\;\; f(r) \to 1 .
\ee
Substituting into the Gross-Pitaevskii equation (\ref{eq:Schrod-eps}) we obtain the differential
equation
\be
\frac{d^2f}{d\eta^2} + \frac{1}{\eta} \frac{df}{d\eta} + \left( 1 - \frac{\sigma^2}{\eta^2} \right) f
- f^3  = 0 ,
\label{eq:ODE-f-eta}
\ee
where we introduced the rescaled radial coordinate $\eta$ and the so-called healing length $\xi$
\cite{pitaevskii2003bose,Rindler-Daller:2011afd},
\be
\eta = \frac{r}{\xi} , \;\;\;
\xi = \frac{\epsilon}{\sqrt{2\lambda\rho_0}} = \frac{\epsilon z_0}{\sqrt{8\pi\rho_0} R_0} .
\label{eq:healing-length}
\ee
We have the asymptotic behaviors
\be
\eta \to 0 : \;\; f \propto \eta^{|\sigma|} , \;\;
\eta \to \infty: \;\; f = 1 - \frac{\sigma^2}{2 \eta^2} + \cdots
\label{eq:asymp-f}
\ee
and we can see that the core radius of the vortex is $\eta_c \sim |\sigma|$, that is,
\be
r_c = |\sigma| \xi = \frac{\epsilon |\sigma|}{\sqrt{2\lambda\rho_0}} .
\label{eq:rc-def}
\ee

The single vortex (\ref{eq:single-vortex-psi}) gives the velocity field
\be
\vec v = \frac{\epsilon \sigma}{r} \vec e_\theta = \epsilon \sigma \frac{\vec e_z \times \vec r}{r^2} ,
\;\;\; v_r = 0 , \;\;\; v_\theta = \frac{\epsilon \sigma}{r} .
\label{eq:v-single-vortex}
\ee
Here, $\vec r = r \vec e_r$ while $\vec e_r$ and $\vec e_\theta$ are the unit radial and azimuthal vectors
in polar coordinates.
We also introduced the unit vertical vector $\vec e_z$, which is convenient for vector operations, when
we embed our 2D system in a 3D space with cylindrical symmetry, where all fields are independent of the
vertical coordinate $z$.
The vorticity reads
\be
\vec\omega = \vec \nabla \times \vec v = 2 \pi \epsilon \sigma \delta_D^{(2)}(\vec r) \vec e_z ,
\label{eq:vorticity-vortex}
\ee
while the circulation $\Gamma(r)$ around a circle ${\cal C}$ of radius $r$ is
\be
\Gamma(r) = \oint_{\cal C} \vec v \cdot \vec{d\ell} = \int_S \vec \omega \cdot \vec{dS} = 2 \pi \epsilon \sigma .
\label{eq:Gamma-sigma}
\ee
Thus the vorticity and the circulation are quantized.

Substituting into the energy functional (\ref{eq:E-psi}) and subtracting the uniform
background at density $\rho_0$ we obtain \cite{pitaevskii2003bose} for the energy
difference between the vortex and static soliton configurations,
$\Delta E_\sigma = E_{\rm vortex} - E_{\rm soliton}$,
\be
\Delta E_\sigma = \epsilon^2 \rho_0 \pi \int_0^{\infty} \!\! d\eta \, \eta \left[ \left(\frac{df}{d\eta}\right)^2
+ \frac{\sigma^2}{\eta^2} f^2 + \frac{1}{2} ( f^2 \! - \! 1 )^2 \right] .
\label{eq:Delta-E_sigma-def}
\ee
From the asymptotic behaviors (\ref{eq:asymp-f}) we can see that the energy is dominated by the
infrared divergence of the angular momentum contribution,
\be
\Delta E_\sigma  \simeq \sigma^2 \pi \rho_0 \epsilon^2 \ln[R_0/( | \sigma | \xi)] ,
\label{eq:Delta-E-L}
\ee
where we cut the integral at the system size, which is of the order of the soliton radius $R_0$.
Thus, vortices of higher spin $\sigma$ have a greater energy.

\subsubsection{Soliton background}

The derivation of the vortex profile and energy in the previous section, which follows standard textbooks
\cite{pitaevskii2003bose,Pethick_Smith_2008}, neglects the radial dependence of the static soliton
profile. However, the latter is needed for the logarithmic cutoff 
$\ln[R_0/(|\sigma|\xi)]$ in Eq.(\ref{eq:Delta-E-L}).

As for the static soliton, we provide in App.~\ref{app:vortex-profile} a detailed derivation
of the global vortex profile in the asymptotic limit $\epsilon \to 0$.
This gives the density falloff (\ref{eq:psi-J-|sigma|-norm}) and (\ref{eq:vortex-center}) at the
center, the generalized Thomas-Fermi regime (\ref{eq:rho-TF-rc}) in the bulk that includes the
orbital barrier associated with the vortex angular momentum, and the boundary layer 
(\ref{eq:vortex-bl}) (for large spin) around the core radius $r_c$.
At larger radii, $r \gtrsim R_0$, the density profile follows the same behavior as the static
soliton, with a modified exponential tail beyond $R_0$.

Next, we present in App.~\ref{app:vortex-energy} a more careful derivation of the excess energy 
associated with the vortex that takes full account of the radial dependence of the density.
We again find that the energy difference $\Delta E_\sigma$ is dominated by the angular momentum 
contribution (\ref{eq:Delta-E-sigma-def}), by a factor $\ln(1/\epsilon)$, and we recover the expression
(\ref{eq:Delta-E-L}).
Thus, we recover the growth with the spin of the excess energy due to the vortex.

\subsection{Several vortices}

For a pair of vortices $\sigma_1$ and $\sigma_2$ that are well separated, in addition to their
individual excess energies (\ref{eq:Delta-E-L}) their interaction energy $E_{\rm int}$
is of the order of
\be
E_{\rm int} \simeq \int d \vec r \, \rho \vec v_1 \cdot \vec v_2 \simeq 
\sigma_1 \sigma_2 2\pi \rho_0 \epsilon^2 \ln(R_0/d_{12}) ,
\ee
where $d_{12}$ is the distance between the two vortices and we assumed $\xi \ll d_{12} \ll R_0$
\cite{Pethick_Smith_2008}.
For $N_v$ vortices of unit spin, we obtain in the limiting case $d_{ij} \sim \xi$
\be
\Delta E_{N_v;\xi} \simeq N_v^2 \pi \rho_0 \epsilon^2 \ln(R_0/\xi) .
\label{eq:E-Nv-1}
\ee
Comparing with Eq.(\ref{eq:Delta-E-L}) we find $\Delta E_{N_v;\xi} \simeq \Delta E_{\sigma}$
for $N_v = | \sigma |$.
On the other hand, if the vortices are uniformly distributed within radius $R_0$, we obtain
in the continuum limit
\be
\Delta E_{N_v;{\rm unif}} \simeq N_v \pi \rho_0 \epsilon^2 \ln(R_0/\xi) 
+ N_v^2 \frac{\pi}{4} \rho_0 \epsilon^2 ,
\label{eq:E-Nv-2}
\ee
which is about a quarter of the single vortex energy (\ref{eq:Delta-E-L}),
for $N_v = | \sigma |$.

The circulation $\Gamma$ of Eq.(\ref{eq:Gamma-sigma}) along a loop that moves with the matter
is conserved by the dynamics. However, a high-spin vortex $\sigma$ may break up into
$N_v = |\sigma|$ vortices of unit spin at fixed $\Gamma$.
Comparing Eq.(\ref{eq:Delta-E-L}) with Eqs.(\ref{eq:E-Nv-1}) and (\ref{eq:E-Nv-2}), 
we find that the excess energy of a single vortex of large spin $\sigma$ is greater than the excess 
energy of a collection of $|\sigma|$ vortices of unit spin,
if $d \gg \xi$ \cite{Pethick_Smith_2008,landau1980statistical}.
As we shall see below in Eq.(\ref{eq:d-Omega-eps}), this is indeed the case in our system
with $\epsilon \ll 1$.
Therefore, it is energetically favorable for a vortex of high spin $\sigma$ to split up into $N_v=|\sigma|$
vortices of unit spin, keeping the total circulation constant.
This explains why in our numerical simulations we only find vortices with $\sigma = \pm 1$, 
in agreement with previous numerical and analytical works \cite{RINDLERDALLER2008,Hui:2020hbq}.
Detailed analytical and numerical analysis of the splitting instability of high-spin vortices
are presented in \cite{Skryabin2000,Kawaguchi2004,Lundh2006,Takeuchi2018,Giacomelli2020,VanAlphen2024}.
In particular, these works show that the instability is present for both finite-size and
infinite-size systems. Because this process is associated with modes that are centered
around the vortex core, which is much smaller than the soliton radius $R_0$, these results also
apply to our case, which only differs by the shape of the confining potential (here self-gravity)
at much larger distance.

Within the outer halo, dominated by interferences between uncorrelated excited modes,
the vortices correspond to points where the wave function happens to vanish.
Near such zeros the Taylor expansion of the wave function is generically of the form
$\psi = a (x-x_0) + b (y-y_0) + \dots$, i.e. it starts at linear order, which corresponds to $\sigma=\pm1$
from Eq.(\ref{eq:asymp-f}), see \cite{Hui:2020hbq}.

\subsection{Vortex lattice}
\label{sec:vortex-network}

\subsubsection{Ansatz for the wave function}

We now consider a set of $N$ vortices of spin $\sigma_j$ at positions $\vec r_j(t)$, associated with
the phases and velocities
\be
S_j(\vec r) = \sigma_j \theta_j , \;\; \theta_j = \widehat{(\vec e_x , \vec r - \vec r_j)} , \;\;
\vec v_j(\vec r) = \epsilon \sigma_j \vec\nabla \theta_j  ,
\ee
which gives
\be
\vec v_j(\vec r) = \epsilon \sigma_j \vec e_z \times \frac{\vec r - \vec r_j}{|\vec r - \vec r_j|^2}
= \epsilon \sigma_j \vec e_z \times \vec\nabla \ln | \vec r - \vec r_j |
\ee
and
\be
\vec\omega_j(\vec r) = 2\pi\epsilon \sigma_j \delta_D^{(2)}(\vec r - \vec r_j) \vec e_z .
\label{eq:omega-j}
\ee
Then, we write for the wave function the ansatz \cite{Fetter-1966,CRESWICK1980}
\be
\psi (\vec r,t) = e^{-i\mu t/\epsilon} \sqrt{\rho_0} \prod_{j=1}^N f(\vec r - \vec r_j) e^{i S_j} ,
\label{eq:psi-ansatz}
\ee
which is the generalization of the single vortex wave function (\ref{eq:single-vortex-psi}).
This assumes that the vortices are well separated, with a typical distance that is much greater than
the healing length (\ref{eq:healing-length}), $|\vec r_{j} - \vec r_{j'}| \gg \xi$.
This gives the density and velocity fields
\be
\rho = \rho_0  \prod_{j=1}^N f^2(\vec r - \vec r_j) , \;\;\;
\vec v = \sum_{j=1}^N \vec v_j , \;\;\; \nabla \cdot \vec v = 0 .
\ee

\subsubsection{Equations of motion}
\label{sec:Creswick}

Following \cite{CRESWICK1980} we substitute the ansatz (\ref{eq:psi-ansatz})
into the hydrodynamic form of the equations of motion. Thus, we obtain the expressions
\be
\frac{\partial\rho}{\partial t} + \nabla \cdot ( \rho \vec v) = \rho \sum_{j=1}^N
\frac{\vec\nabla f^2(\vec r - \vec r_j)}{f^2(\vec r - \vec r_j)} \cdot \left[ \vec v - \dot{\vec r}_j \right]
\label{eq:continuity-ansatz}
\ee
\be
\frac{\partial \vec v}{\partial t} + ( \vec v \cdot \vec\nabla) \vec v = \sum_{j=1}^N
( [\vec v - \dot{\vec r}_j] \cdot \vec \nabla ) \vec v_j .
\label{eq:Euler-ansatz}
\ee
The terms in the right-hand sides are dominated by the neighborhoods of the vortices, $\vec r \simeq \vec r_j$.
Therefore, the continuity and Euler equations are satisfied at leading order for
\be
\dot{\vec r}_{j} = \sum_{j' \neq j} \vec v_{j'} ( \vec r_j) .
\ee
Here we used the fact that the component $\vec v_j$ of $\vec v$ in the $j$-term in the right-hand side in
Eqs.(\ref{eq:continuity-ansatz}) does not contribute, as $\vec\nabla f^2(\vec r - \vec r_j) \cdot \vec v_j = 0$,
whereas its contribution in Eq.(\ref{eq:Euler-ansatz}),
$- \epsilon^2 s_j^2 (\vec r - \vec r_j) /| \vec r - \vec r_j|^4$
vanishes if we integrate over angles over a small region centered on $\vec r_j$.
Thus, each vortex follows the flow generated by the other vortices
\cite{Fetter-1966,CRESWICK1980,LUND1991}.

\subsubsection{Effective action}
\label{sec:effective-action}

A more general and elegant approach is to substitute our ansatz into the
action (\ref{eq:action-hydro}).
Thus, we write the wave function as
\be
\psi(\vec r,t) = \sqrt{\rho} e^{i s} \prod_{j=1}^N e^{i \sigma_j \theta_j} ,
\label{eq:psi-ansatz-rho-s}
\ee
where again
\be
\theta_j(\vec r) = \widehat{(\vec e_x, \vec r - \vec r_j)} , \;\;\; \vec\nabla \theta_j
= \vec e_z \times \vec\nabla \ln | \vec r - \vec r_j | .
\ee
Here $\rho(\vec r,t)$ and $s(\vec r,t)$ are smooth functions and we neglect the width of the vortices
and their impact on the density, only keeping track of their large-scale effect on the velocity.
When there are no vortices we recover the usual Madelung expression (\ref{eq:Madelung}).
Thus, the total phase $S$ and velocity $\vec v$ read
\be
S = s + \sum_{j=1}^N \sigma_j \theta_j , \;\;\;
\vec v = \epsilon \vec\nabla s + \sum_{j=1}^N \vec v_j ,
\label{eq:S-s-theta-j}
\ee
which defines $\vec v$.
Substituting into the action (\ref{eq:action-hydro}) gives
\ba
{\cal S}[\rho,s,\vec r_j] & \!\! = \!\! & \!\! \int \!\! d\vec r dt \biggl \lbrace - \epsilon \rho
\biggl[ \frac{\partial s}{\partial t} - \sum_j \sigma_j \vec\nabla \theta_j \cdot \dot{\vec r}_j \biggl]
- \frac{\epsilon^2}{8\rho} ( \vec\nabla \rho)^2  \nonumber \\
&& - \frac{1}{2} \rho ({\vec v})^2  - \rho \Phi_N - {\cal V}_I \biggl \rbrace  .
\label{eq:action-vortices}
\ea
The variation of the action with respect to $s$ gives back the continuity equation (\ref{eq:continuity}).
The variation with respect to $\rho$ gives the modified Hamilton-Jacobi equation
\be
\epsilon \frac{\partial s}{\partial t} - \sum_j \vec v_j \cdot \dot{\vec r}_j + \frac{(\vec v)^2}{2}
= - \Phi_Q - \Phi_N - \Phi_I .
\ee
Taking its gradient gives the modified Euler equation
\be
\frac{\partial\vec v}{\partial t} + (\vec v \cdot \vec\nabla) \vec v + \sum_j ( \vec v - \dot{\vec r}_j) \times
\vec\omega_j = - \vec\nabla( \Phi_Q + \Phi_N + \Phi_I ) ,
\label{eq:Euler-omega}
\ee
where we used
\be
\vec\omega = \vec\nabla \times \vec v = \sum_j \vec\omega_j ,
\ee
as we assumed that the component $s$ of the phase is regular.
The last term in the left-hand side in Eq.(\ref{eq:Euler-omega}) is singular, as the vorticities
contain Dirac distributions, $\vec \omega_j \propto \delta_D^{(2)}(\vec r - \vec r_j) \vec e_z$,
whereas the other terms do not contain Dirac distributions by assumption.
Requiring that both the singular and regular parts vanish gives both the Euler equation (\ref{eq:Euler})
and the equations
\be
\dot{\vec r}_j = \vec v(\vec r_j) .
\label{eq:dot-rj-v}
\ee
Here the contribution of the vortex $j$ to its velocity is taken to vanish. This agrees with a
coarse-graining on a small region centered on $\vec r_j$ as we do not model the internal structure
of the vortices in the ansatz (\ref{eq:psi-ansatz-rho-s}).
This also agrees with the fact that a vortex does not move by itself, as seen in the single
vortex solution described in Sec.~\ref{sec:single-vortex}.
Thus, we recover the property that vortices follow the matter flow as in classical hydrodynamics
of ideal fluids, which obey Kelvin's circulation theorem \cite{Batchelor_2000}.

Finally, the variation of the action with respect to $\vec r_j$ gives
\be
\int d \vec r \left[ - \frac{\partial}{\partial t} ( \rho \vec v_j )
- \rho ( \dot{\vec r}_j \cdot \vec\nabla ) \vec v_j
+ \rho ( \vec v \cdot \vec\nabla ) \vec v_j \right] = 0 .
\ee
Using the continuity equation (\ref{eq:continuity}) and integration by parts we can see that these
equations are automatically satisfied.

An advantage over the approach presented in Sec.~\ref{sec:Creswick} is that, as expected,
we recover in addition that the velocity field also obeys the Euler equation (\ref{eq:Euler}),
sourced by the continuous forces associated with the quantum pressure, the gravitational potential,
and the self-interaction effective pressure.

The key difference with the hydrodynamical equations presented in Sec.~\ref{sec:hydro}
is that we no longer have $\vec\nabla \times \vec v = 0$.
This is because the phase $S$ in (\ref{eq:S-s-theta-j}) is singular, so that $\vec v = \vec\nabla S$
no longer implies $\vec\nabla \times \vec v = 0$, as explicitly seen in Sec.~\ref{sec:single-vortex}
for a single vortex.
Thus, the velocity field still evolves according to the Euler equation but it is no longer curl-free.
The system remains described by standard hydrodynamical equations but its velocity field now
obtains all degrees of freedom of hydrodynamical flows, containing both potential and rotational
components.

\section{Continuum limit}
\label{sec:continuum}

\subsection{Rotating soliton}
\label{sec:rotating-soliton}

Whereas the static soliton (\ref{eq:rho-TF-0}) was a minimum of the energy (\ref{eq:E-rho-v}) at fixed
mass, we now look for solutions at fixed mass and angular momentum to obtain rotating solitons.
Indeed, both mass and angular momentum are conserved by the Gross-Pitaevskii dynamics.
Because the velocity (\ref{eq:Madelung}) is the gradient of a phase, as in superfluids, the system
cannot develop rotation and vorticity in a regular manner.
Instead, as pointed out by Feynman \cite{FEYNMAN1955},
to accommodate rotation the system  develops singularities,
the vortices that carry the vorticity at discrete locations as in Eq.(\ref{eq:omega-j}).
As seen in Sec.~\ref{sec:effective-action}, the density and velocity field remain governed by the
continuity and Euler equations, but the velocity field now includes a singular rotational component.

In the limit $\epsilon \to 0$, each vortex (\ref{eq:omega-j}) carries an infinitesimal vorticity,
so that a fixed amount of total rotation (i.e., angular momentum) requires a number of vortices
that grows without bound as $1/\epsilon$.
In the continuum limit, the discrete set of vortices becomes irrelevant and we simply have
a smooth vorticity field $\vec\omega$ that can be nonzero at all points in space.
Therefore, we still have the hydrodynamical equations of motion (\ref{eq:continuity})-(\ref{eq:Euler})
and the energy functional (\ref{eq:E-rho-v}) but the velocity field is now free to include
a rotational component.

Thus, we look for configurations where the first variation
\be
\delta^{(1)} \left( E - \mu M - \Omega L_z \right) = 0  ,
\label{eq:delta1-E-M-Lz}
\ee
vanishes, where $M$ and $L_z$ are the total mass and angular momentum,
\be
M = \int d\vec r \, \rho , \;\;\; L_z = \int d\vec r \, \rho ( x v_y - y v_x ) ,
\ee
and $\mu$ and $\Omega$ are Lagrange multipliers.
In agreement with the limit $\epsilon \to 0$, we consider the Thomas-Fermi regime where the quantum
pressure is negligible, that is, we neglect the term
$\epsilon^2 (\vec\nabla \sqrt\rho)^2/2$ in the energy (\ref{eq:E-rho-v}).
Taking first the variations (\ref{eq:delta1-E-M-Lz}) with respect to $\delta v_x$ and $\delta v_y$
gives
\be
v_x = - \Omega y, \;\; v_y = \Omega x, \;\; \mbox{hence} \;\; \vec v = r \Omega \, \vec e_{\theta} .
\label{eq:v-solid-rotation}
\ee
This is a solid-body rotation at rate $\dot\theta = \Omega$.
Substituting this velocity field and taking the variation (\ref{eq:delta1-E-M-Lz}) with respect to
$\rho$ gives
\be
\Phi_N + \Phi_I - \frac{r^2\Omega^2}{2} = \mu .
\label{eq:mu-Omega}
\ee
This is the equation (\ref{eq:TF-static}) obtained in the static case with the addition of the
term $r^2\Omega^2/2$ due to the rotation.
Taking the Laplacian and solving the linear second-order differential equation over $\rho(r)$,
we find that the static density profile (\ref{eq:rho-TF-0}) is modified into
\be
\rho_{\rm TF,\Omega}(r) = \left( \rho_{0} - \frac{\Omega^2}{2\pi} \right) J_0( z_0 r/R_0)
+ \frac{\Omega^2}{2\pi} .
\label{eq:rho-TF-Omega}
\ee

The radius $R_\Omega$ of the soliton corresponds to the first zero crossing of the density profile
(\ref{eq:rho-TF-Omega}). At first order in $\Omega^2$ we obtain at fixed $\rho_0$
\be
R_{\Omega} = R_0 + \frac{\Omega^2}{2\pi\rho_0} \frac{R_0}{z_0 J_1(z_0)} + \cdots
\ee
As expected, we find that the rotation flattens and expands the soliton, because of the
centrifugal force \cite{Fetter_2001}.
The mass and angular momentum within a radius $0\leq r \leq R_{\Omega}$ are given by
\be
M(<r) = \frac{r}{2} \left[ r \Omega^2 + \frac{2 R_0 (2\pi\rho_0 - \Omega^2)}{z_0} J_1(z_0 r/R_0) \right]
\ee
and
\ba
L_z(<r) & = & \frac{r^4 \Omega^3}{4} + \frac{r^2 R_0 \Omega (2\pi\rho_0 - \Omega^2)}{z_0^2} \nonumber \\
&& \times \left[ 2 R_0 J_2(z_0 r/R_0) - r z_0 J_3(z_0 r/R_0)  \right] . \;\;\;
\label{eq:Lz-r}
\ea
For the mass of the system to be finite, the expression (\ref{eq:rho-TF-Omega}) must vanish
before or at the first minimum of the Bessel function. Otherwise, the density will remain strictly
positive at all radii, giving an infinite mass. Using $J_0'(z)=-J_1(z)$, we can see that the first
minimum of expression (\ref{eq:rho-TF-Omega}) occurs at $r_1=z_1 R_0/z_0$, where $z_1$ is the
first zero of the first-order Bessel function, $J_1(z_1)=0$ and $z_1 \simeq 3.832$.
Therefore, the mass is finite provided Eq.(\ref{eq:rho-TF-Omega}) is negative at $r=r_1$.
This gives the condition
\be
| \Omega | \leq \Omega_{\max} , \;\;
\Omega_{\max} = \sqrt{ \frac{2 \pi J_0(z_1) \rho_0}{J_0(z_1)-1} } \simeq 1.343 \sqrt{\rho_0} .
\label{eq:bound-Omega}
\ee
Thus, solitons of a given central density can only support rotation rates below $\Omega_{\max}$,
with a radius below $R_{\max}$ with
\be
R_{\max} = z_1 R_0/z_0 \simeq 1.593 \, R_0 .
\label{eq:R-max}
\ee
While $\Omega_{\max}$ grows with $\rho_0$ the radius $R_{\max}$ is independent of $\rho_0$.
Thus, independently of their mass and rotation rate, all soliton radii fall in the finite
range $R_0 \leq R \leq R_{\max}$.
The maximum angular momentum associated with $\Omega_{\max}$ and $R_{\max}$ reads
\be
L_{z,\max} \simeq 1.78 \, \rho_0^{3/2} R_0^4 \simeq 0.51 \, M^{3/2} R_0 .
\label{eq:Lz-max}
\ee

\subsection{Dynamical stability}

The solitons (\ref{eq:rho-TF-Omega}) are stable if they correspond to a minimum of the energy
(i.e., not merely a saddle-point or a maximum).
Thus, we need to show that the second variation of the energy is positive.
The linear variation with respect to $\delta \vec v$ and $\delta\rho$ around the soliton equilibrium
$(\rho_0,\vec v_0)$ vanishes, from its definition (\ref{eq:delta1-E-M-Lz}).
The quadratic variation of the energy reads
\be
\delta^{(2)}E = \int d {\vec r} \left[ \frac{1}{2} \rho_0 (\delta \vec v)^2
+ \frac{1}{2} \delta\rho \delta\Phi_N + \frac{\lambda}{2} \delta\rho^2 \right] ,
\label{eq:d2E}
\ee
where we used that we consider variations at fixed angular momentum.
For this quantity to be positive it is sufficient to have
$(1/2) \int d {\vec r} \, [ \delta\rho \delta\Phi_N + \lambda \delta\rho^2 ] \geq 0$.
Thus, the quadratic form $(1/2) \int d {\vec r}d {\vec r}^{\,'}  K({\vec r},{\vec r}^{\,'})
\delta\rho({\vec r}) \delta\rho({\vec r}^{\,'})$ must be positive, where the symmetric
operator $K$ reads
\be
K \cdot \delta\rho = 4\pi \Delta^{-1} \delta\rho + \lambda .
\ee
Therefore, it is sufficient to check that all eigenvalues of the operator $K$ are positive.
Looking for eigenvectors of the form $\delta\rho = f(r) e^{i\ell\theta}$, we obtain for the
eigenvalue problem $K \cdot \delta\rho = \nu \delta\rho$ the differential equation
\be
\frac{d^2f}{dr^2} + \frac{1}{r} \frac{df}{dr} + \left( \kappa - \frac{\ell^2}{r^2} \right) f = 0 ,
\;\; \kappa = \frac{4\pi}{\lambda-\nu} .
\label{eq:stability-ODE}
\ee
The soliton is stable if $\nu \geq 0$, that is, $\kappa > 4\pi/\lambda$ or $\kappa  < 0$.
The instability appears when $\nu=0$, that is, $\kappa= 4\pi/\lambda$.
The solutions of Eq.(\ref{eq:stability-ODE}) that are regular at the origin are
$f(r) = J_{\ell}(\sqrt{\kappa} r)$.

Let us first consider the modes $\ell \geq 1$.
These modes automatically conserve the total mass, $\delta M=0$, by integration over the polar angle.
We can impose the boundary condition $\delta\rho(R)=0$ at a radius $R>R_{\Omega}$ beyond the soliton
radius, where $\rho_0$ identically vanishes.
Then, the eigenvectors are $J_{\ell}\left(\sqrt{\kappa_n^{(\ell)}}r\right)$ with
$\sqrt{\kappa_n^{(\ell)}} R = x_n^{(\ell)}$,
where $x_n^{(\ell)}$ is the n$^{\rm th}$ zero of $J_{\ell}(x)$.
Such a mode is stable if $\kappa_n^{(\ell)} > 4\pi/\lambda$.
From the ordering of the zeros of the Bessel functions, we find that the soliton is stable
with respect to all these modes provided $(z_1/R)^2 > 4\pi/\lambda$, which gives the condition
\be
R_{\Omega} < z_1 \sqrt{\frac{\lambda}{4\pi}} = \frac{z_1}{z_0} R_0 .
\label{eq:stable-ell-gtr-1}
\ee

Let us now consider the modes $\ell=0$, $f(r) = J_0(\sqrt{\kappa} r)$.
This gives the mass perturbation $\delta M(<r) = 2\pi r J_1(\sqrt{\kappa} r)/\sqrt{\kappa}$.
We now impose the boundary condition $\delta M(<R) = 0$ for a given radius $R$ beyond the soliton radius,
as the mass of the system is constant. This gives the eigenvectors $\sqrt{\kappa_n^{(0)}} R = x_1^{(n)}$
and we find again that all these modes are stable provided the condition (\ref{eq:stable-ell-gtr-1})
is satisfied.
Comparing with the upper bound (\ref{eq:R-max}), we can see that all solitons of radius
$R_{\Omega} < R_{\max}$, that is, with a rotation rate $|\Omega| < \Omega_{\max}$, are dynamically
stable.

Our results are consistent with the analysis of Ref.\cite{Rindler-Daller:2011afd}, who found
that in rotating ellipsoids the formation of a vortex is energetically favored in the Thomas-Fermi
regime. In the deep Thomas-Fermi regime that we consider in this paper many vortices form,
in agreement with Ref.\cite{RINDLERDALLER2008}.
This allows us to perform a continuum analysis. As we shall check in Sec.~\ref{sec:simulations},
our numerical simulations show that the continuum limit provides a good approximation as soon
as $\epsilon \lesssim 0.03$.

\subsection{Uniform density of vortices}

Within the soliton (\ref{eq:rho-TF-Omega}) of solid-body rotation rate $\Omega$, the circulation
$\Gamma(r)$ along the circle of radius $r$ reads
\be
\Gamma(r) = \oint_{\cal C} \vec v \cdot \vec{d\ell} = 2\pi r^2 \Omega .
\ee
On the other hand, as each vortex (\ref{eq:vorticity-vortex}) carries a vorticity quantum
$2\pi\epsilon\sigma$, the circulation also reads
\be
\Gamma(r) = \int_S \vec \omega \cdot \vec{dS} = 2 \pi \epsilon N_v(<r) ,
\ee
where $N_v(<r)$ is the total number of vortices within radius $r$, weighted by their spin,
$N_v = \sum_\sigma \sigma N_{v,\sigma}$.
This gives
\be
N_v(<r) = \frac{r^2\Omega}{\epsilon} , \;\;\; n_v(\vec r) = \frac{\Omega}{\pi \epsilon} ,
\label{eq:vortex-density}
\ee
hence a constant vortex number density $n_v(\vec r)$.
We can also directly obtain this constant number density without assuming axisymmetry.
From the vortex number density
\be
n_v(\vec r) = \sum_j \sigma_j \delta_D^{(2)}(\vec r - \vec r_j) ,
\ee
we obtain the velocity field as
\be
\vec v = \epsilon \vec e_z \times \int d\vec r^{\,'} n_v(\vec r^{\,'}) \vec\nabla
\ln |\vec r - \vec r^{\,'}| .
\ee
This can be inverted as
\be
\nabla \cdot ( \vec e_z \times \vec v ) = - 2 \pi \epsilon \, n(\vec r) ,
\ee
and substituting the solid-body rotation velocity (\ref{eq:v-solid-rotation}) we obtain the constant vortex
density (\ref{eq:vortex-density}) in the continuum limit.
Thus, as pointed out by Feynman \cite{FEYNMAN1955} for rotating superfluid $^4{\rm He}$,
a uniform lattice of vortices develops to mimic a solid-body rotation \cite{Fetter-2009}.

Since $\nabla \cdot \vec v=0$, the velocity field (\ref{eq:v-solid-rotation}) inside the soliton
is entirely due to the vortices. The smooth phase component $s$ in Eq.(\ref{eq:psi-ansatz-rho-s})
is uniform, such as $s=-\mu t/\epsilon$ with a constant $\mu$, and $\vec\nabla s = 0$.

From the vortex number density (\ref{eq:vortex-density}) we obtain the typical distance $d$
between vortices,
\be
d = \sqrt{\pi\epsilon/\Omega} ,
\label{eq:d-Omega-eps}
\ee
which does not depend on the system size.
Thus, in the semi-classical limit $\epsilon\to 0$ the healing length introduced in
Eq.(\ref{eq:healing-length}) decreases much faster than the inter-vortex distance $d$.
This means that our assumption of well-separated vortices, where we can neglect the internal
structure of the vortices, is well justified in the limit $\epsilon\to 0$ that we consider
in this paper.

\subsection{Comparison with angular momentum eigenstates}

It is sometimes proposed to extend the static spherically symmetric soliton presented in
Sec.~\ref{sec:static-soliton} to rotating configurations by looking for eigenstates of
the Schr\"odinger equation of the form
\be
\psi_{\ell}(\vec x,t) = e^{-i\mu t/\epsilon} f(r) e^{i\ell\theta} , \;\;\;
v_\theta = \frac{\epsilon \ell}{r} ,
\label{eq:orbital-eigenstate}
\ee
as for the single vortex state (\ref{eq:single-vortex-psi}) but with a large orbital
quantum number, $\ell \sim 1/\epsilon$, to generate a macroscopic angular momentum.
(In 3D this corresponds to expansions over the spherical harmonics,
$\psi_{\ell m}(\vec x,t) = e^{-i\mu t/\epsilon} f(r) Y_{\ell}^m(\theta,\varphi)$
\cite{Hertzberg:2018lmt,Guzman:2019gqc} and one can also consider combinations
of various eigenstates.)
Substituting in the Gross-Pitaevskii equation (\ref{eq:Schrod-eps}) and neglecting the
radial derivative in the Thomas-Fermi regime, we obtain
\be
\Phi_N + \Phi_I + \frac{\epsilon^2\ell^2}{2r^2} = \mu .
\ee
Here we keep the leading-order angular derivative because $\ell$ can be large.
This is the generalization of the hydrostatic equilibrium (\ref{eq:hydrostatic-full})
to the case of nonzero angular momentum, $l_z=r v_\theta=\epsilon \ell$.
Instead of the regular quadratic term obtained in Eq.(\ref{eq:mu-Omega}), associated with
the solid-body rotation generated by a vortex lattice, the angular term $\ell^2/r^2$ now gives rise
to an orbital angular momentum barrier.
In this Thomas-Fermi limit, this means that the density vanishes close to the origin
and the soliton takes the shape of a ring, with finite nonzero minimum and maximum
radii $r_{\min} < r < r_{\max}$.
In the case of attractive self-interactions, as for axions, it has been proposed
\cite{Banik:2013rxa} that such configurations could explain the existence of dark matter caustic
rings as suggested by some observations.
This is further motivated by the expectation that for attractive self-interactions vortices
merge to form a single big vortex at the center of the galaxy \cite{Banik:2013rxa}.
In this paper we consider instead repulsive self-interactions, as in dilute gas BEC,
and as in superfluid experiments we will find in our numerical simulations that the vortices
arrange on a regular lattice to generate a solid-body rotation, as in (\ref{eq:mu-Omega}),
and a soliton density profile that peaks at the center, instead of the large-$\ell$
eigenstate (\ref{eq:orbital-eigenstate}).

In the case $\epsilon \sim 1$, where a single vortex would have a galactic size,
the model (\ref{eq:orbital-eigenstate}) with $\ell=1$ could contribute to the
rotation curve of the galaxy \cite{Boehmer:2007um,Brook:2009ku,Kain:2010rb}.
However, this is not the regime that we consider in this paper.

\section{Numerical simulations}
\label{sec:simulations}

\subsection{Initial conditions and simulation setup}

\begin{figure*}
\centering
\includegraphics[height=5cm,width=0.3\textwidth]{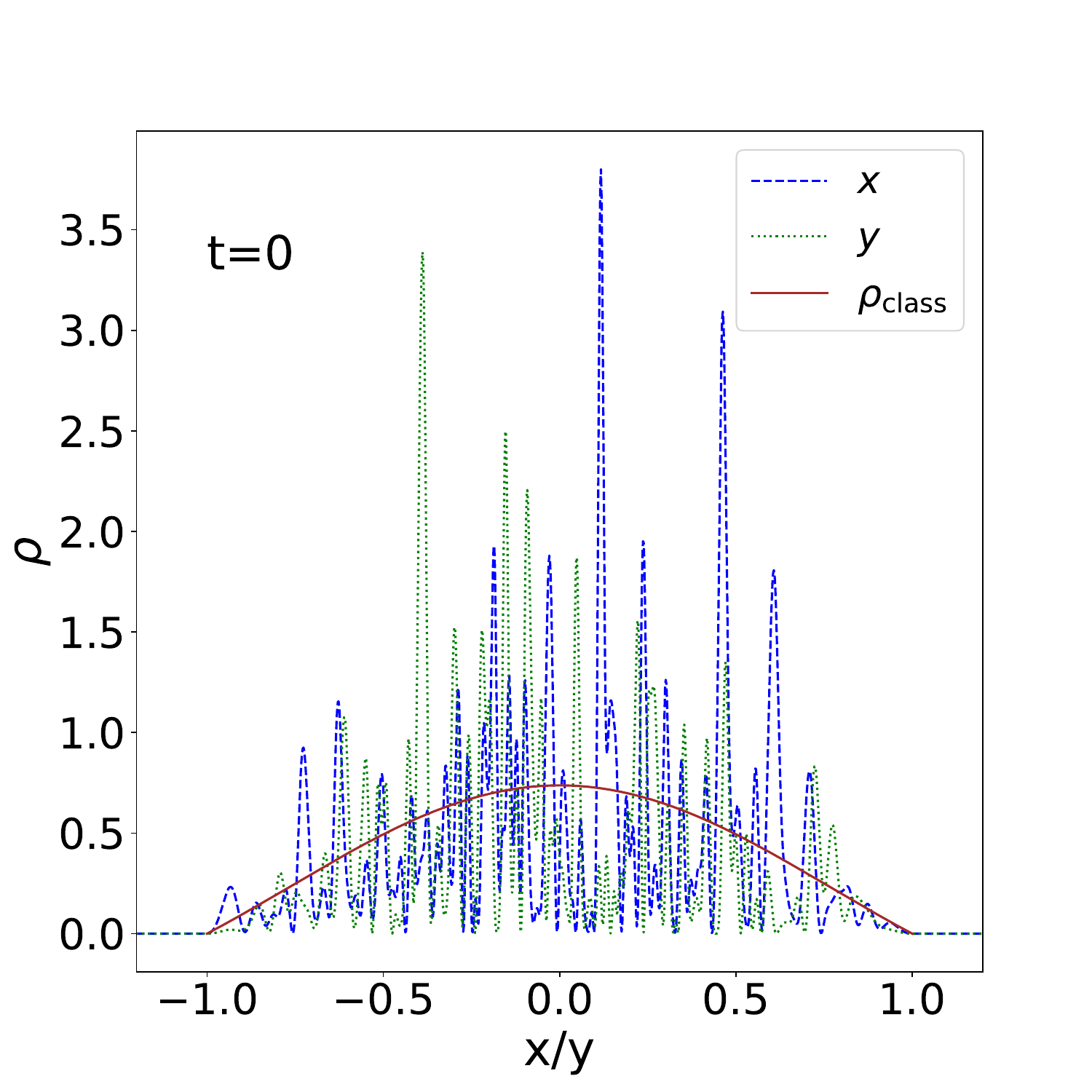}
\includegraphics[height=5cm,width=0.34\textwidth]{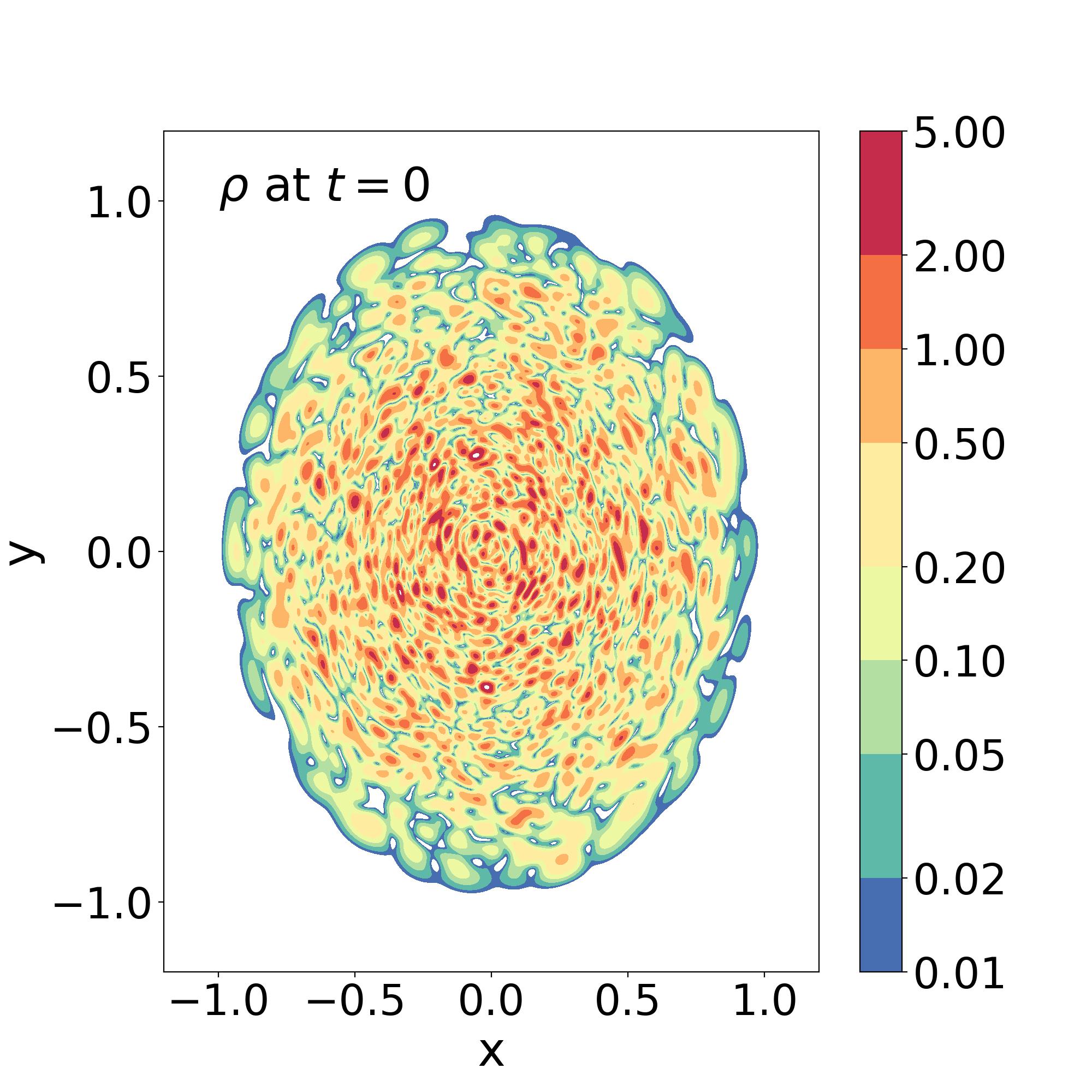}
\includegraphics[height=5cm,width=0.34\textwidth]{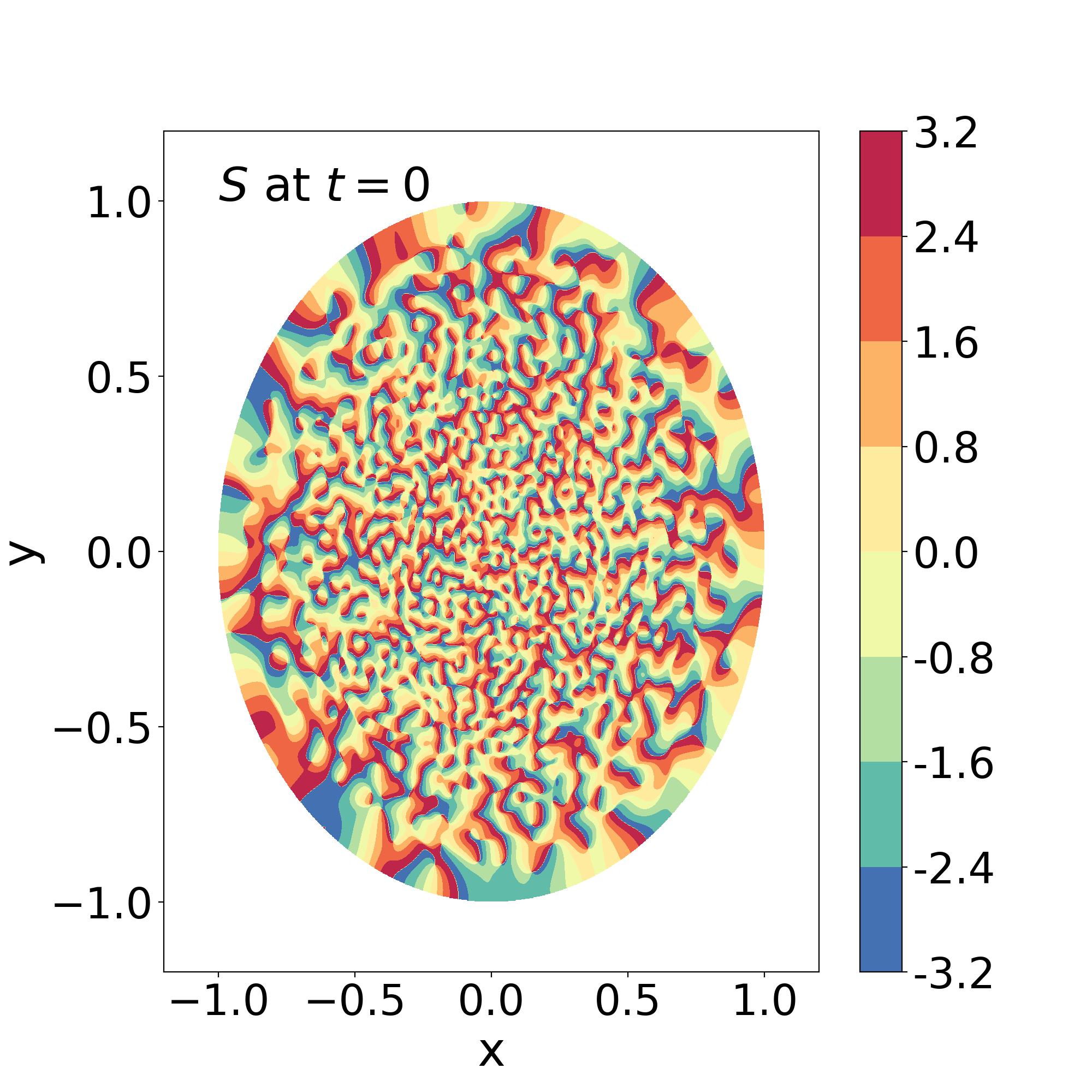}
\caption{
Initial condition of our simulations for the case $[\epsilon=0.01, \alpha=1]$.
{\it Left panel:} density profiles along the $x$ and $y$ axis (dashed and dotted lines with large
fluctuations) and classical density profile $\rho_{\rm class}$ (smooth brown line)
of Eq.(\ref{eq:rho-class}).
{\it Middle panel:} 2D map of the density $\rho(x,y)$.
{\it Right panel:} 2D map of the phase $S(x,y)$, defined over $]-\pi,\pi]$.
}
\label{fig:initial}
\end{figure*}

\subsubsection{Expansion over eigenfunctions}
\label{sec:eigenfunctions}

As in \cite{Garcia:2023abs,GalazoGarcia:2024fzq},
we start our simulations without a central soliton since we are interested in the formation
of the solitons and their generic properties.
Thus, we start with stochastic initial conditions associated with a collisionless virialized halo in the
semiclassical limit.
We choose a target classical density profile of the same form as Eq.(\ref{eq:rho-TF-0}),
\be
0 < r < R : \;\;\; \rho_{\rm class}(r) = \rho_{0} J_0( z_0 r/R)  ,
\label{eq:rho-class}
\ee
where $R$ is the halo radius, which is greater than the expected soliton radius.
This is a simple model for compact halos with a flat core and total mass
\be
M = \rho_0 2 \pi R^2 J_1(z_0)/z_0 .
\label{eq:M-halo}
\ee
The associated gravitational potential reads
\be
\Phi_N(r) = \Phi_{N0} J_0( z_0 r/R)  , \;\;\; \Phi_{N0} = -4\pi \rho_0 R^2/z_0^2 .
\label{eq:PhiN-class}
\ee
Then, we take for the initial wave function \citep{Lin:2018whl,Yavetz:2021pbc,Garcia:2023abs}
a sum with random coefficients $a_{n\ell}$ over the eigenmodes
$\hat \psi_{n\ell}(\vec r)$ of the Schr\"odinger equation defined by this target gravitational
potential $\Phi_{N}$,
\be
\psi(\vec r) = \sum_{n\ell} a_{n\ell} \hat\psi_{n \ell}(\vec r) , \;\;
\hat\psi_{n \ell}(\vec r) = {\cal R}_{n | \ell |}(r) e^{i\ell\theta} ,
\label{eq:psi-halo-a_nl}
\ee
where the radial parts satisfy the radial Schr\"odinger equation
\be
\left[ - \frac{\epsilon^2}{2} \frac{1}{r} \frac{d}{dr} \left( r \frac{d}{dr} \right) + \frac{\epsilon^2 \ell^2}{2 r^2}
+ \Phi_N \right] {\cal R}_{n | \ell |} = E_{n | \ell |} {\cal R}_{n | \ell |}
\ee
and obey the normalization conditions
\be
\int_0^{\infty} dr r \, {\cal R}_{n_1 | \ell |} {\cal R}_{n_2 | \ell |} = \delta_{n_1,n_2} .
\ee
The coefficients $a_{n\ell}$ have deterministic amplitudes but random phases $\Theta_{n\ell}$
that are uncorrelated with a uniform distribution over $0 \leq \Theta < 2\pi$,
\be
a_{n\ell} = | a_{n\ell} | e^{i\Theta_{n\ell}} , \;\;\; \langle a_{n_1\ell_1}^{\star} a_{n_2\ell_2} \rangle
= | a_{n_1\ell_1} |^2 \delta_{n_1,n_2} \delta_{\ell_1,\ell_2} ,
\ee
where the statistical average $\langle\dots\rangle$ is taken over the random phases
$\Theta_{n\ell}$.
The mean density is then
\be
\langle \rho \rangle = \langle |\psi|^2 \rangle = \sum_{n\ell} |a_{n\ell}|^2 |\hat\psi_{n \ell}|^2
\ee
and its variance reads
\be
\langle ( \rho - \langle \rho \rangle )^2 \rangle = \langle \rho \rangle^2 ,
\label{eq:rho-variance}
\ee
which shows that it displays large random fluctuations.

Using the WKB approximation and the change of variable $L=\epsilon\ell$, we obtain in the continuum limit
\be
\langle \rho \rangle = \frac{1}{\pi\epsilon^2} \int dE dL \frac{| a_{n\ell} |^2}
{\sqrt{2 r^2 (E-\Phi_N) - L^2}} ,
\label{eq:rho-average}
\ee
where $E$ and $L$ are the energy and angular momentum.
On the other hand, from the expression of the current associated with the wave function $\psi$,
\be
\vec J = \rho \vec v = \frac{i\epsilon}{2} ( \psi \vec\nabla \psi^\star - \psi^\star \vec\nabla \psi )
\ee
we obtain $\langle \rho v_r \rangle = 0$ and
\be
\langle \rho v_{\theta} \rangle = \frac{1}{\pi\epsilon^2} \int dE dL
\frac{| a_{n\ell} |^2}{\sqrt{2 r^2 (E-\Phi_N) - L^2}} \frac{L}{r} .
\label{eq:Jtheta-average}
\ee

For a classical system of collisionless particles, we have
\be
E = \frac{\vec v^{\,2}}{2} + \Phi_N , \;\;\; L = r v_\theta ,
\ee
\be
\rho_{\rm class} = 2 \int dE dL \frac{f(E,L)}{\sqrt{2 r^2 (E-\Phi_N) - L^2}} ,
\label{eq:rho-f}
\ee
and
\be
\rho_{\rm class} v_{\theta} = 2 \int dE dL \frac{f(E,L)}{\sqrt{2 r^2 (E-\Phi_N) - L^2}} \frac{L}{r} ,
\label{eq:Jtheta-f}
\ee
where $f(E,L)$ is the classical phase-space distribution.
Comparing Eqs.(\ref{eq:rho-average})-(\ref{eq:Jtheta-average}) with
Eqs.(\ref{eq:rho-f})-(\ref{eq:Jtheta-f}), we can see that we recover the classical target density
and angular velocity if we take for the coefficients $a_{n\ell}$ the amplitude
\be
| a_{n\ell} |^2 = 2\pi\epsilon^2 f(E,L) .
\label{eq:anl-f}
\ee

\subsubsection{Isotropic initial conditions}

For isotropic initial conditions the classical phase-space distribution does not depend
on the angular momentum,
\be
f(E,L) = f_0(E) .
\ee
Then, we obtain
\be
\rho = 2\pi \int_{\Phi_N}^0 dE \, f_0(E) ,
\label{eq:rho-f0E}
\ee
which can be inverted to give the isotropic distribution function
\be
\Phi_{N0} < E < 0 : \;\;\; f_0(E) = \frac{z_0^2}{8 \pi^2 R^2} .
\label{eq:f0-isotropic}
\ee
Thus, the distribution function is a constant, which depends on the halo radius but not on its
density or its mass. The dependence on the mass arises through the lower bound $\Phi_{N0}$
of its support $\Phi_{N0} < E < 0$, and from Eq.(\ref{eq:rho-f0E}) we recover
Eqs.(\ref{eq:rho-class}) and (\ref{eq:PhiN-class}).

\subsubsection{Anisotropic initial conditions}

For anisotropic initial conditions the phase-space distribution depends on $L$.
In particular, the mean angular rotation becomes nonzero if $f(E,L)$ depends on the
sign of $L$. In this paper we take the simple choice
\be
f(E,L) = f_0(E) [ 1 + \alpha \, {\rm sign}(L) ] , \;\;\; -1 \leq \alpha \leq 1 ,
\label{eq:f-EL-mu}
\ee
where $f_0(E)$ is the isotropic distribution (\ref{eq:f0-isotropic}).
This recovers the same target density (\ref{eq:rho-class}) but with a different proportion
of particles with positive and negative angular momentum.
If $\alpha =\pm 1$ we only keep the particles that have a positive/negative angular momentum.
From Eq.(\ref{eq:Jtheta-f}) we obtain the mean angular velocity
\be
v_\theta = \alpha \frac{4}{3\pi} \sqrt{-2\Phi_N} .
\label{eq:v-theta-r-class}
\ee
This gives for the initial angular momentum of the system
\be
L_{z,\rm init} \simeq 0.27 \alpha M^{3/2} R .
\label{eq:Lz-init}
\ee

\subsubsection{Simulation parameters}

In this paper we consider the cases $\epsilon=0.005, 0.01$ and $0.03$, as we focus on the
semi-classical regime, and we mostly illustrate our results with the intermediate case $\epsilon=0.01$.
For the initial wave function we take $R=1$ and $M=1$ for the target halo radius and mass
in Eqs.(\ref{eq:rho-class})-(\ref{eq:M-halo}).
For the anisotropy parameter of Eq.(\ref{eq:f-EL-mu}) we consider the cases
$\alpha=0, 0.5$ and $1$.

We show in Fig.~\ref{fig:initial} our initial condition for the case $\epsilon=0.01$
and $\alpha=1$.
As in \cite{Garcia:2023abs,GalazoGarcia:2024fzq}, where we studied 3D isotropic systems,
the initial density field shows strong fluctuations of order unity around the target
classical density (\ref{eq:rho-class}), as seen in the left two panels and in agreement
with Eq.(\ref{eq:rho-variance}).
The spatial width of the fluctuations is set by the de Broglie wavelength (\ref{eq:de-Broglie-rescaled}).
From the wave function $\psi$ we also obtain the phase $S$ as in Eq.(\ref{eq:Madelung}),
which we define in the interval $]-\pi,\pi]$.
It again shows strong fluctuations, on the same scale as the density.
The interferences between the many modes in the sum (\ref{eq:psi-halo-a_nl}) give rise to many
points inside the halo where the amplitude $|\psi|$ vanishes and the phase $S$ is ill-defined.
They typically correspond to a vortex of spin $\sigma= \pm 1$ \citep{Hui:2020hbq},
where the phase is singular as it rotates by a multiple of $2\pi$ in a small circle around
that point.

Even though locally we can always perform the Madelung transform (\ref{eq:Madelung})
to go to the hydrodynamical picture, within any region that does not encircle a singularity,
the fast fluctuations of the phase, associated with large gradients for the local velocity,
mean that the hydrodynamical picture is meaningless. Instead, the system mimics
a gas of collisionless particles, as seen from the construction in Sec.~\ref{sec:eigenfunctions}.
In the regime where the quartic self-interactions are negligible (e.g., in the outer halo outside
the central soliton) the dynamics must be described by the Vlasov equation rather than
hydrodynamical equations
\cite{Widrow:1993qq,Mocz:2018ium,GalazoGarcia:2022nqv,Garcia:2023abs,Liu:2024pjg}.

We take for the self-interaction coupling constant $\lambda$ in Eq.(\ref{eq:Poisson-eps})
the value associated with a static soliton radius $R_0=0.5$ in Eq.(\ref{eq:rho-TF-0}).
Thus, the halo will typically collapse to form a soliton, with a radius that is one half
of the initial halo, embedded within a remaining virialized envelope made of many excited states
as in (\ref{eq:psi-halo-a_nl}).

\subsection{Numerical algorithm}
\label{sec:numerical-method}

\begin{figure*}
\centering
\includegraphics[height=5cm,width=0.3\textwidth]{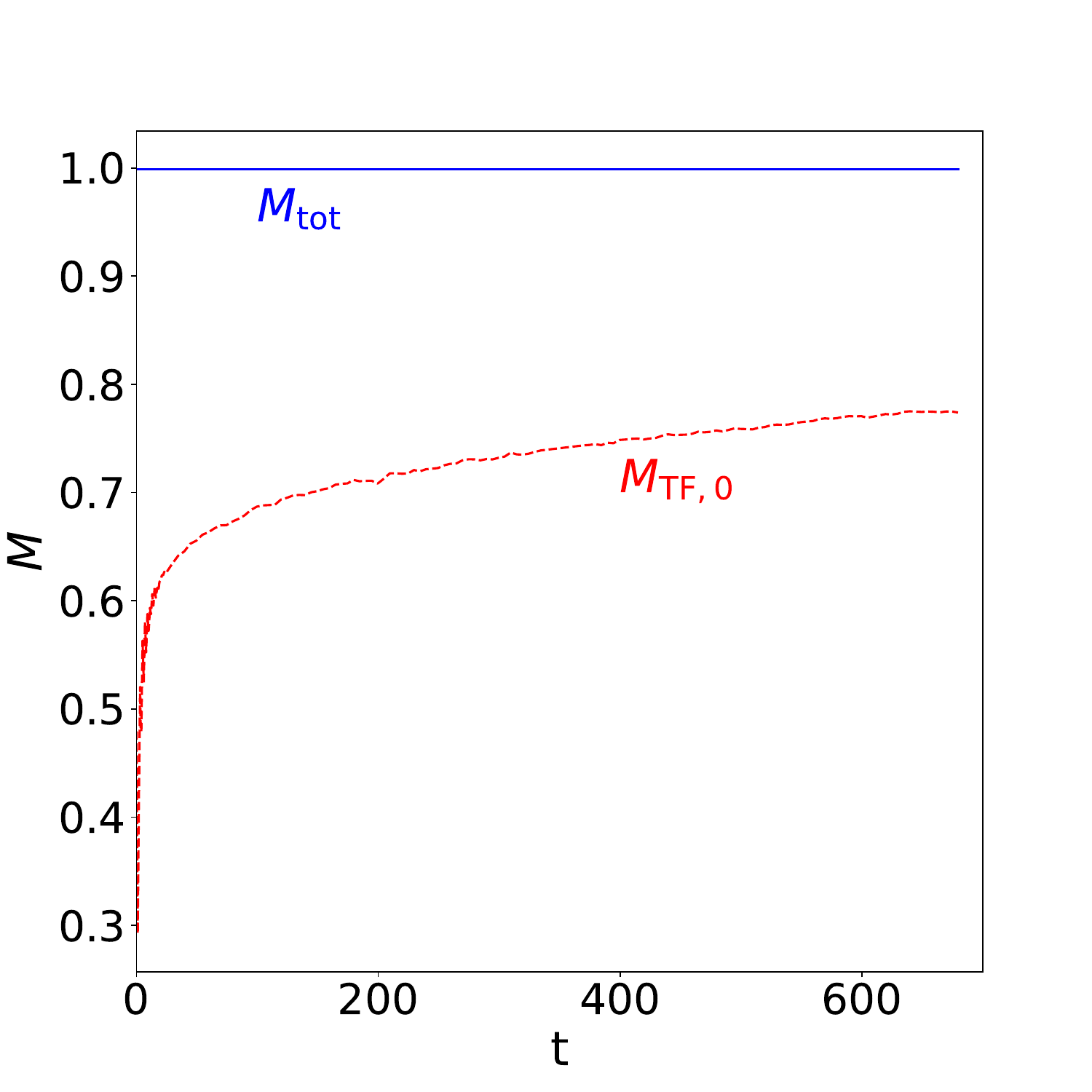}
\includegraphics[height=5cm,width=0.34\textwidth]{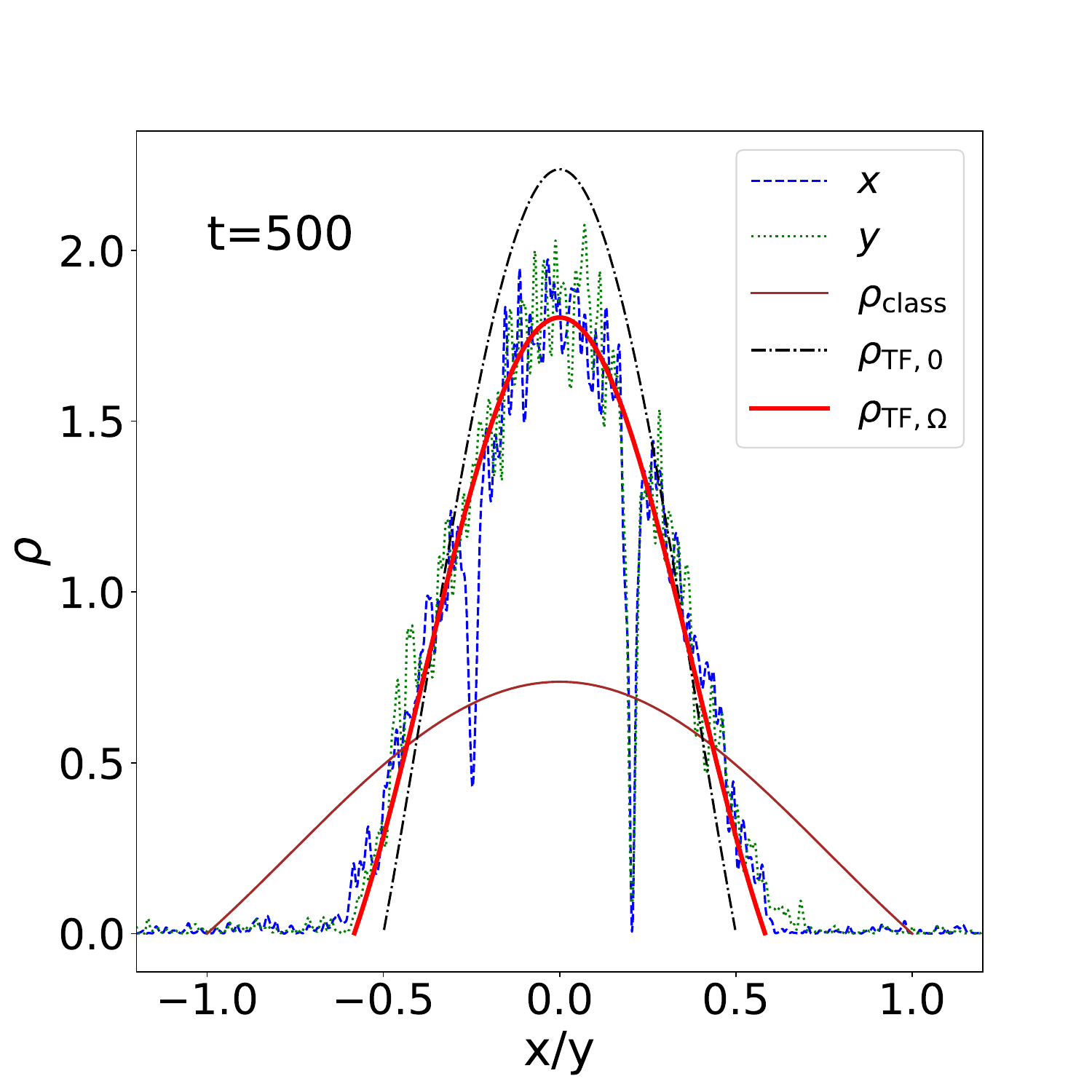}
\includegraphics[height=5cm,width=0.34\textwidth]{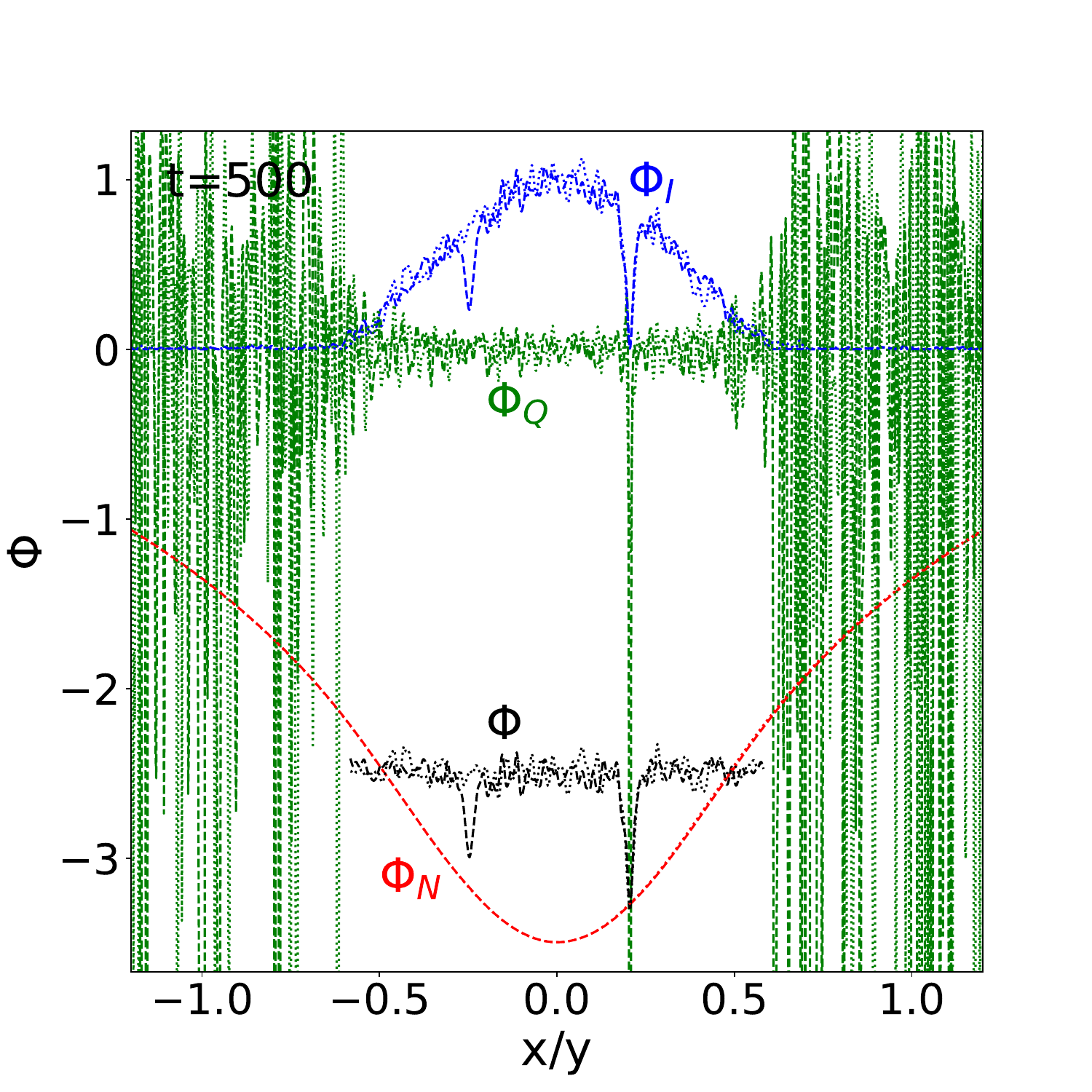}
\caption{
$[\epsilon=0.01, \alpha=1]$.
{\it Left panel:} growth of the soliton mass with time (red dashed line) and total mass of the
system (blue solid line).
{\it Middle panel:} density profiles along the $x$ (blue dashed line) and $y$ (green dotted line) axis,
at time $t=504$.
We also show the initial target profile (\ref{eq:rho-class}) (brown solid line $\rho_{\rm class}$),
the profile (\ref{eq:rho-TF-0}) of a static soliton (black dot-dashed line $\rho_{\rm TF,0}$),
and the profile (\ref{eq:rho-TF-Omega}) of a rotating soliton (red solid line $\rho_{\rm TF,\Omega}$).
{\it Right panel:} potentials $\Phi_Q$ (green dashed/dotted line),
$\Phi_I$ (blue dashed/dotted line), $\Phi_N$ (red dashed/dotted line)
and the sum $\Phi = \Phi_N+\Phi_I-r^2\Omega^2/2$ (black dashed/dotted line within the soliton radius
$R_{\Omega}$), along the $x/y$ axis.
}
\label{fig:evol-mu1-0p01}
\end{figure*}

To solve the Gross-Pitaevskii equation (\ref{eq:Schrod-eps})
we use a pseudo-spectral method with a split-step algorithm
\citep{Pathria-1990,Zhang-2008,Edwards-2018}.
The wave function after a time step $\Delta t$ is obtained as
\ba
&&\psi(\vec{x},t+ \Delta t) = \exp\left[-\frac{i \Delta t}{2\epsilon}
\Phi(\vec{x},t+ \Delta t)\right] \times \nonumber \\
&& \mathcal{F}^{-1} \exp\left[-\frac{i \epsilon\Delta t}{2} k^2\right]\mathcal{F}
\exp\left[-\frac{i \Delta t }{2\epsilon}\Phi(\vec{x},t)\right]\psi(\vec{x},t) \qquad .
\label{eq:step}
\ea
where the sequence of the operations is from right to left.
Here $\mathcal{F}$ and $\mathcal{F}^{-1}$ are the discrete Fourier
transform and its inverse, $k$ is the wavenumber in Fourier space and $\Phi=\Phi_N+\Phi_I$.
We also solve the Poisson equation (\ref{eq:Poisson-eps}) by Fourier transforms
and we apply periodic boundary conditions to our simulation box.

At each time, to compute and display in the figures below various quantities, we define the
center of the system as the location of the minimum of the gravitational potential.
This is more stable than choosing the location of the maximum density, as the density typically
shows non-negligible random fluctuations on top of its mean equilibrium profile, which are averaged
out in the gravitational potential.
It also removes the effects associated with the fluctuations of the position of the soliton.
The 1D profiles, as in Fig.~\ref{fig:evol-mu1-0p01} below, correspond to the $x$ and $y$ axis that run
through this center. The number of vortices $N_v(<r)$ and the angular momentum $L_z(<r)$
within radius $r$ are also computed with respect to this center.

\subsection{Results for $\epsilon=0.01$ and $\alpha=1$}

\subsubsection{Mass, density and velocity profiles}

\begin{figure}
\centering
\includegraphics[height=5cm,width=0.4\textwidth]{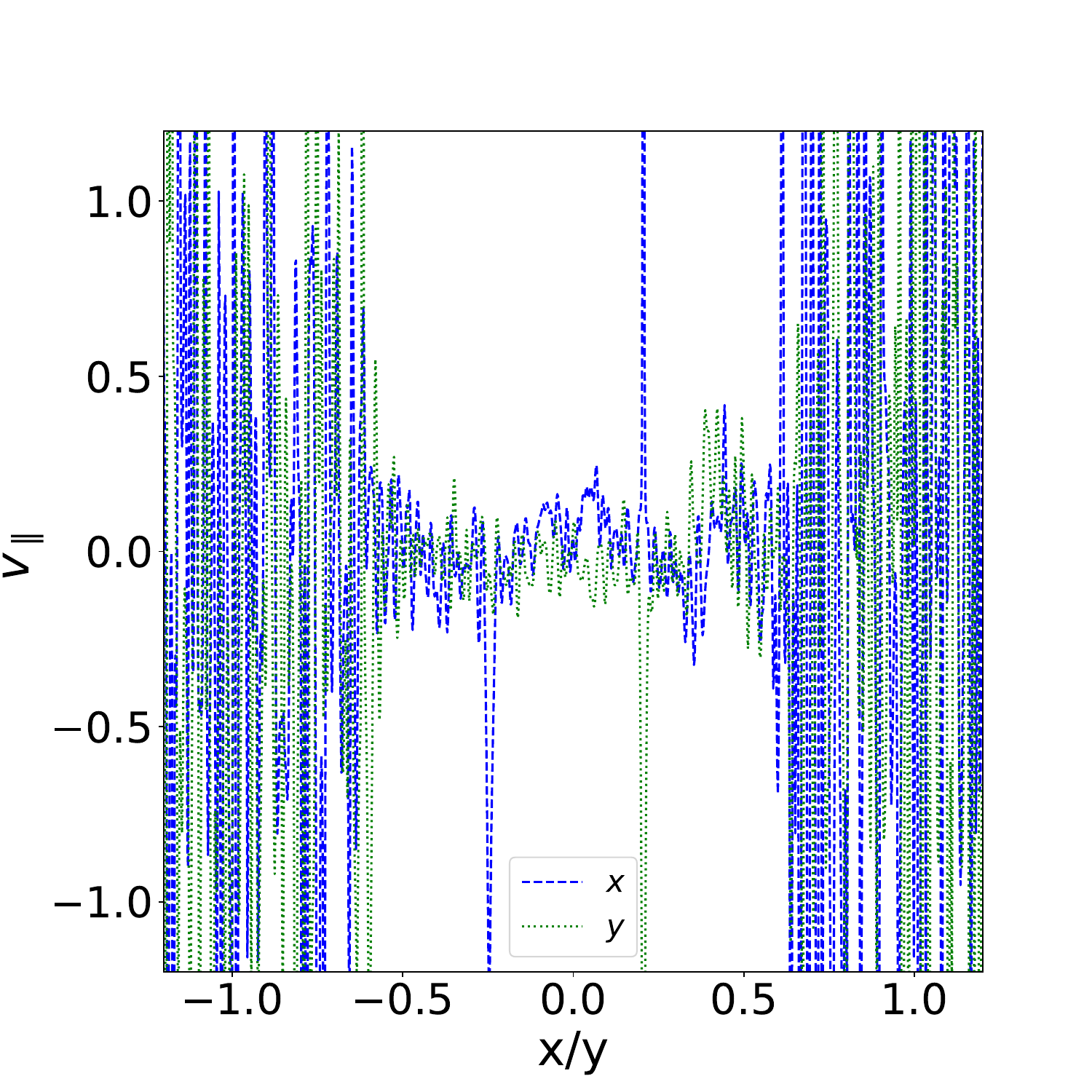}
\includegraphics[height=5cm,width=0.4\textwidth]{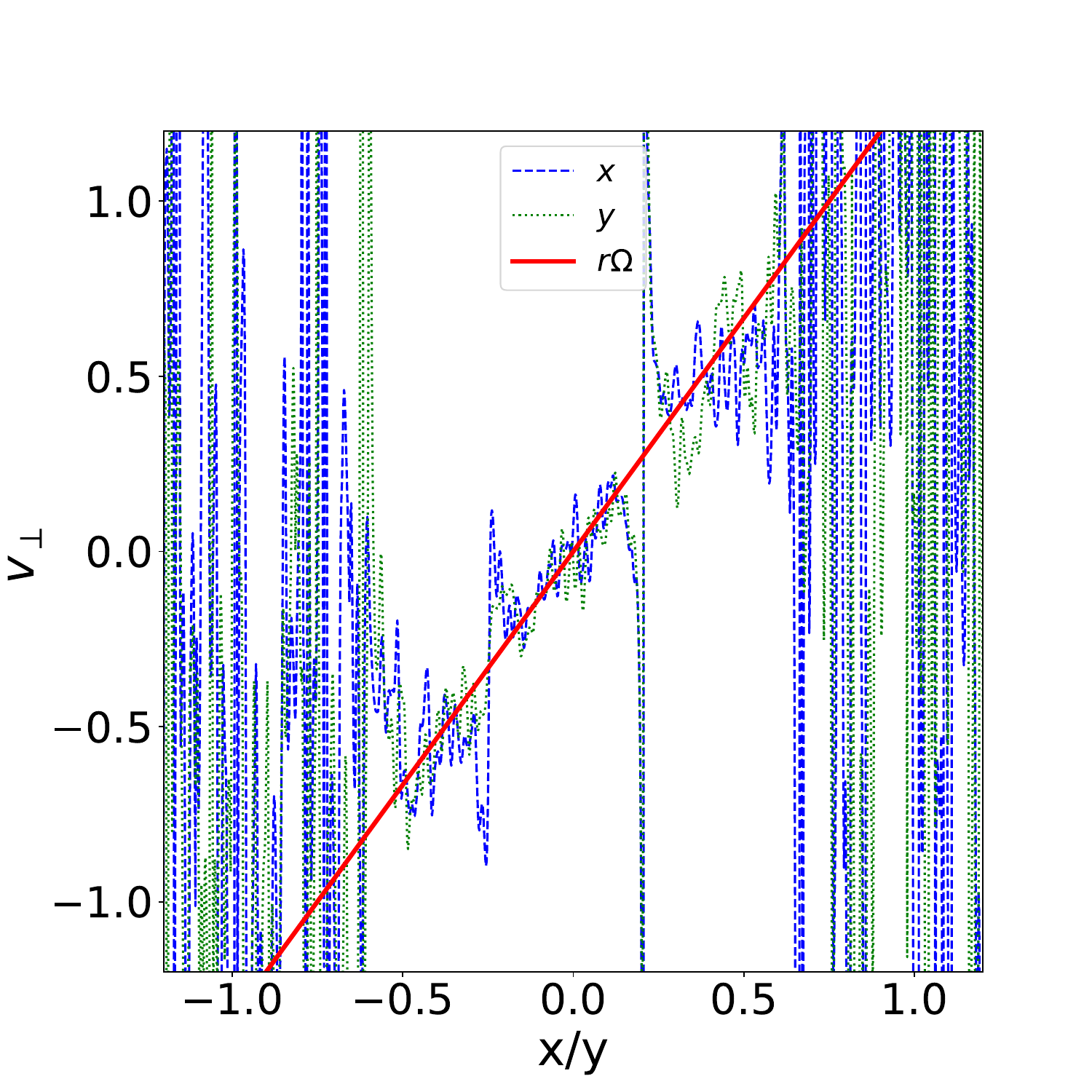}
\caption{
$[\epsilon=0.01, \alpha=1]$.
{\it Upper panel:} parallel velocities ($v_x/v_y$) along the $x/y$ axis.
{\it Lower panel:} transverse velocities ($v_y/-v_x$) along the $x/y$ axis.
The red solid line is the best fit $\Omega r$ in the central region, which provides our measurement
of the rotation rate $\Omega$.
}
\label{fig:v-mu1-0p01}
\end{figure}

\begin{figure*}
\centering
\includegraphics[height=4.8cm,width=0.32\textwidth]{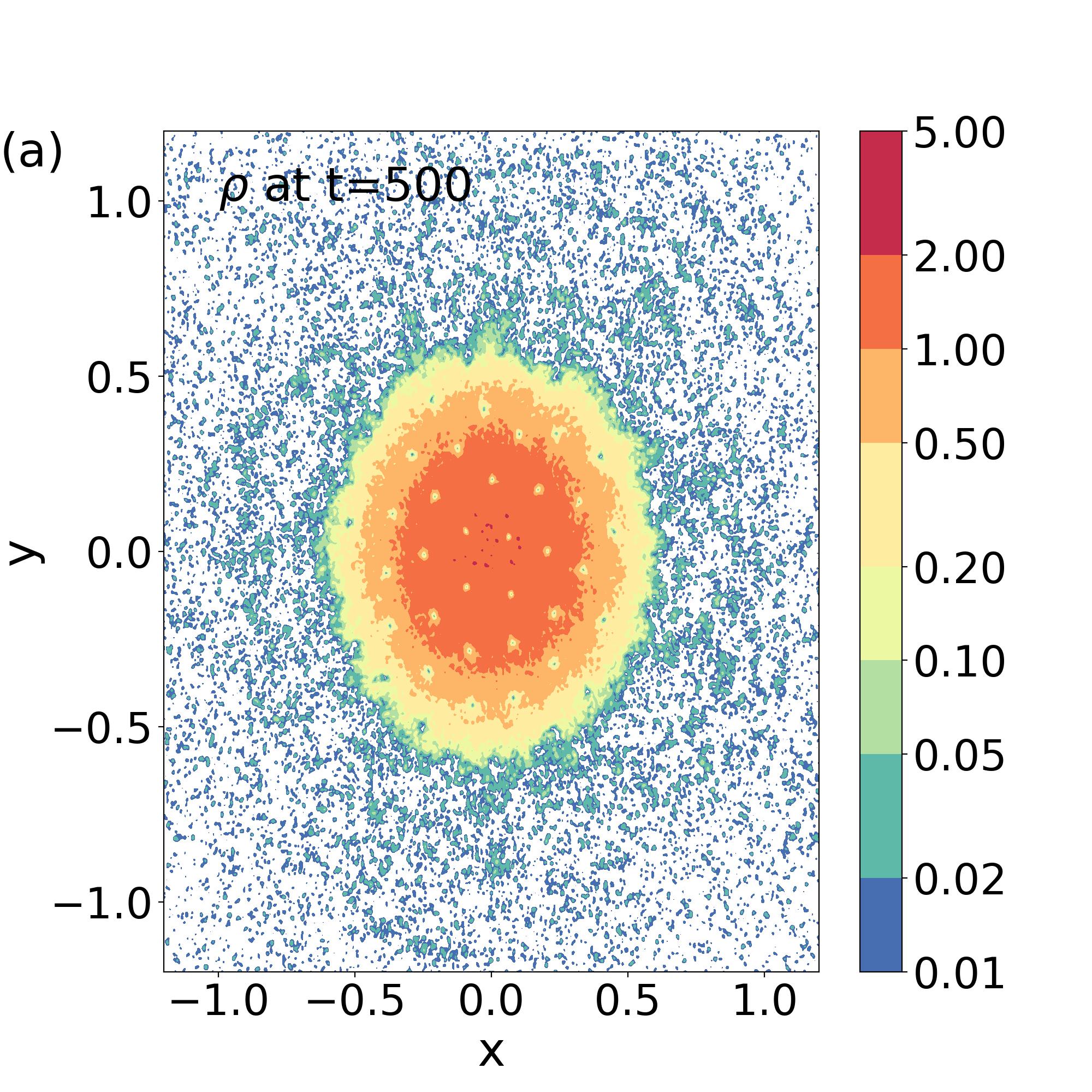}
\includegraphics[height=4.8cm,width=0.32\textwidth]{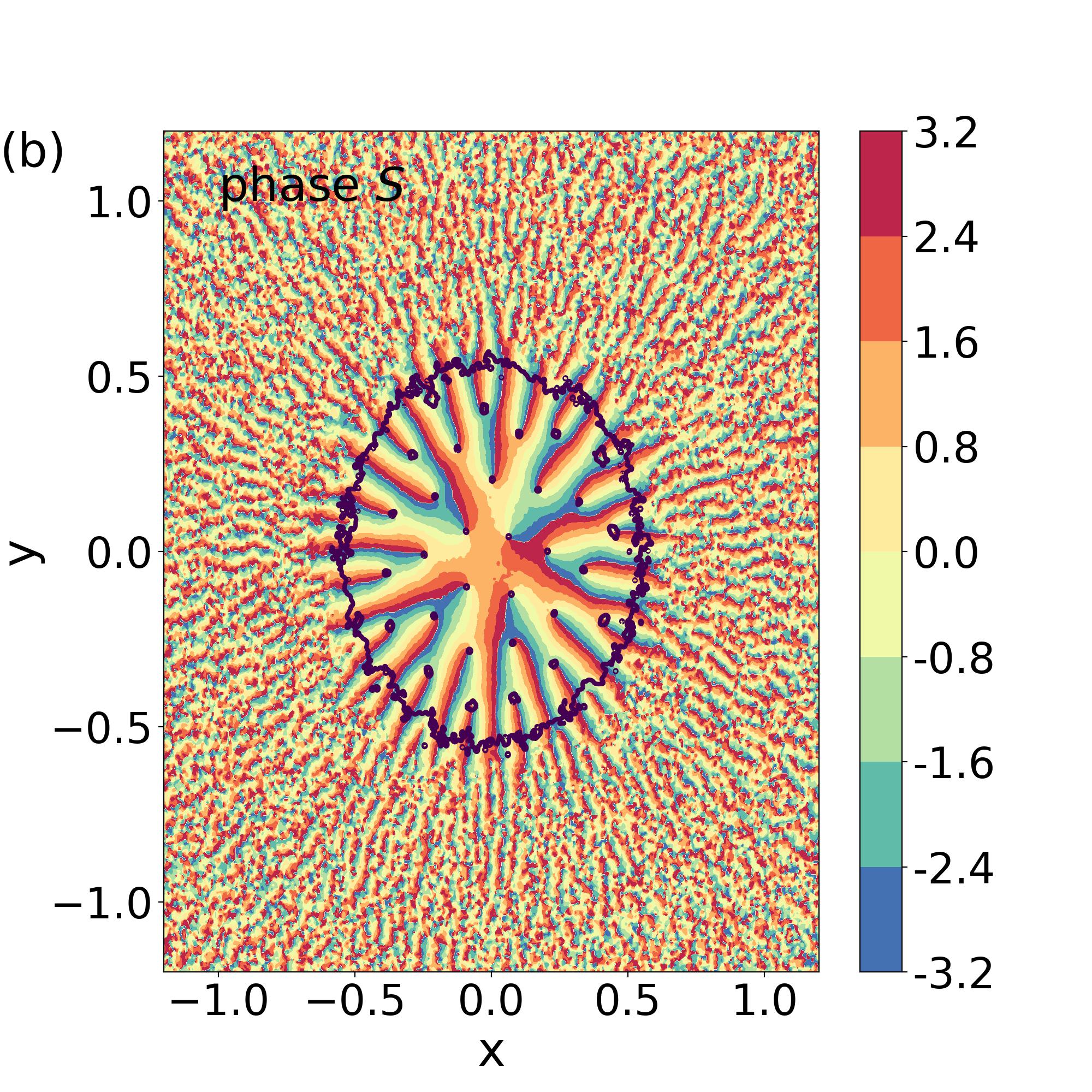}
\includegraphics[height=4.8cm,width=0.32\textwidth]{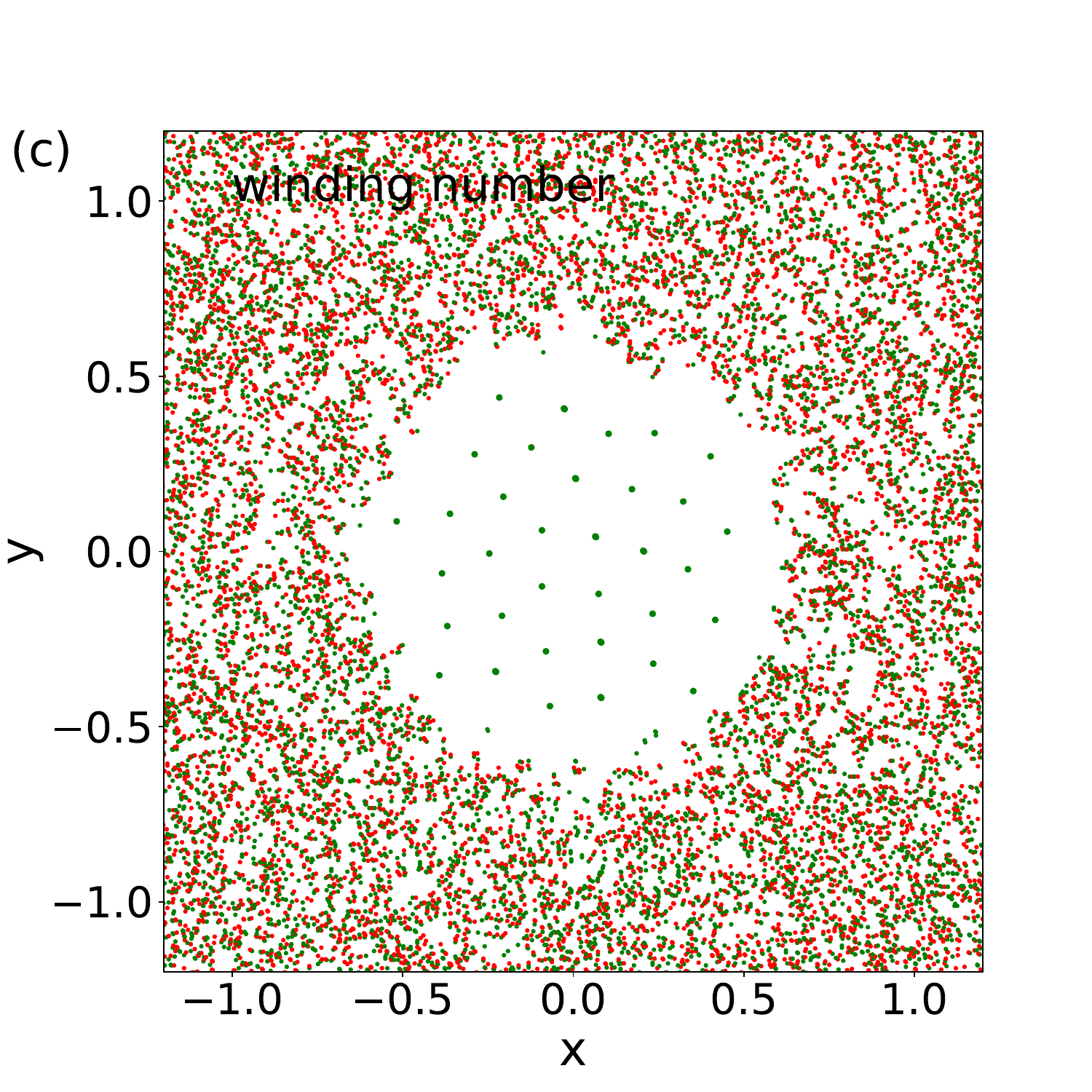}\\
\includegraphics[height=4.8cm,width=0.32\textwidth]{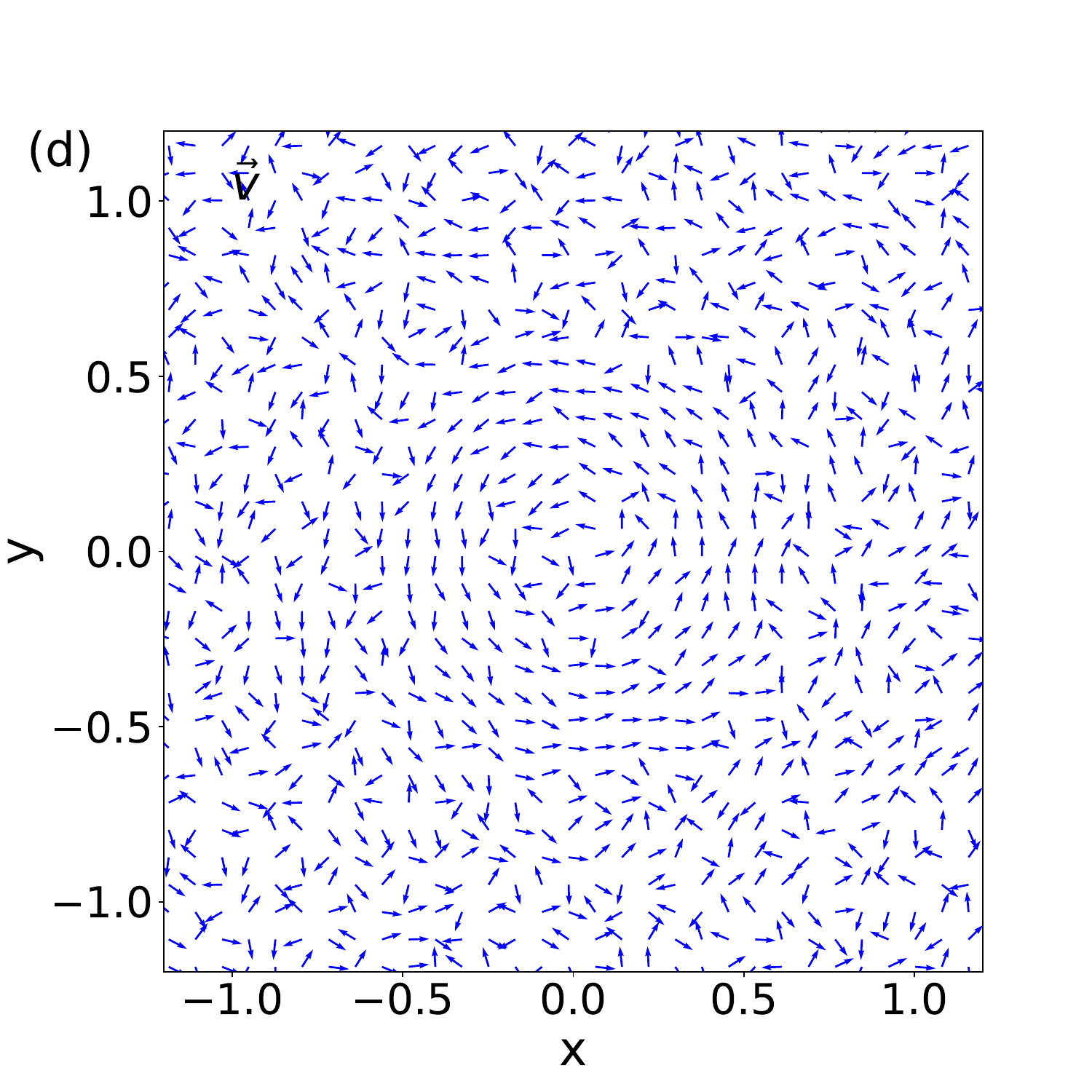}
\includegraphics[height=4.8cm,width=0.32\textwidth]{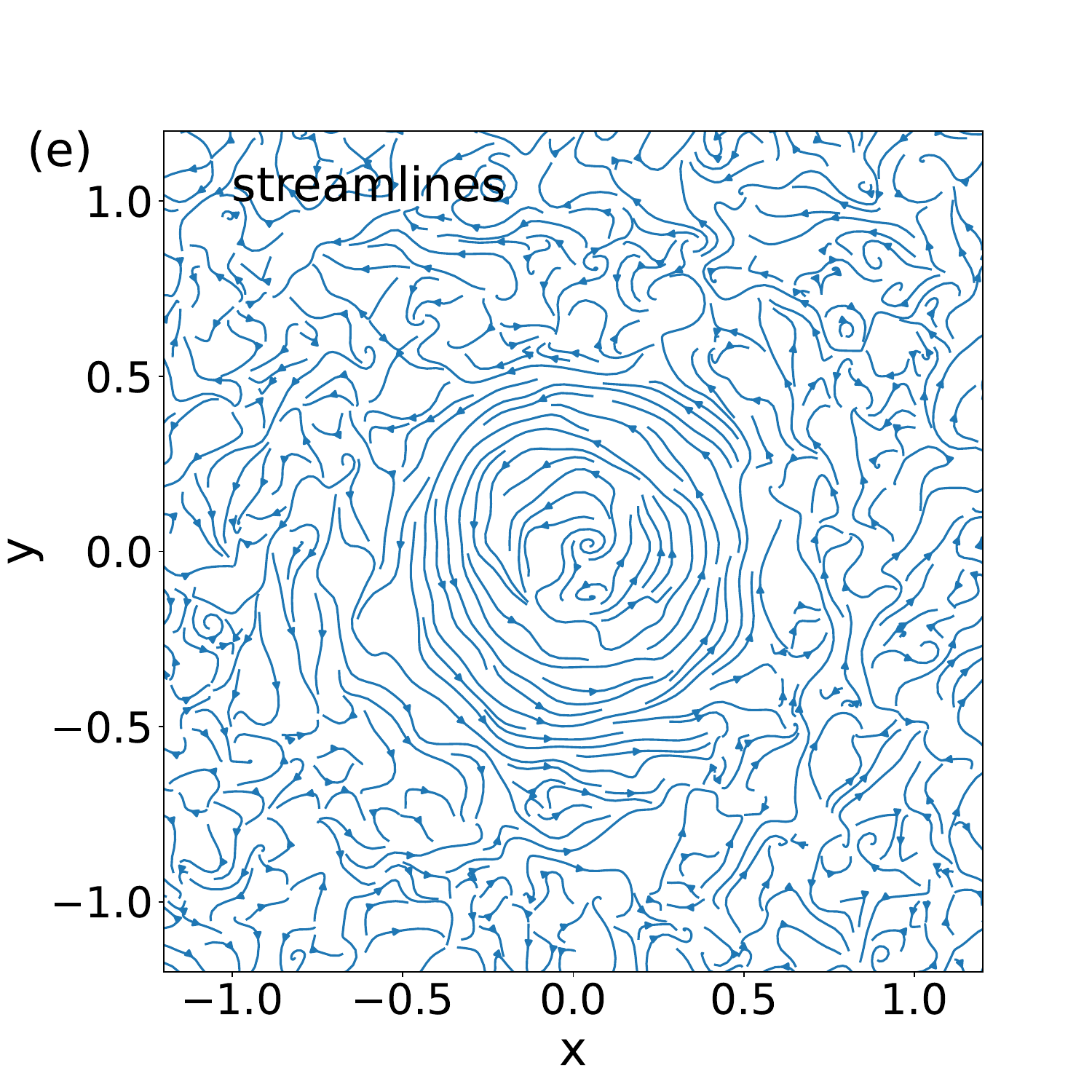}
\includegraphics[height=4.8cm,width=0.32\textwidth]{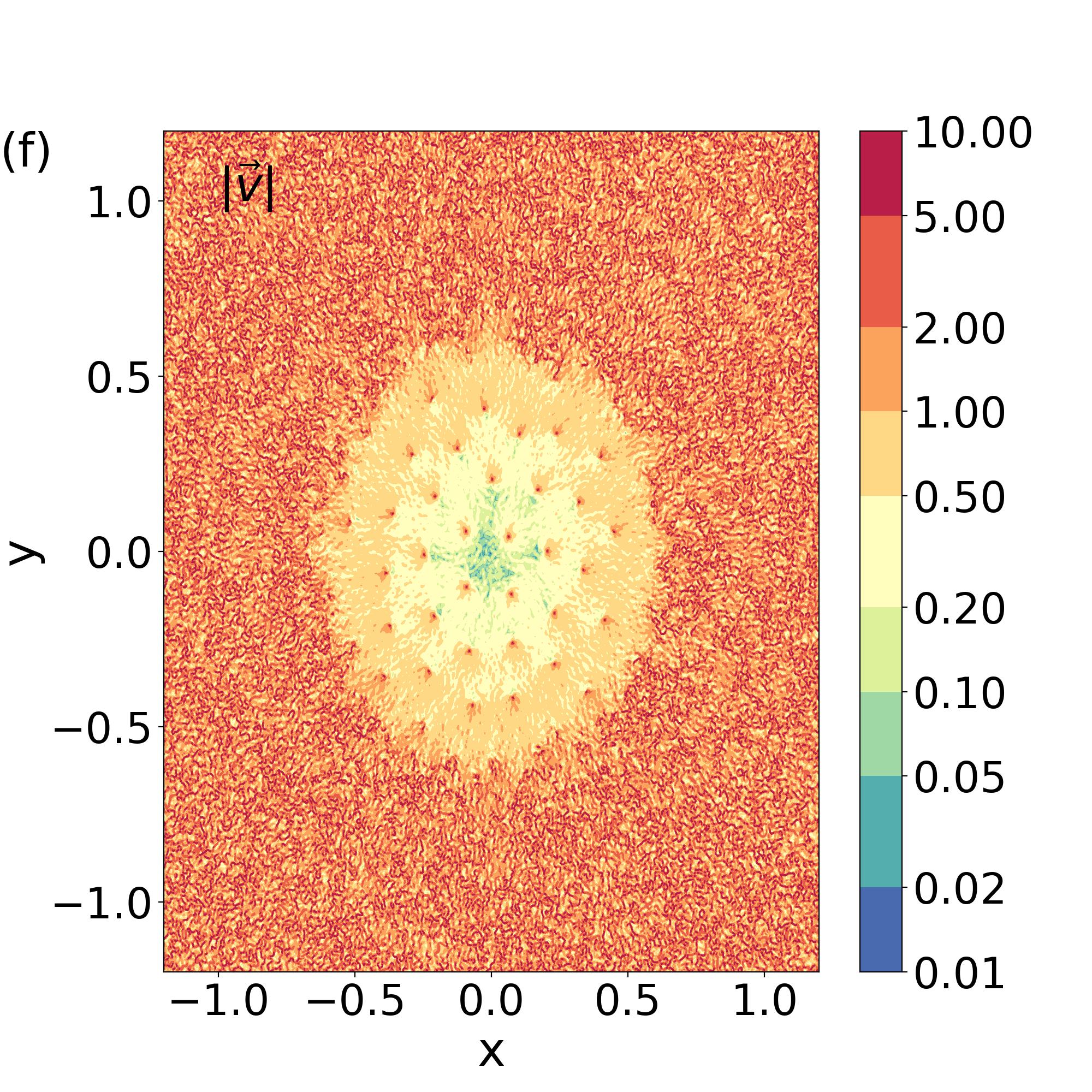}
\caption{
$[\epsilon=0.01, \alpha=1]$.
{\bf (a)} map of the 2D density field $\rho(\vec r)$ at time $t=500$.
{\bf (b)} map of the phase $S(\vec r)$ of the wave function. The black solid line is the density
isocontour $\rho=0.2$.
{\bf (c)} map of the winding number $w(\vec r)$. Green dots correspond to $w=1$ and red dots
to $w=-1$. Each dot corresponds to a singularity of the phase, i.e. a vortex.
{\bf (d)} map of the normalized velocity field $\vec v/|\vec v|$.
{\bf (e)} streamlines of the velocity field $\vec v(\vec r)$.
{\bf (f)} map of the velocity amplitude $|\vec v(\vec r)|$.
}
\label{fig:2D-mu1-0p01}
\end{figure*}

\begin{figure}
\centering
\includegraphics[height=4.5cm,width=0.35\textwidth]{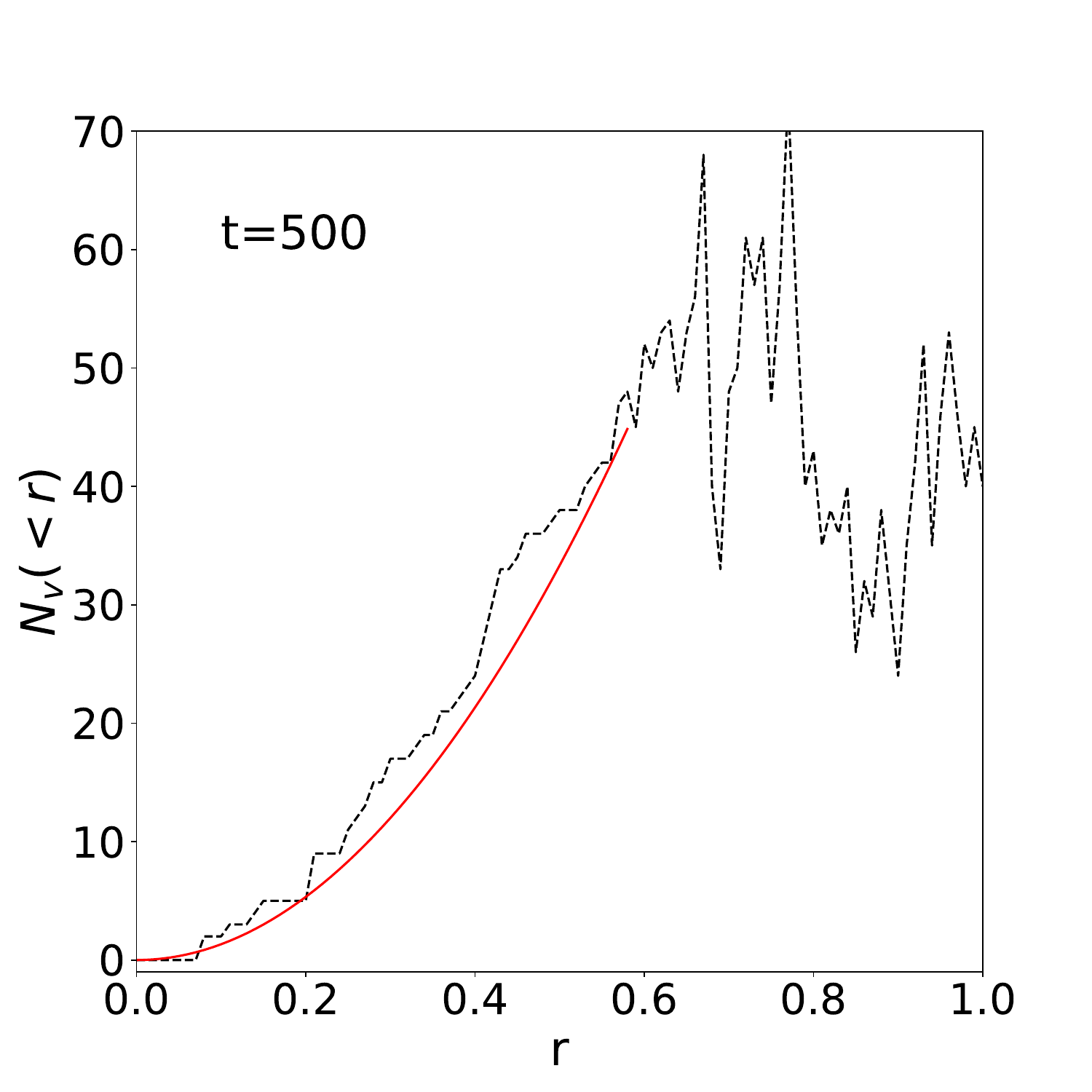}
\includegraphics[height=4.5cm,width=0.35\textwidth]{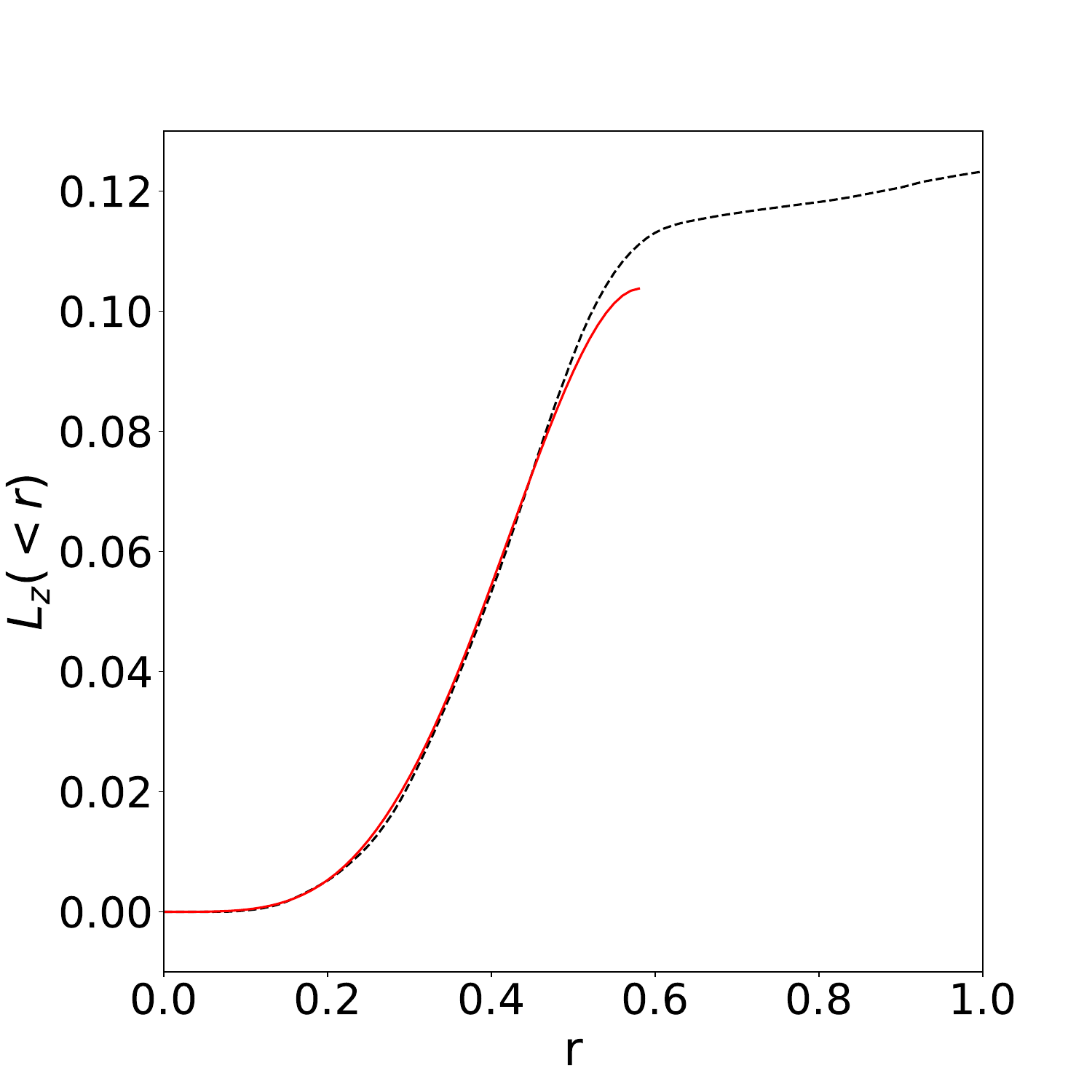}
\caption{
$[\epsilon=0.01, \alpha=1]$.
{\it Upper panel:} number of vortices $N_v(<r)$ within radius $r$ (black dashed line).
The red solid line is the prediction (\ref{eq:vortex-density}), with the value of $\Omega$
obtained from the lower panel in Fig.~\ref{fig:v-mu1-0p01}.
{\it Lower panel:} angular momentum $L_z(<r)$ within radius $r$ (black dashed line).
The red solid line is the prediction (\ref{eq:Lz-r}).
}
\label{fig:Lz-r-mu1-0p01}
\end{figure}

For the case $\epsilon=0.01$ and $\alpha=1$, we show in Fig.~\ref{fig:evol-mu1-0p01} how the system
evolves from the initial condition (\ref{fig:initial}).
As in the 3D isotropic simulations presented in \cite{Garcia:2023abs},
in a few dynamical times a soliton quickly forms at the center of the system.
As seen in the left panel, where we show the mass $M_{\rm TF,0}$ within the radius $R_0$
of Eq.(\ref{eq:rho-TF-0})
(which still gives the radius of rotating solitons within a factor $1.6$ as seen in Eq.(\ref{eq:R-max})),
the system quickly collapses to form a central soliton that contains about $65\%$ of the initial mass.
After this violent relaxation, the soliton keeps growing until the end of the simulation at an increasingly
small rate.
As a numerical check on the simulation, we can also see that the total mass of the system is conserved.

In the middle panel we compare the density profile obtained in the simulation with the initial classical profile
(\ref{eq:rho-class}), the static soliton profile (\ref{eq:rho-TF-0}) associated with the measured mass
within $R_0$, and the rotating soliton profile (\ref{eq:rho-TF-Omega}), fitted to the measured density
$\rho_0$ around the center and a value of $\Omega$ measured from the velocity field (as explained below).
We can clearly see the deviation of the profile from the static prediction (\ref{eq:rho-TF-0}) and
the  very good agreement with the rotating prediction (\ref{eq:rho-TF-Omega}).
The two downward spikes seen in the figure correspond to two vortices that happen to be located close to
the $x$ axis, since the density vanishes at the center of the vortices as seen in Sec.~\ref{sec:single-vortex}.

In the right panel we show the potentials $\Phi_Q$, $\Phi_I$ and $\Phi_N$.
We also show the sum $\Phi=\Phi_N+\Phi_I-r^2\Omega^2/2$ of Eq.(\ref{eq:mu-Omega}) over the
extent of the soliton, within the radius $R_\Omega$.
We can see that inside the soliton the quantum pressure is negligible whereas gravity is balanced
by the self-interactions and the rotation.
In particular, we can check that the sum $\Phi_N+\Phi_I-r^2\Omega^2/2$ is constant,
which is a signature of a soliton with solid-body rotation.
Outside of the soliton there is a low-density virialized envelope with strong density fluctuations.
The quantum pressure dominates over the gravitational and self-interaction potentials but
this outer halo is not described by an hydrodynamical equilibrium. Instead, it is built of many excited
states as in (\ref{eq:psi-halo-a_nl}) and corresponds to a virialized halo of collisionless particles in the
semi-classical limit, supported by rotation and velocity dispersion
\cite{Widrow:1993qq,Mocz:2018ium,GalazoGarcia:2022nqv,Garcia:2023abs,Liu:2024pjg}.

We show in Fig.~\ref{fig:v-mu1-0p01} the profiles of the parallel and transverse velocities,
along the $x$ and $y$ axis. In agreement with the solid-body rotation (\ref{eq:v-solid-rotation}),
we find that $v_\parallel$ fluctuates around zero whereas $v_{\perp}$ fluctuates around a linear
slope $\Omega r$.
A least-squared fit over the central region of radius $R_0$ to a straight line provides the best-fit
parameter $\Omega \simeq 1.3$. This is shown by the red solid line, which indeed provides a good fit
to the mean transverse velocity.
The two velocity spikes correspond to the two vortices that were already visible in the density profile
in Fig.~\ref{fig:evol-mu1-0p01}, as the velocity diverges as $1/r$ at the center of vortices,
as seen in (\ref{eq:v-single-vortex}).
It is this value of $\Omega$ that we used in the middle panel in Fig.~\ref{fig:evol-mu1-0p01}
to compute the density profile (\ref{eq:rho-TF-Omega}) of the rotating soliton.
Thus, we can see that both the density and velocity profiles agree with the rotating soliton
obtained in Sec.~\ref{sec:rotating-soliton}.

\subsubsection{2D maps}

We show in Fig.~\ref{fig:2D-mu1-0p01} the 2D maps of the system at time $t=500$,
when the system has relaxed to a rotating central soliton with an outer virialized halo.
We can clearly see in panel (a) the central high-density soliton, with a circular shape,
surrounded by a low-density virialized halo made of many ``granules'', that is, strong density fluctuations
on a scale set by the de Broglie wavelength (\ref{eq:de-Broglie-rescaled}).
In addition, inside the soliton we can see a regular lattice of density troughs.
They correspond to the vortices, with vanishing density at their center.
In agreement with the scalings (\ref{eq:healing-length}) and (\ref{eq:d-Omega-eps}), we can check that
for small $\epsilon$ we are in the dilute regime, where the healing length $\xi$ of
Eq.(\ref{eq:healing-length}) is much smaller than the distance between neighbouring vortices.

In panel (b) we show the phase $S$, obtained from the wave function by the Madelung
transform (\ref{eq:Madelung}).
We can see that it is smooth inside the soliton, except at the positions of the vortices and along
the cuts originating from the vortices where it jumps from $-\pi$ to $\pi$.
The black line is the isodensity contour $\rho=0.2$. It shows that the singularities of the phase
are precisely located at the points where the density vanishes, as each phase singularity inside the
soliton is also located within a tight low-density isocontour.
Outside of the soliton, also marked by the circular outer density isocontour, the phase is very noisy.
This is because of the incoherent interferences between the many excited states that dominate the
outer halo, as was also the case in the initial condition displayed in Fig.~\ref{fig:initial}.

In panel (c) we show the map of winding numbers $w$. For each point on the numerical grid,
we draw a surrounding square of side twice the grid step, and we measure the phase difference
along this curve, $w = (1/2\pi) \oint_{\cal C} \vec{d\ell} \cdot \vec \nabla S$.
In a regular region we have $w=0$ along this closed loop, but around a vortex $w$ is equal to the
spin $\sigma$ of the vortex. We can check that we only find spins $\sigma = \pm 1$.
Inside the soliton we only have positive spins, $\sigma=1$ along a regular lattice.
This is because the initial condition has a positive angular momentum, as $\alpha=1$, which gives
rise to a solid-body rotation $\Omega>0$ supported by positive spin vortices.
The regularity of the vortex lattice is the discrete representation of the uniform vortex density
(\ref{eq:vortex-density}) obtained in the continuum limit. It implies that all neighbouring vortices
are roughly separated by the same distance $d$.
This behavior is similar to the regular vortex lattices observed in rotating BEC of a gas of cold
atoms \cite{Abo-Shaeer-2001}.

Outside of the soliton, in the virialized halo associated with incoherent granules, there are
many disordered vortices of both signs, $\sigma=\pm 1$.
They continuously form and annihilate following the zeros of the wave function that arise from
the interference between the different eigenmodes \cite{Hui:2020hbq}.
In 2D, the two conditions that the real and imaginary parts of the wave function vanish define a
set of points in the plane (whereas they would define a set of vortex lines in 3D).
This halo is mostly supported by its velocity
dispersion, rather than by a coherent rotation as in the soliton.

We show the map of the normalized velocity field $\vec v/|\vec v|$ and of the streamlines in panels (d)
and (e). Inside the soliton, we can clearly see the smooth solid-body rotation (\ref{eq:v-solid-rotation}).
The linear slope $|\vec v| \propto r$ was already seen in Fig.~\ref{fig:v-mu1-0p01}.
There are some fluctuations around the solid-body rotation because of the vortices and the incomplete
relaxation of the system.
Outside of the soliton, we find a disordered velocity field, associated with the many vortices of any sign,
which supports the halo by its dispersion rather than rotation.

In panel (f) we show the map of the amplitude of the velocity $|\vec v|$. It is mostly smooth and of the
order of unity inside the soliton, in agreement with the solid-body rotation (\ref{eq:v-solid-rotation}).
However, it diverges as $1/r$ at the location of the vortices, in agreement with Eq.(\ref{eq:v-single-vortex}).
We can check that we recover the same vortex lattice as in the density and winding maps shown in
panels (a) and (c).
Outside the soliton, the disordered velocity field has a large amplitude, because the phase varies
on the de Broglie wavelength $\lambda_{\rm dB}$, which leads to large gradients
$\vec v = \epsilon\vec\nabla S$.

\subsubsection{Radial angular momentum profile}

\begin{figure}
\centering
\includegraphics[height=4.cm,width=0.235\textwidth]{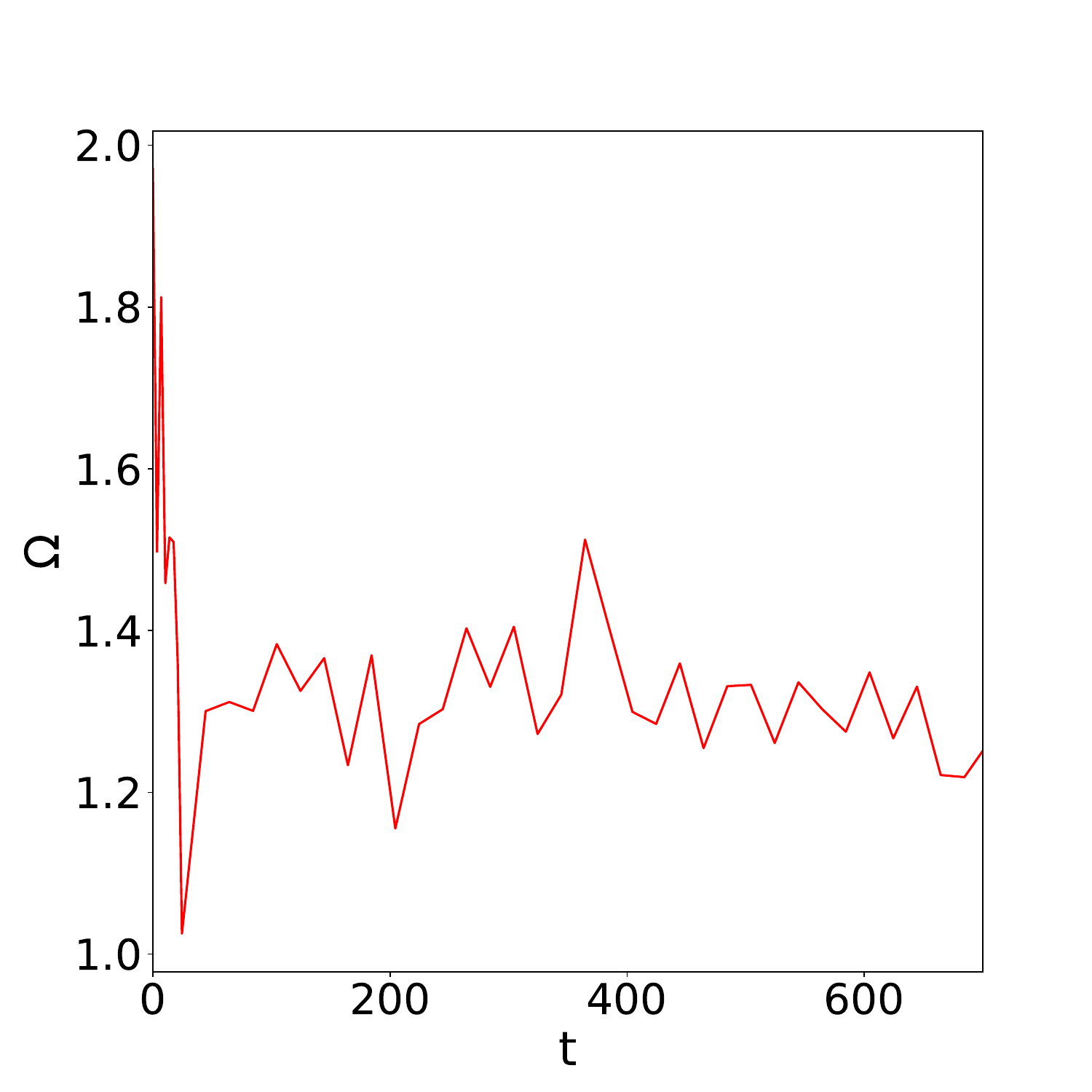}
\includegraphics[height=4.cm,width=0.235\textwidth]{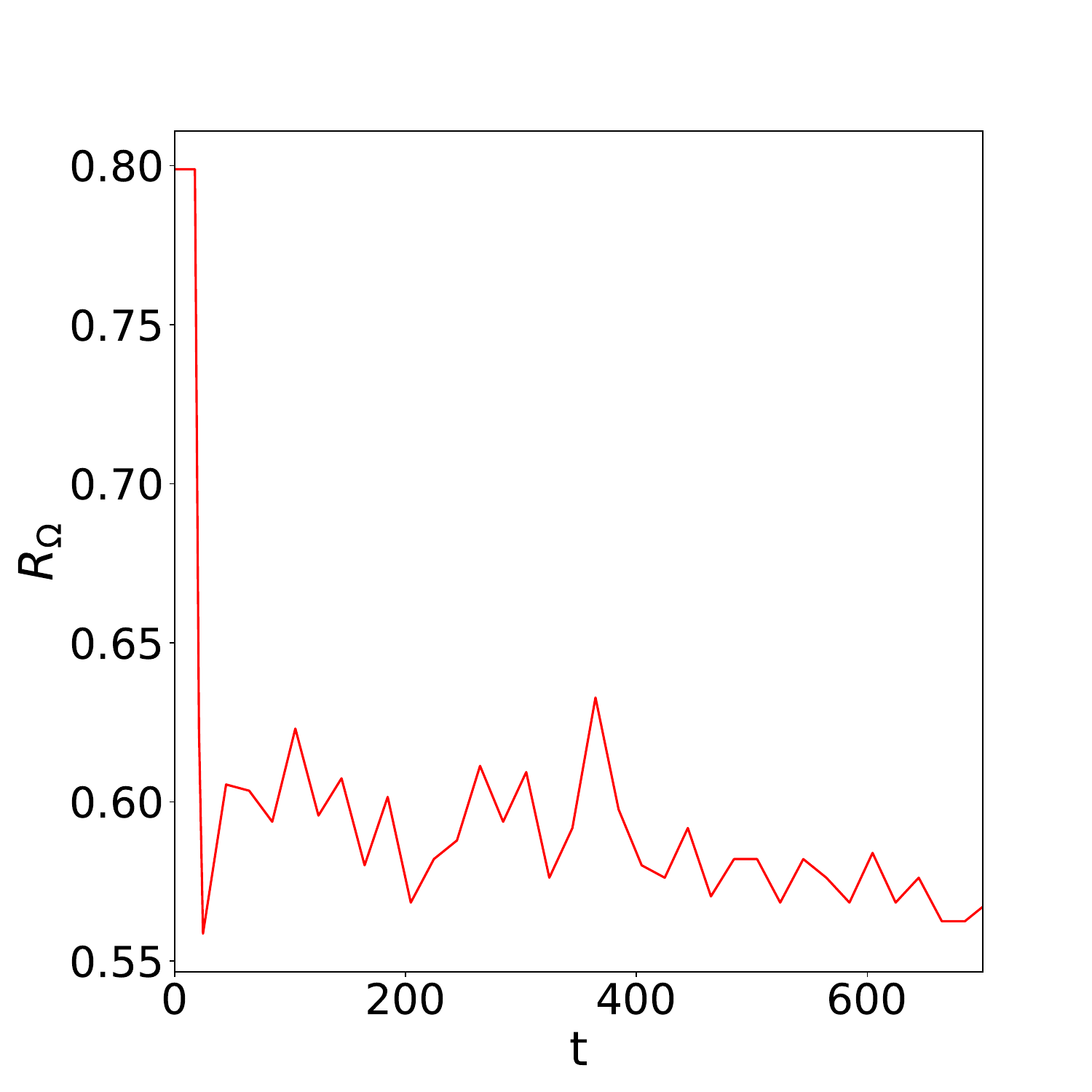}
\includegraphics[height=4.cm,width=0.235\textwidth]{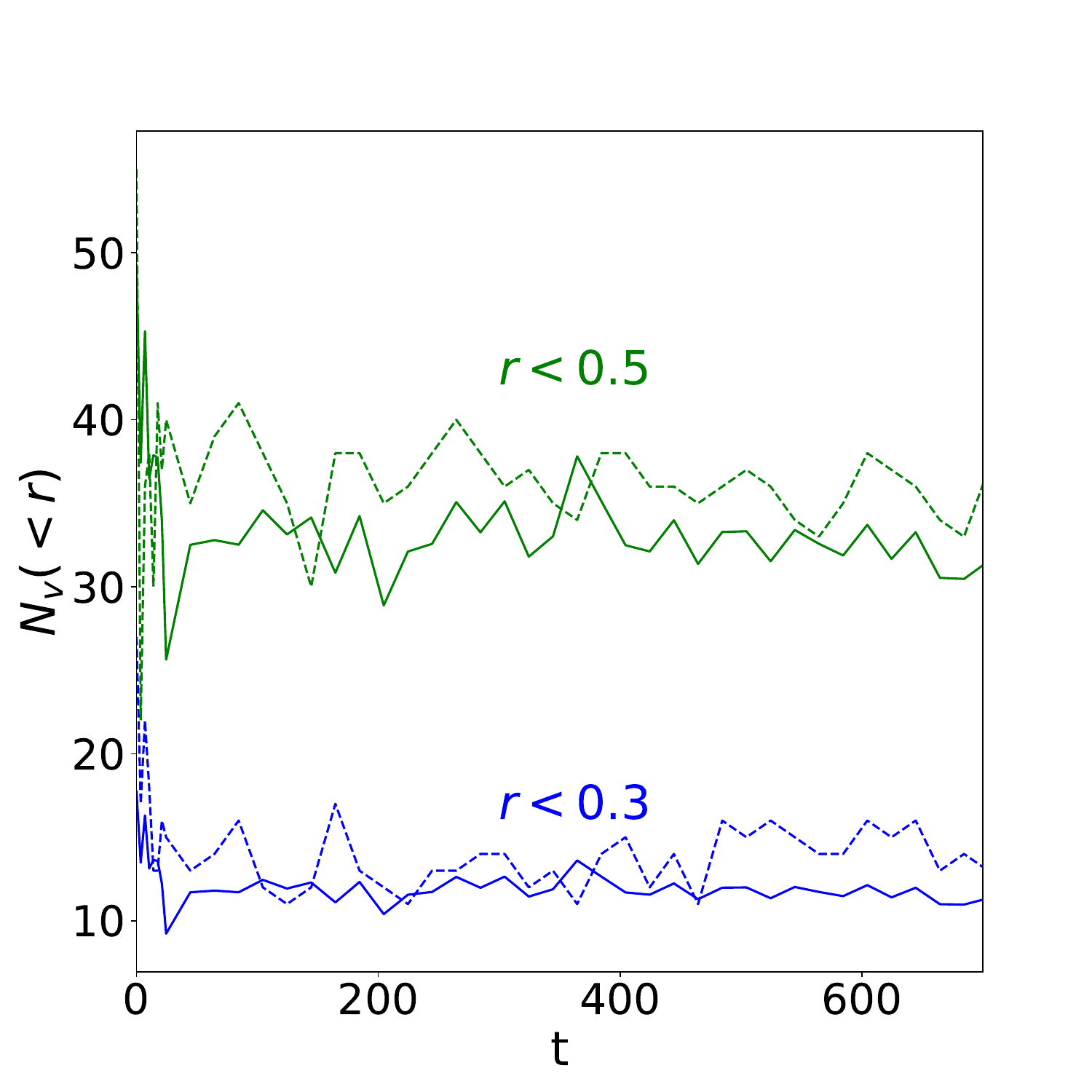}
\includegraphics[height=4.cm,width=0.235\textwidth]{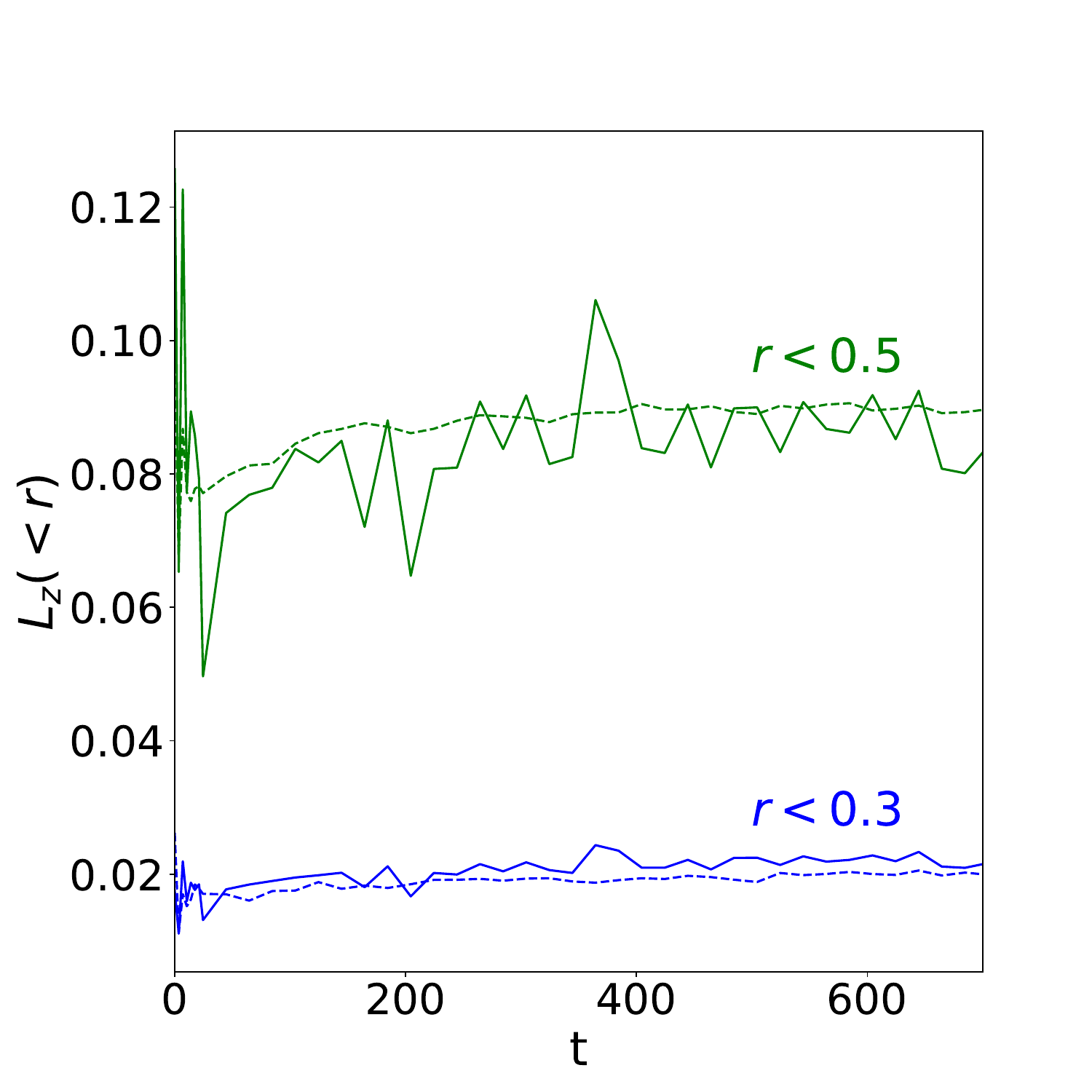}
\caption{
$[\epsilon=0.01, \alpha=1]$.
{\it Upper left panel:} rotation rate $\Omega(t)$ as a function of time.
{\it Upper right panel:} rotating soliton radius $R_{\Omega}(t)$.
{\it Lower left panel:} number of vortices $N_v(<r,t)$ as a function of time, within the two
radii $r=0.3$ and $r=0.5$ (dashed lines). The solid lines are the prediction (\ref{eq:vortex-density}).
{\it Lower right panel:} angular momentum $L_z(<r,t)$ as a function of time, within the two
radii $r=0.3$ and $r=0.5$ (dashed lines). The solid lines are the prediction (\ref{eq:Lz-r}).
}
\label{fig:Om-t-mu1-0p01}
\end{figure}

\begin{figure}
\centering
\includegraphics[height=5.5cm,width=0.41\textwidth]{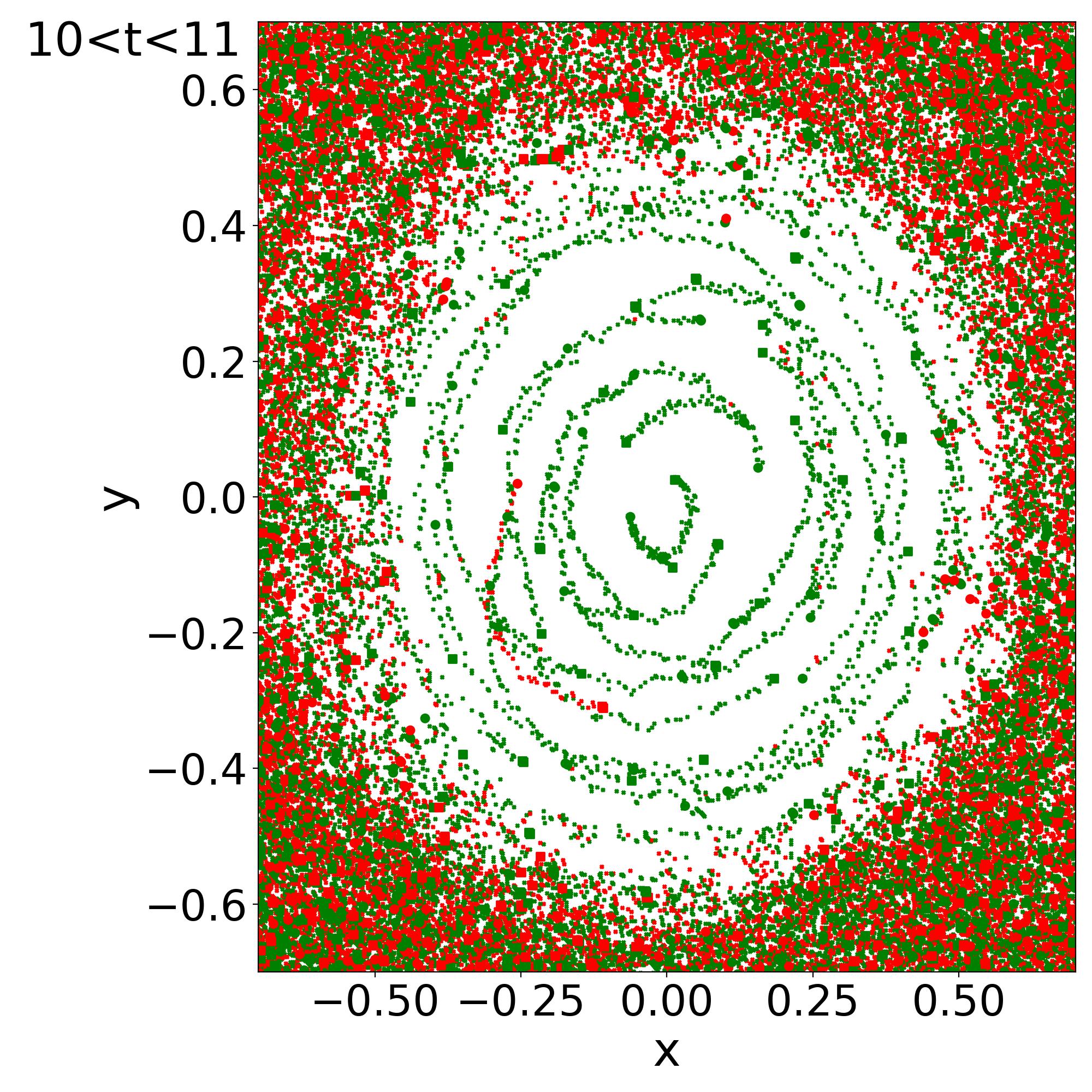}
\includegraphics[height=5.5cm,width=0.41\textwidth]{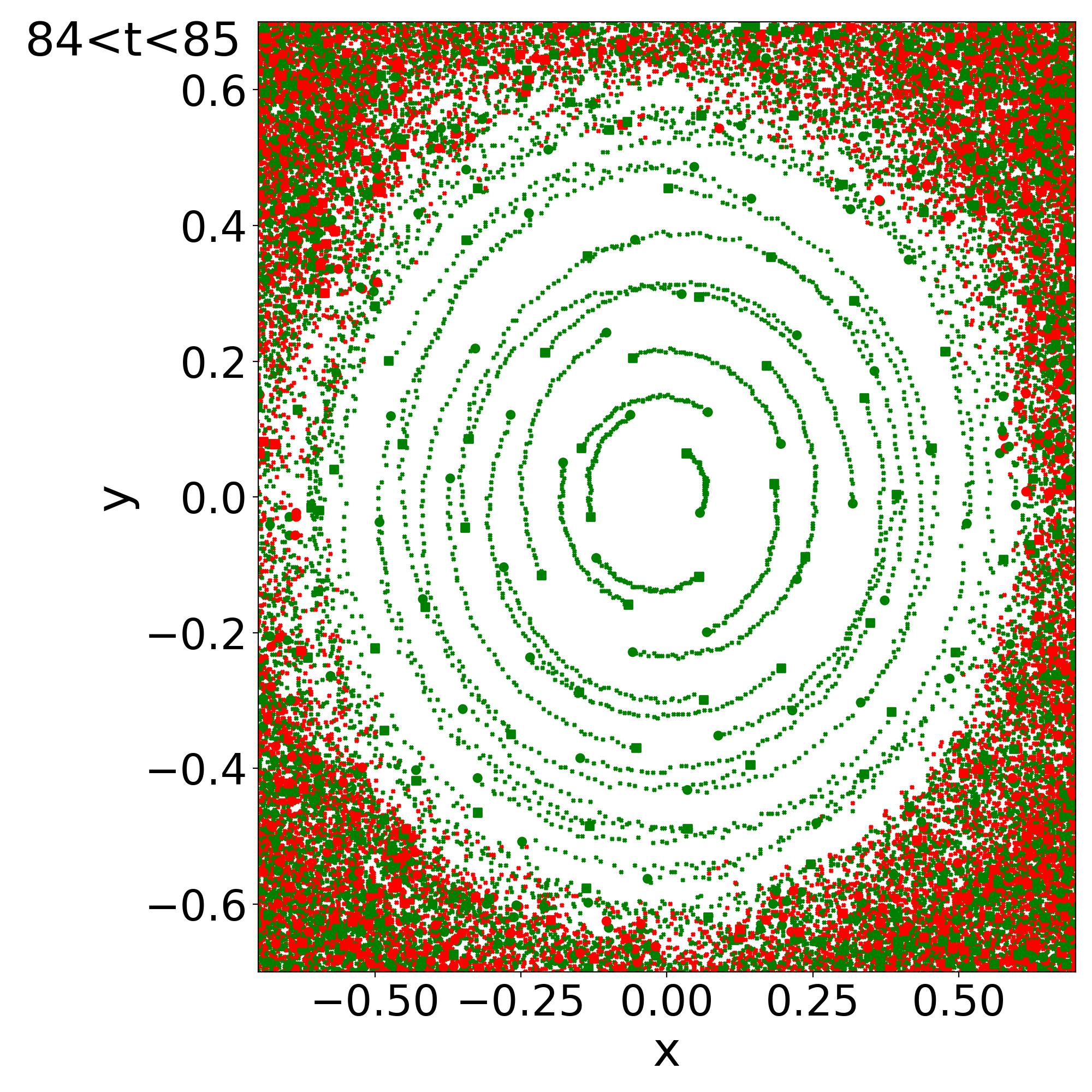}
\caption{
$[\epsilon=0.01, \alpha=1]$.
{\it Upper panel:} superposition of many snapshots between times $t=10$ and $t=11$ of the maps
of the vortices, as in panel (c) in Fig.~\ref{fig:2D-mu1-0p01}.
The filled circles/squares are the positions at the initial/final time.
{\it Lower panel:} superposition of many snapshots between times $t=84$ and $t=85$.
}
\label{fig:wind-mu1-0p01}
\end{figure}

We show in Fig.~\ref{fig:Lz-r-mu1-0p01} the number of vortices (weighted by their spin $\sigma$)
and the angular momentum within radius $r$.
The red solid lines are the analytical predictions (\ref{eq:vortex-density}) and (\ref{eq:Lz-r}),
plotted within the radius $R_\Omega$.
We can check the good agreement between our predictions and the numerical results inside
the soliton.
In the outer halo there are about the same number of vortices of either sign.
We can see that about half of the initial angular momentum (\ref{eq:Lz-init}) of the system
is contained inside the soliton, as we have $L_z \simeq 0.12$ within the soliton radius whereas
the initial angular momentum was $L_{z,\rm init} \simeq 0.27$.
At early times the other half of the angular momentum is transferred
to large radii, where the lower amount of collective angular rotation is compensated by the larger radii.

The amount of angular momentum contained in the soliton can be estimated from the initial
profile defined by Eqs.(\ref{eq:rho-class}) and (\ref{eq:v-theta-r-class}).
Indeed, we can see from the left panel in Fig.~\ref{fig:evol-mu1-0p01} that the soliton initially
forms in a few dynamical times with a mass $M_{\rm TF} \simeq 0.65$.
Assuming the radial ranking of matter shells is not too much modified during the collapse,
this mass comes from shells within radius $r\lesssim 0.61$ in the initial halo (\ref{eq:rho-class}).
With the initial angular velocity (\ref{eq:v-theta-r-class}) this matter carried a total angular
momentum $L_z \simeq 0.12$. We can see in Fig.~\ref{fig:Lz-r-mu1-0p01} that this is roughly the
total angular momentum of the soliton.

It is interesting to compare this with the maximum angular momentum (\ref{eq:Lz-max}) that can
be carried by the soliton.
With $M\simeq 0.65$ and $R_0=0.5$ we obtain $L_{z,\max} \simeq 0.13$.
Thus, we can see that the soliton obtained in this simulation is actually close to the upper
bound (\ref{eq:Lz-max}).

\subsubsection{Evolution of the rotation rate}

We show in Fig.~\ref{fig:Om-t-mu1-0p01} the evolution with time of the rotation of the system.
We can see that after the quick relaxation of the system and the formation of the central soliton,
in a few dynamical times, the rotation rate (measured from the slope of the transverse
velocity along the $x$ and $y$ axis in the central region) settles around $\Omega \simeq 1.3$.
There remain modest fluctuations, due to the discrete vortices and incomplete relaxation.

The soliton radius $R_\Omega$ is obtained from the measurement of the central density
$\rho_0$ and of $\Omega$ as the first zero crossing of the density profile (\ref{eq:rho-TF-Omega}).
It also quickly settles to $R_\Omega \simeq 0.58$, following the fluctuations of $\rho_0$ and
$\Omega$.

As seen in the lower panels, the number of vortices and the angular momentum within radii
$r=0.3$ and $r=0.5$ inside the soliton agree well with the analytical predictions
(\ref{eq:vortex-density}) and (\ref{eq:Lz-r}).
In particular, the angular momentum within the soliton is roughly constant, in agreement with the
conservation of angular momentum by the Gross-Pitaevskii equation (\ref{eq:Schrod-eps}).
Thus, once the soliton is formed and stabilized, there is little exchange of angular momentum with the
outer halo.
This justifies the analysis in Sec.~\ref{sec:continuum}, where we obtained the rotating soliton as the
minimum of the energy at fixed mass and angular momentum.

\subsubsection{Trajectories of the vortices}

\begin{figure*}
\centering
\includegraphics[height=4.8cm,width=0.32\textwidth]{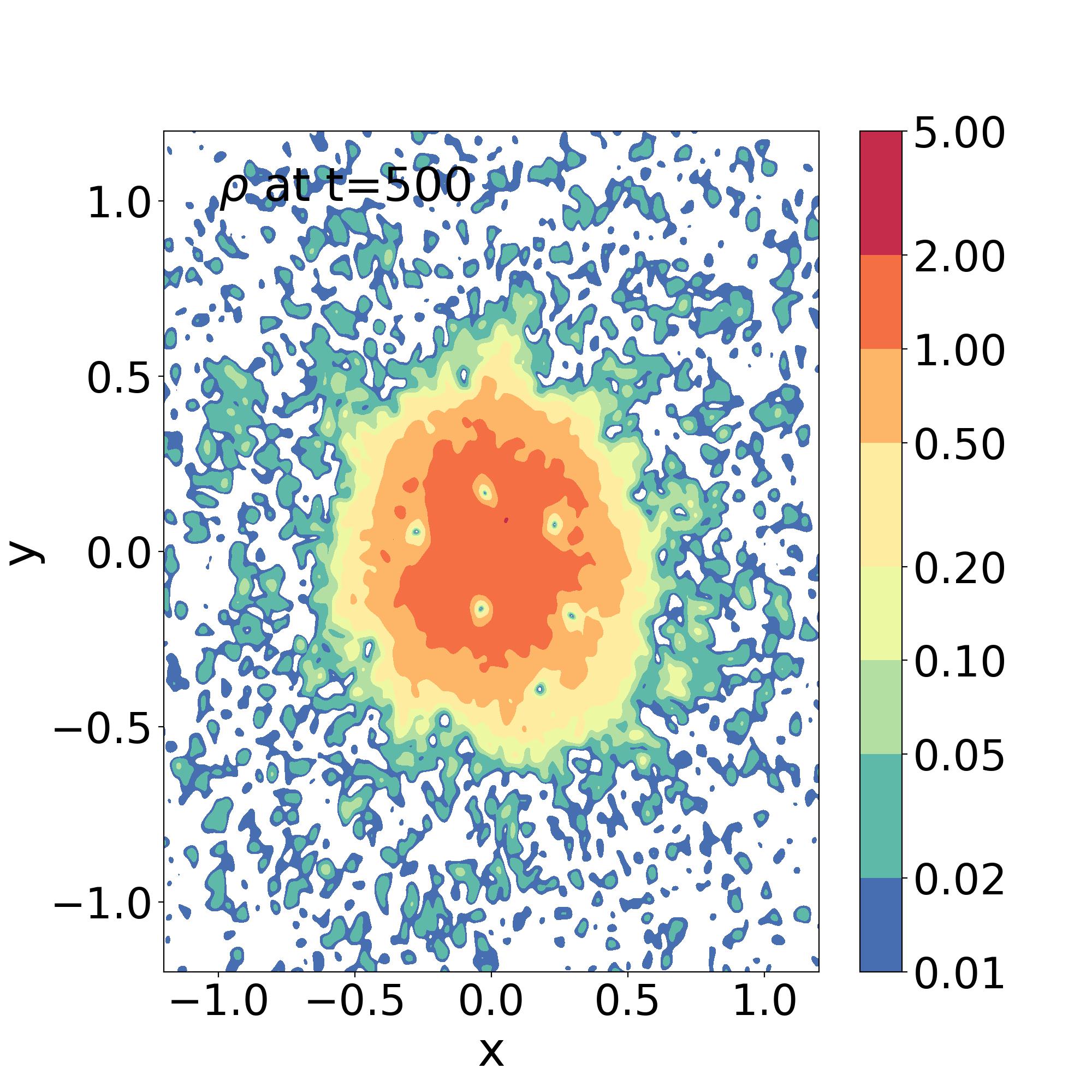}
\includegraphics[height=4.8cm,width=0.32\textwidth]{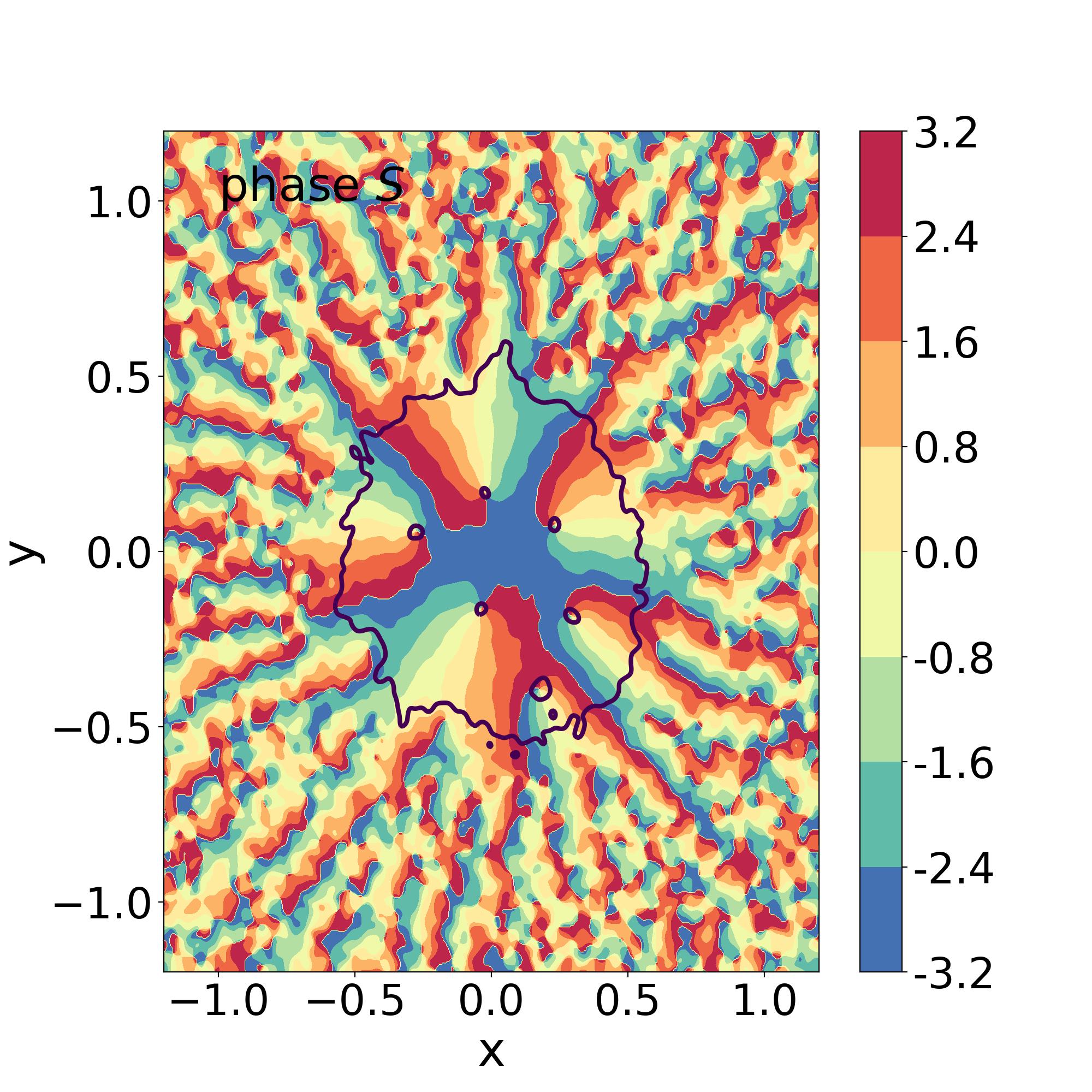}
\includegraphics[height=4.8cm,width=0.32\textwidth]{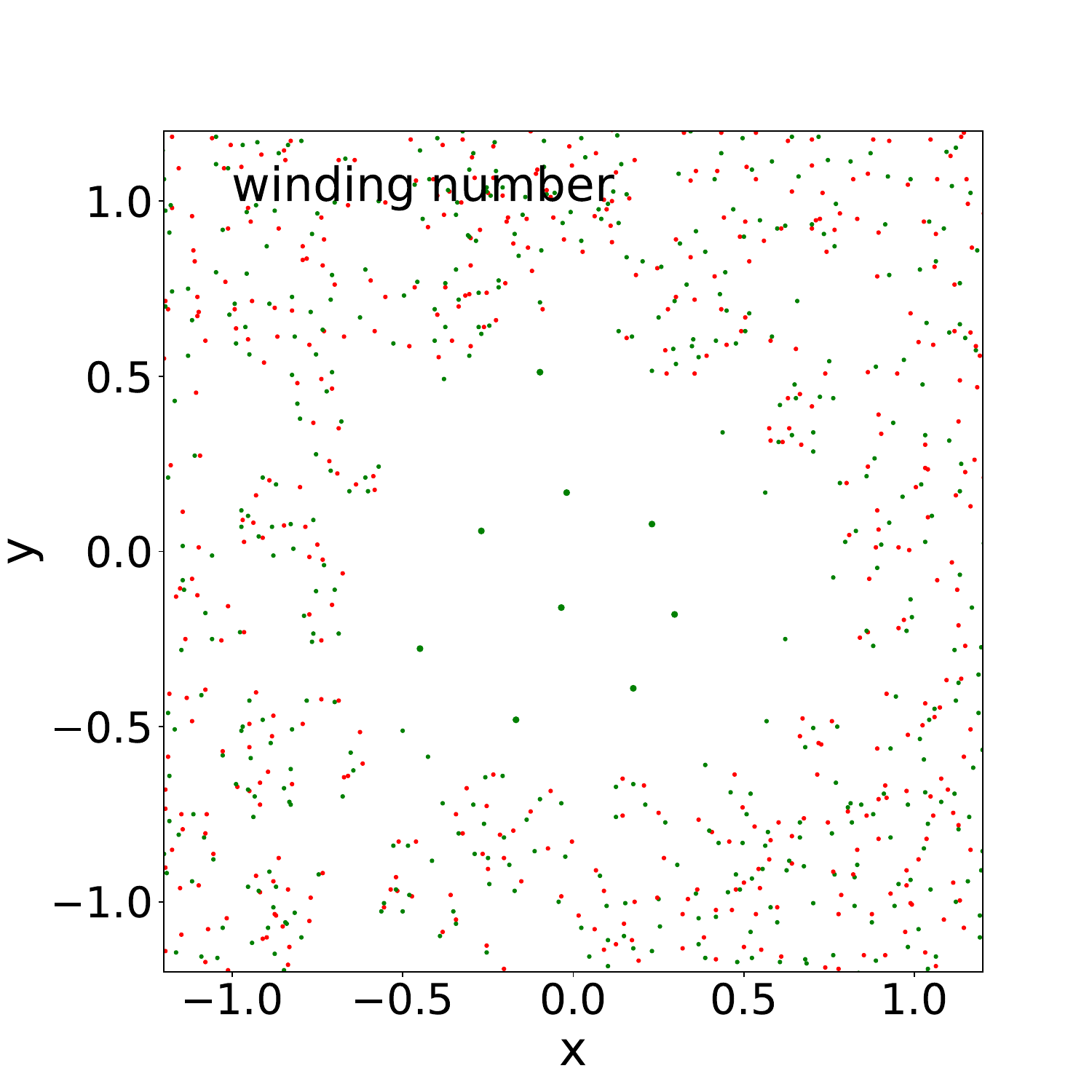}\\
\includegraphics[height=4.8cm,width=0.32\textwidth]{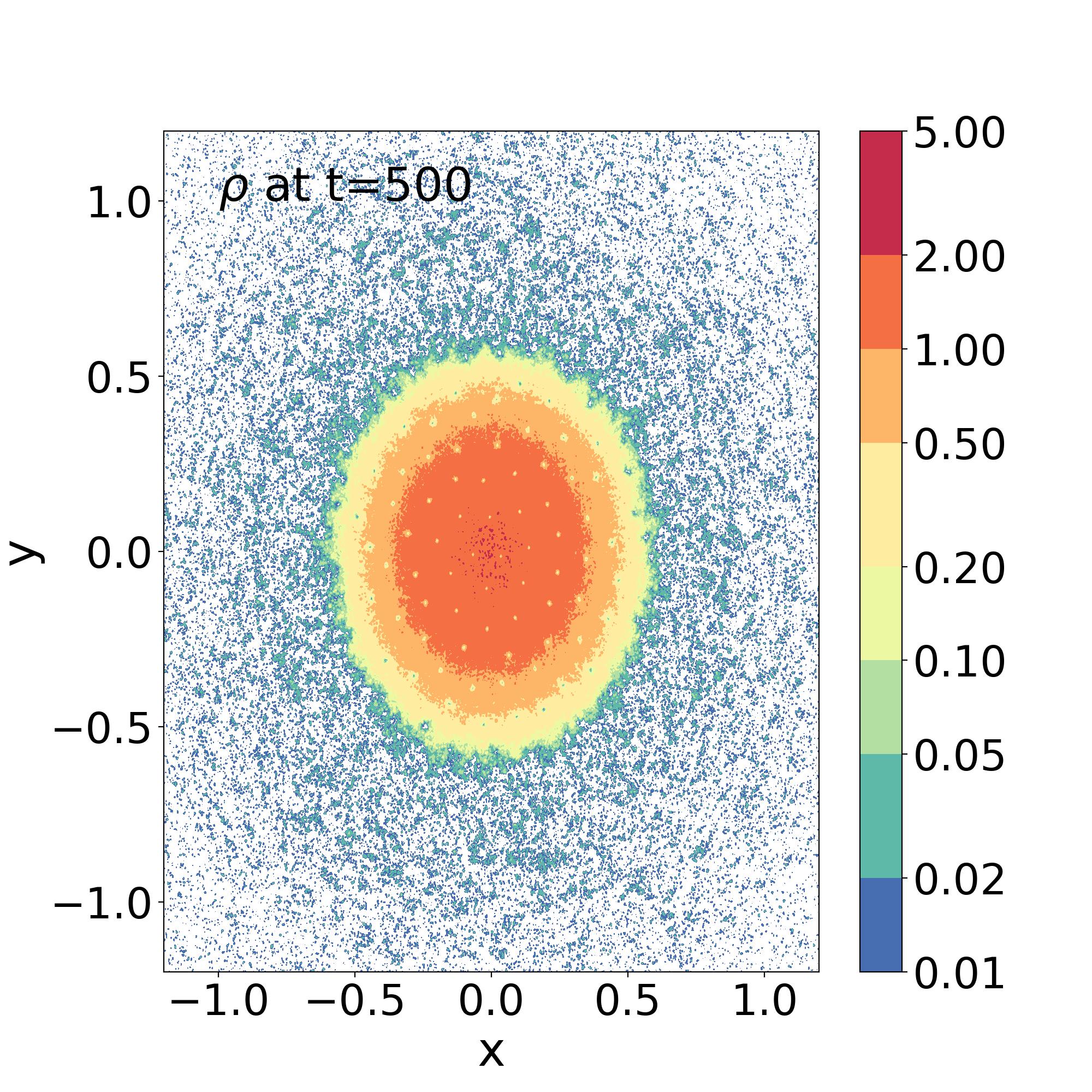}
\includegraphics[height=4.8cm,width=0.32\textwidth]{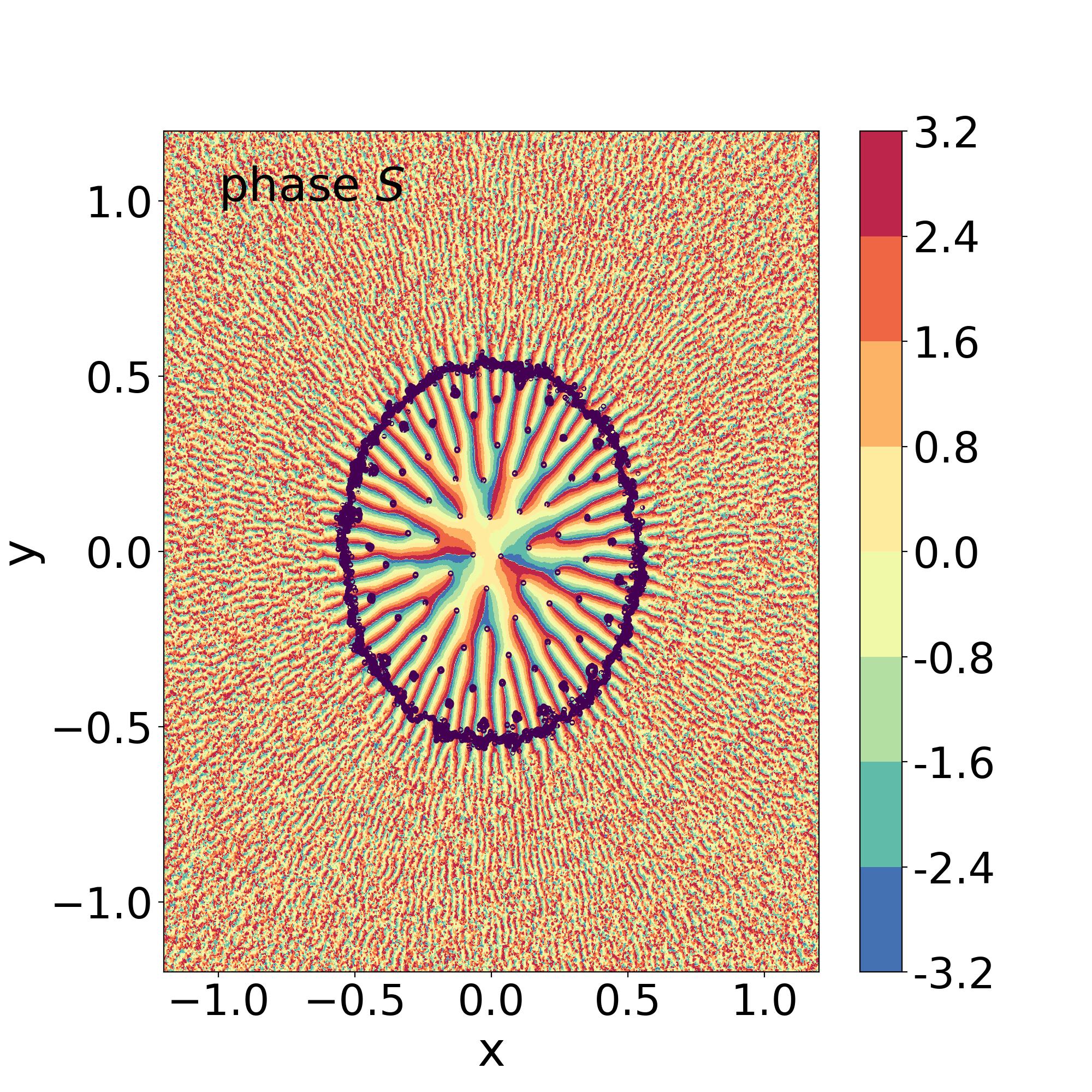}
\includegraphics[height=4.8cm,width=0.32\textwidth]{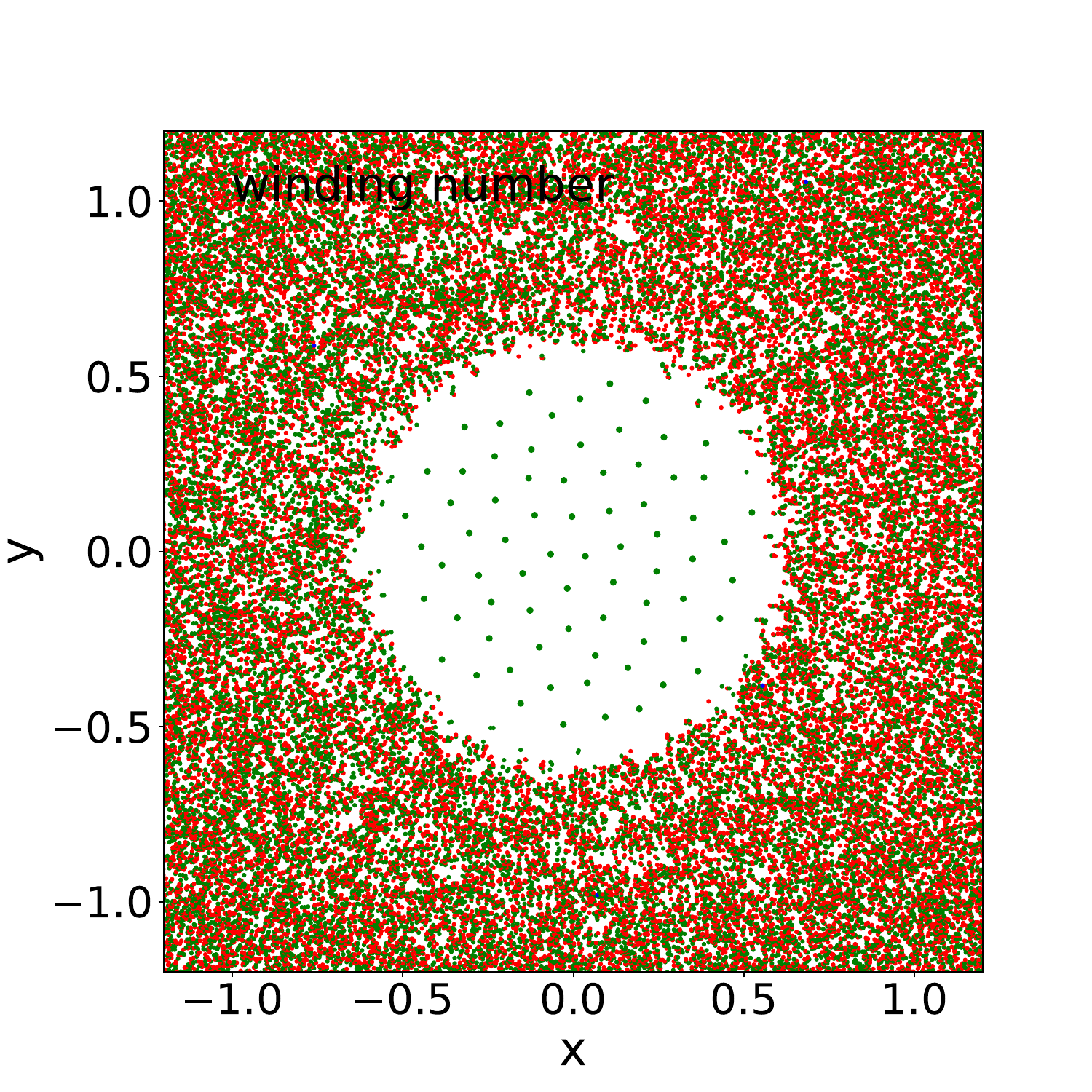}
\caption{
{\it Upper row:} for the case $[\epsilon=0.03, \alpha=1]$, maps of the 2D density field $\rho$,
of the phase $S$ of the wave function, and of the winding number $w$,
at time $t=500$, as in the upper row in Fig.~\ref{fig:2D-mu1-0p01}.
{\it Lower row:} same plots for the case $[\epsilon=0.005, \alpha=1]$.
}
\label{fig:2D-rho-mu1}
\end{figure*}

\begin{figure*}
\centering
\includegraphics[height=4.8cm,width=0.32\textwidth]{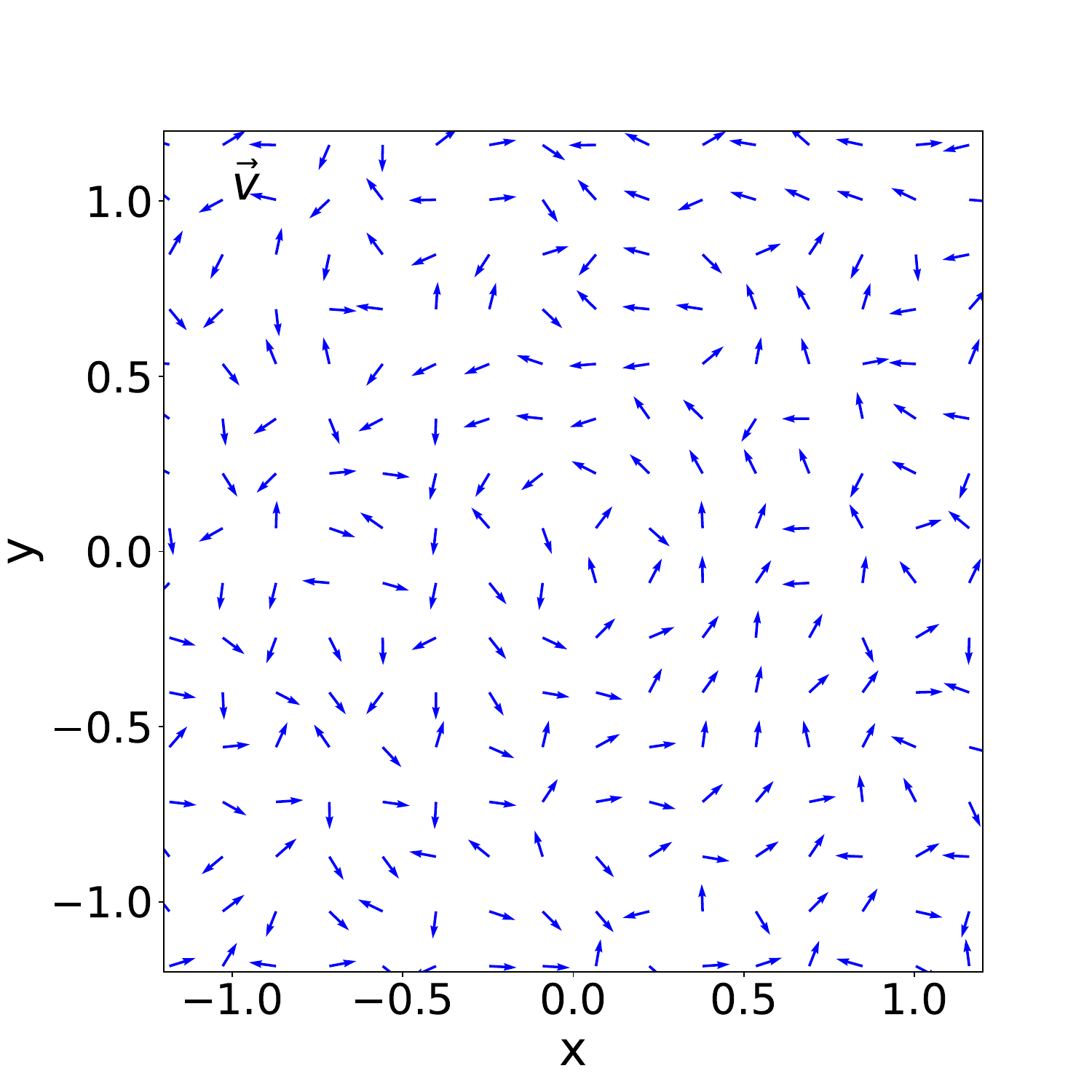}
\includegraphics[height=4.8cm,width=0.32\textwidth]{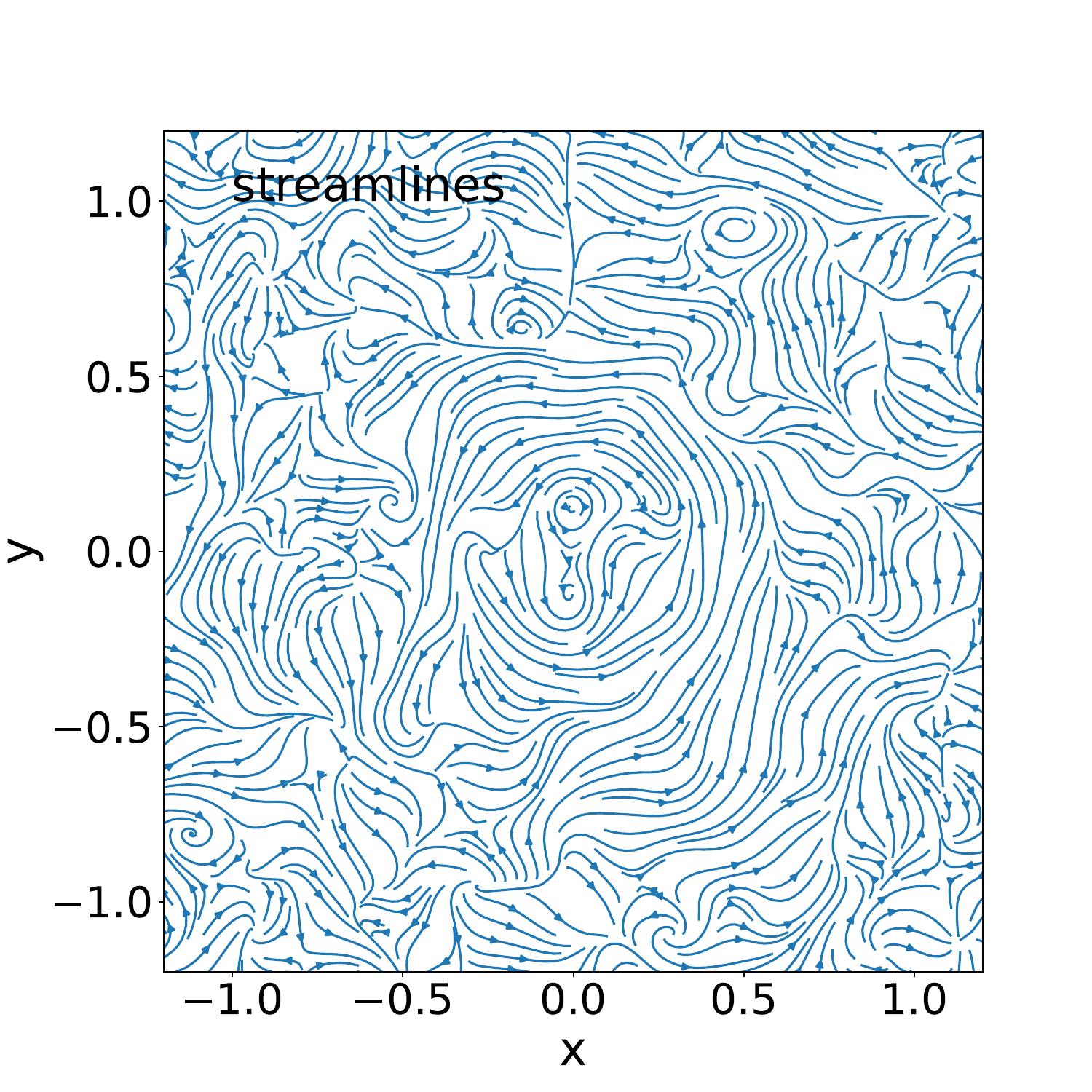}
\includegraphics[height=4.8cm,width=0.32\textwidth]{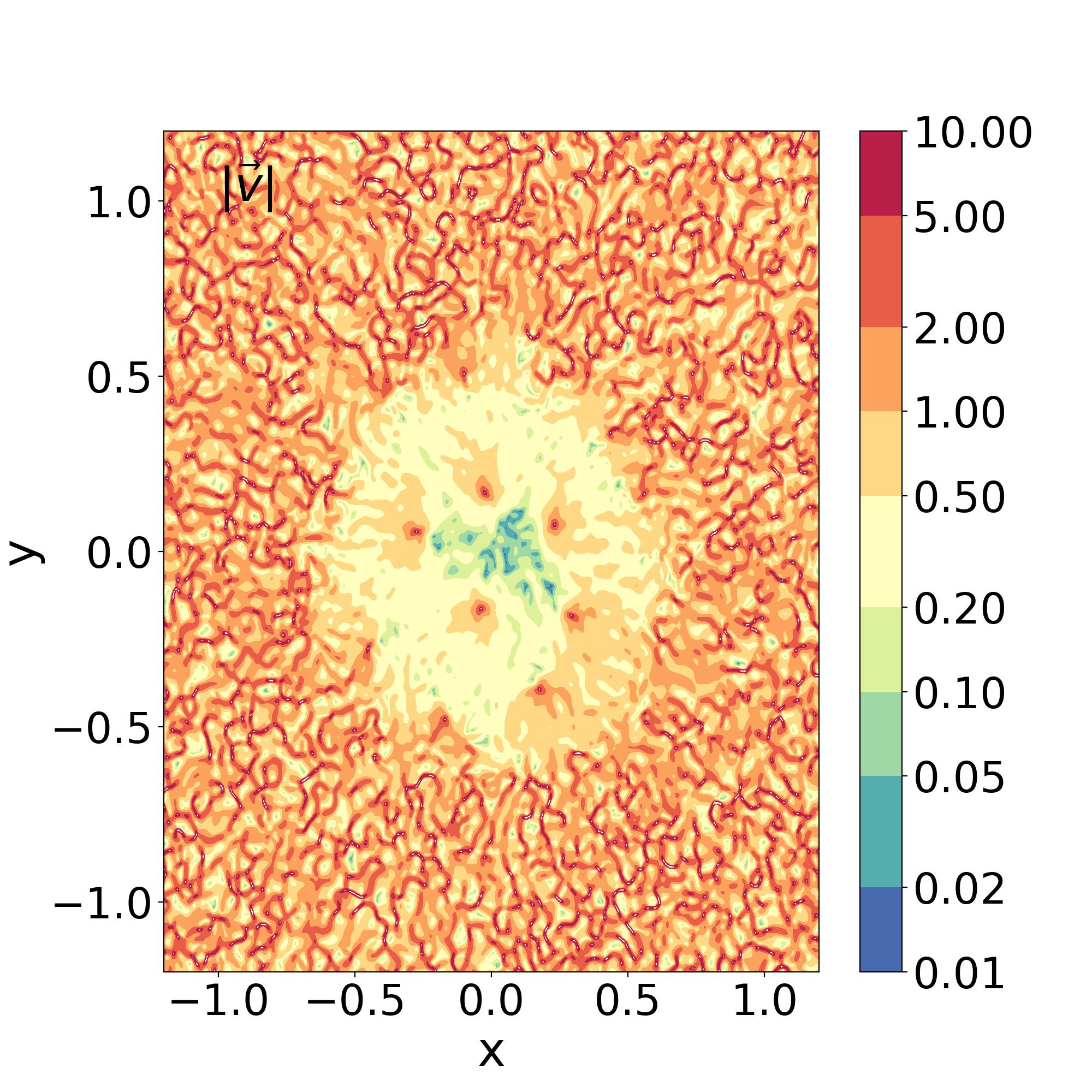}\\
\includegraphics[height=4.8cm,width=0.32\textwidth]{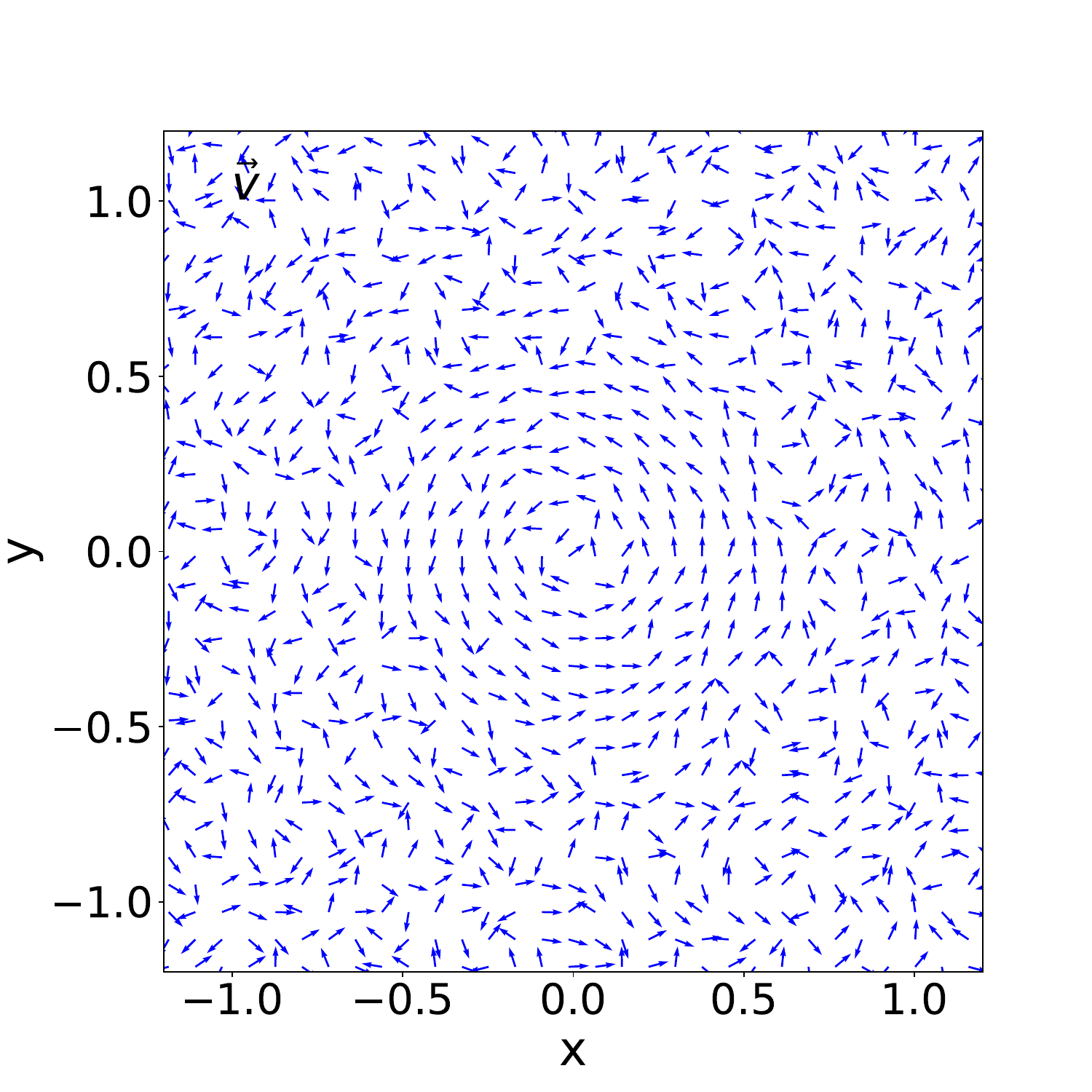}
\includegraphics[height=4.8cm,width=0.32\textwidth]{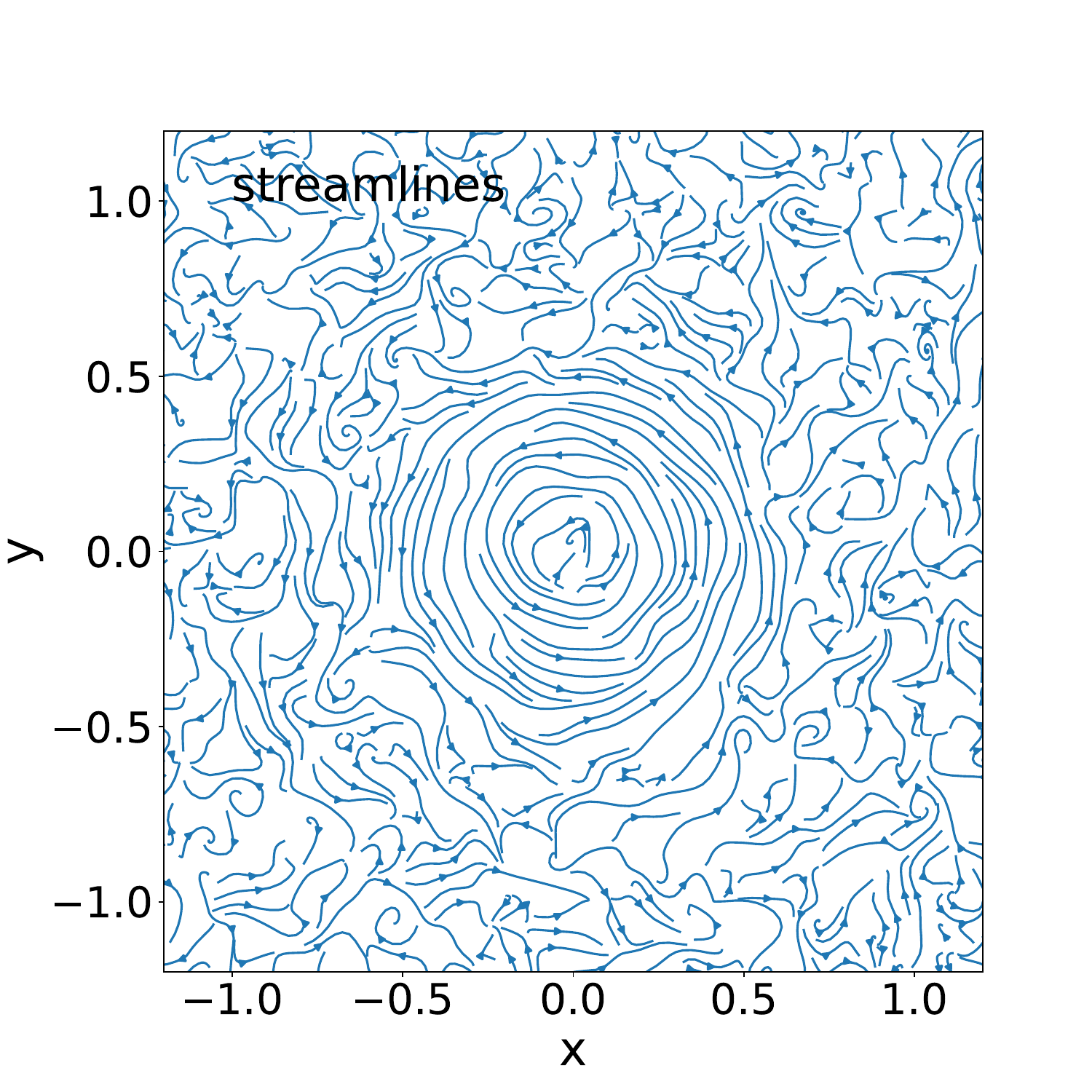}
\includegraphics[height=4.8cm,width=0.32\textwidth]{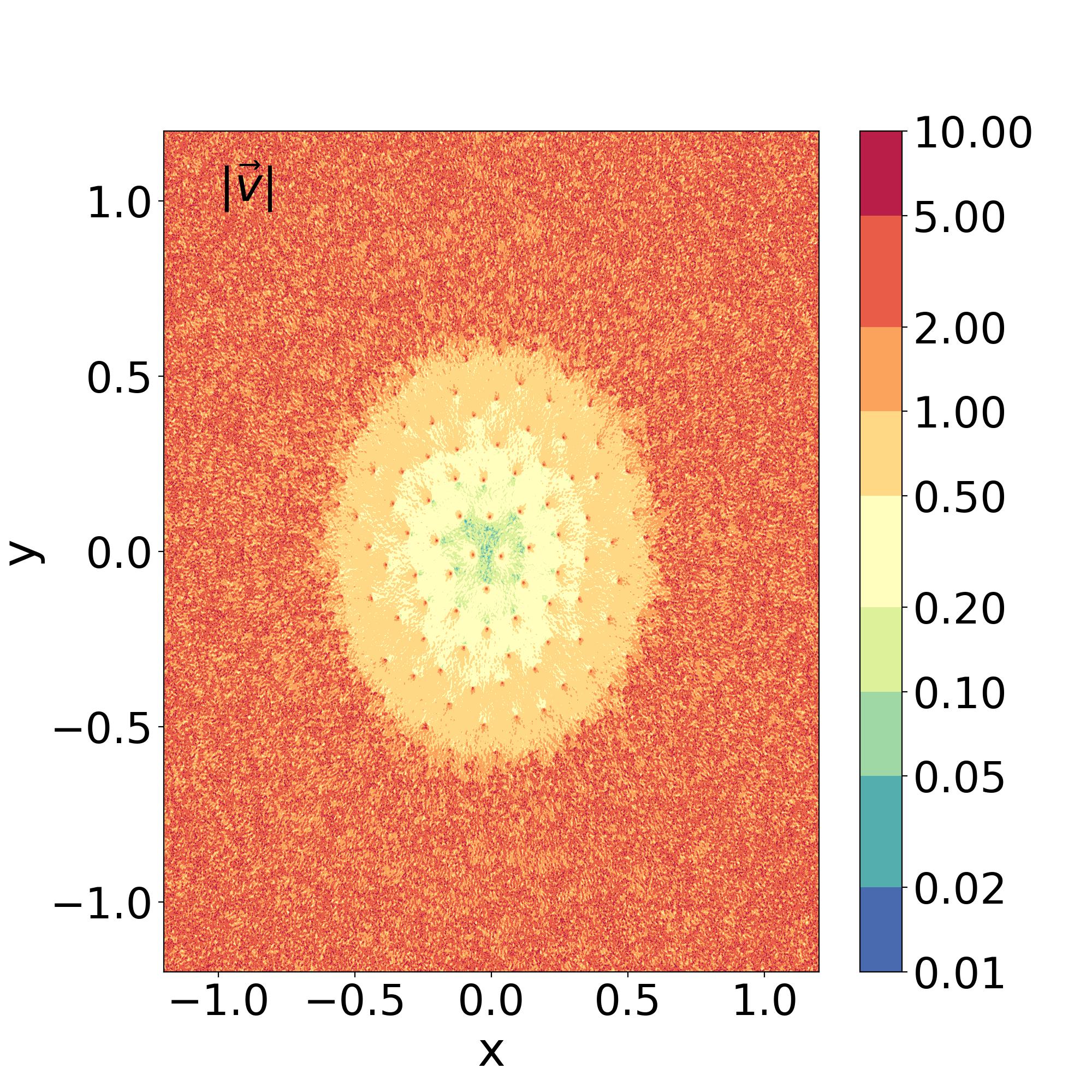}
\caption{
{\it Upper row:} for the case $[\epsilon=0.03, \alpha=1]$, maps of the 2D normalized velocity field
$\vec v/|\vec v|$, of the streamlines and of the velocity amplitude $|\vec v|$,
at time $t=500$, as in the lower row in Fig.~\ref{fig:2D-mu1-0p01}.
{\it Lower row:} same plots for the case $[\epsilon=0.005, \alpha=1]$.
}
\label{fig:2D-v-mu1}
\end{figure*}

We show in Fig.~\ref{fig:wind-mu1-0p01} a superposition of snapshots of the locations of the vortices,
at times $10<t<11$ and $84<t<85$.
The filled circles are the positions at the initial time while the filled squares are the positions at the final time.
We can clearly see the permanence and the circular trajectories of the vortices inside the soliton,
in agreement with the analysis in Sec.~\ref{sec:vortex-network} and Eq.(\ref{eq:dot-rj-v}).
Outside the soliton, the vortices are continuously annihilated and created, as there are many vortices
of either spin, and they follow random trajectories without clear collective rotation.
These features agree with the velocity field and the streamlines displayed in Fig.~\ref{fig:2D-mu1-0p01}.

At the earlier times, $10<t<11$, the soliton has already formed but the system has not yet completely
relaxed to the solid-body rotation equilibrium. Thus, the trajectories are not perfect circles and there still
remain a few negative-spin vortices inside the soliton.
At the later times, $84<t<84$, there are no more negative vortices left and the velocity field has
relaxed to the solid-body rotation, associated with a uniform grid of the vortices.
Thus, the vortices follow regular circles with the common angular velocity $\Omega$.

\subsection{Dependence on $\epsilon$}

We now consider how the numerical results obtained in the previous section vary with $\epsilon$,
the parameter that measures the ratio of the de Broglie wavelength to the size of the system
as in Eq.(\ref{eq:epsilon-de-Broglie}).

We performed simulations for the cases $\epsilon=0.03$ and $\epsilon=0.005$.
The main behaviors remain the same as for the case $\epsilon=0.01$ studied in the previous section.
The 1D profiles are similar to those shown in Figs.~\ref{fig:evol-mu1-0p01}
and \ref{fig:v-mu1-0p01}, but the fluctuations are broader for $\epsilon=0.03$ and narrower for
$\epsilon=0.005$, in agreement with the linear scaling over $\epsilon$ of the de Broglie wavelength
(\ref{eq:de-Broglie-rescaled}).

\begin{figure}
\centering
\includegraphics[height=4.cm,width=0.235\textwidth]{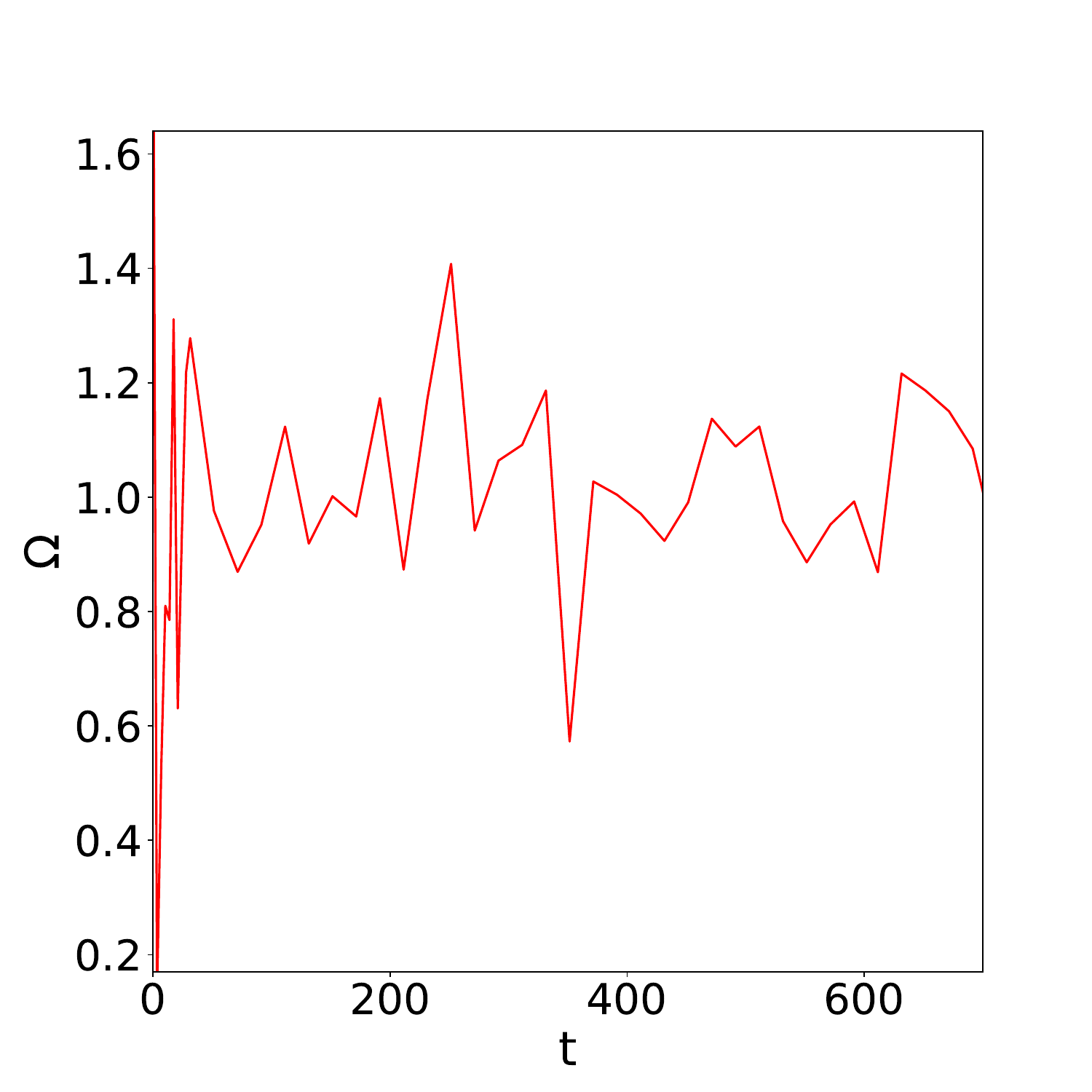}
\includegraphics[height=4.cm,width=0.235\textwidth]{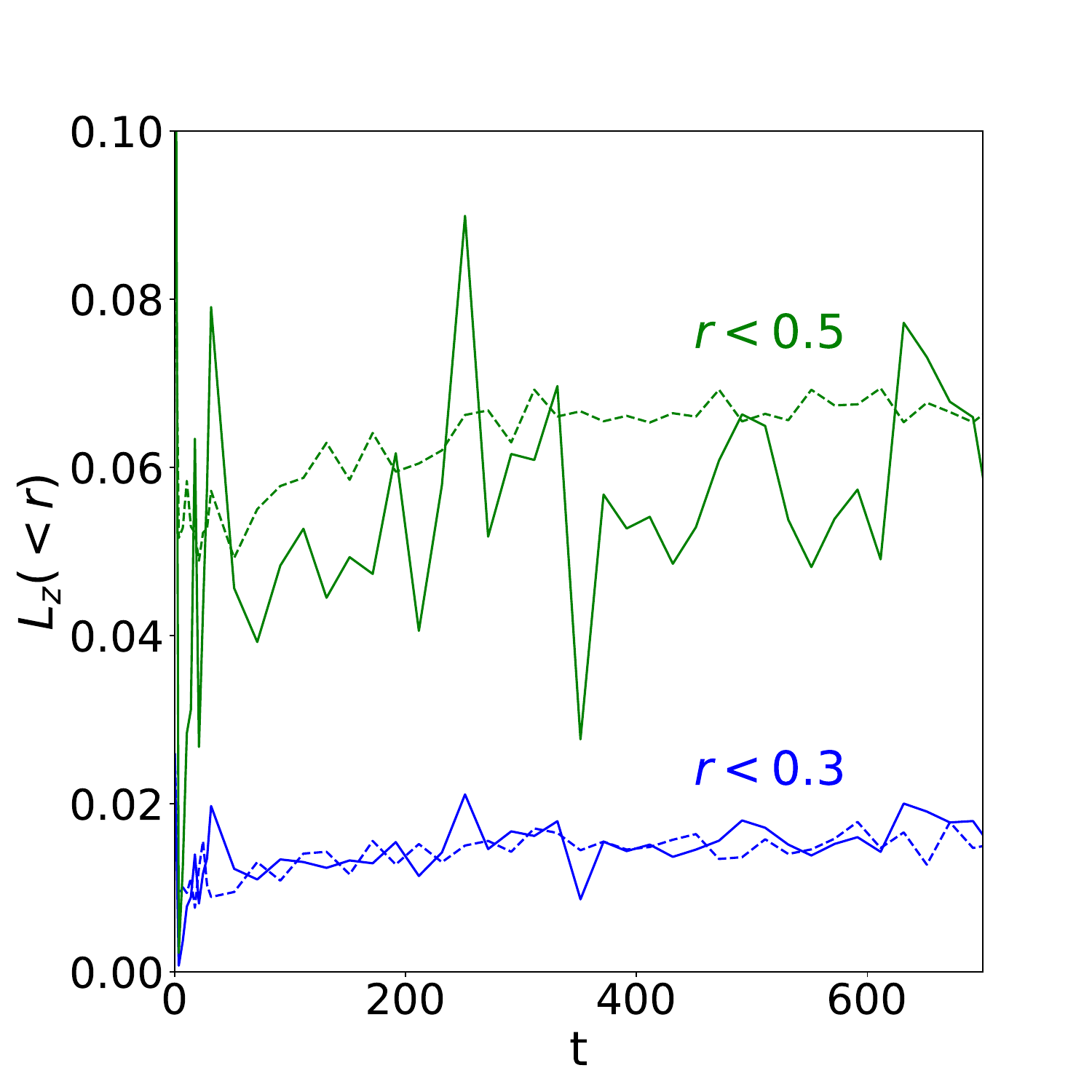}
\includegraphics[height=4.cm,width=0.235\textwidth]{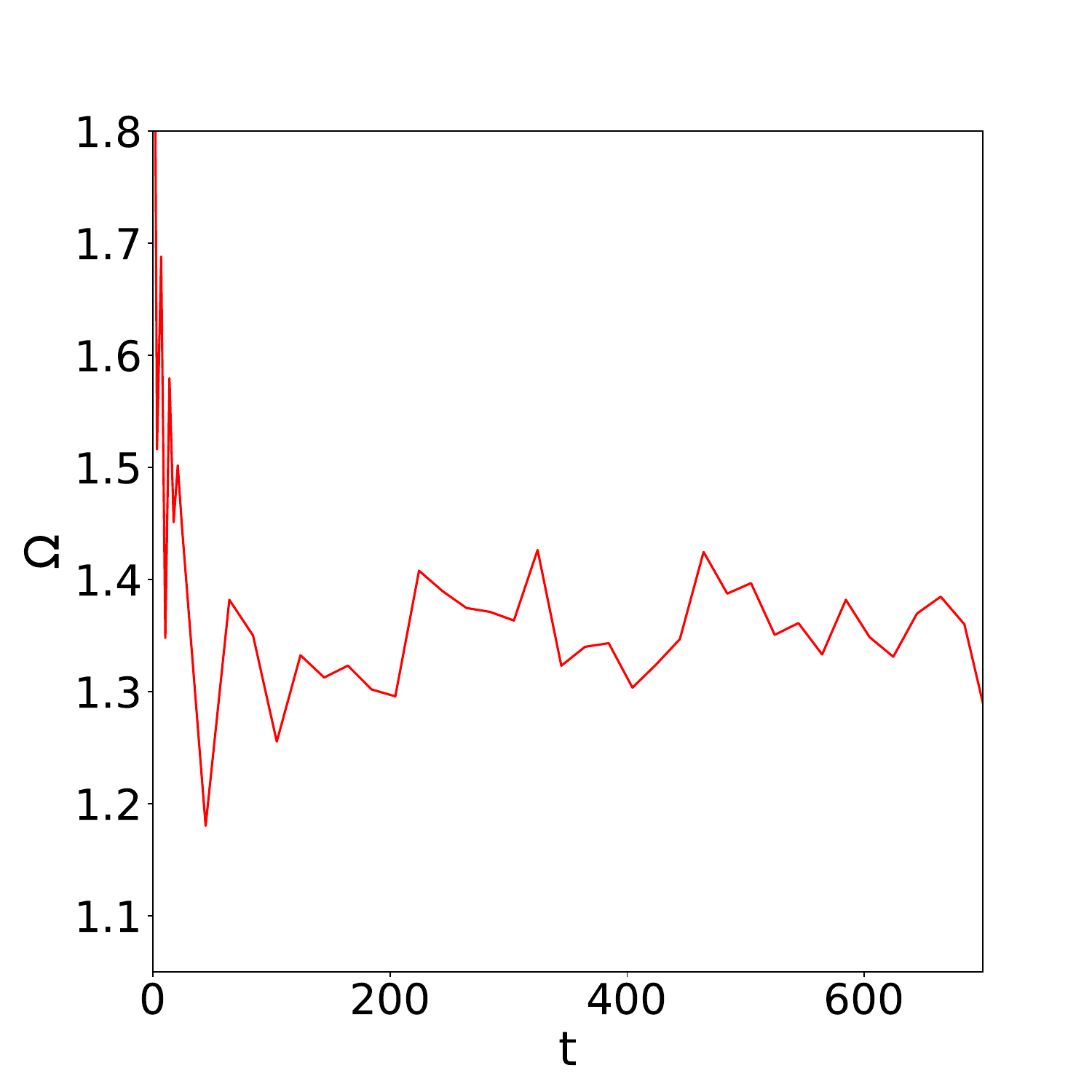}
\includegraphics[height=4.cm,width=0.235\textwidth]{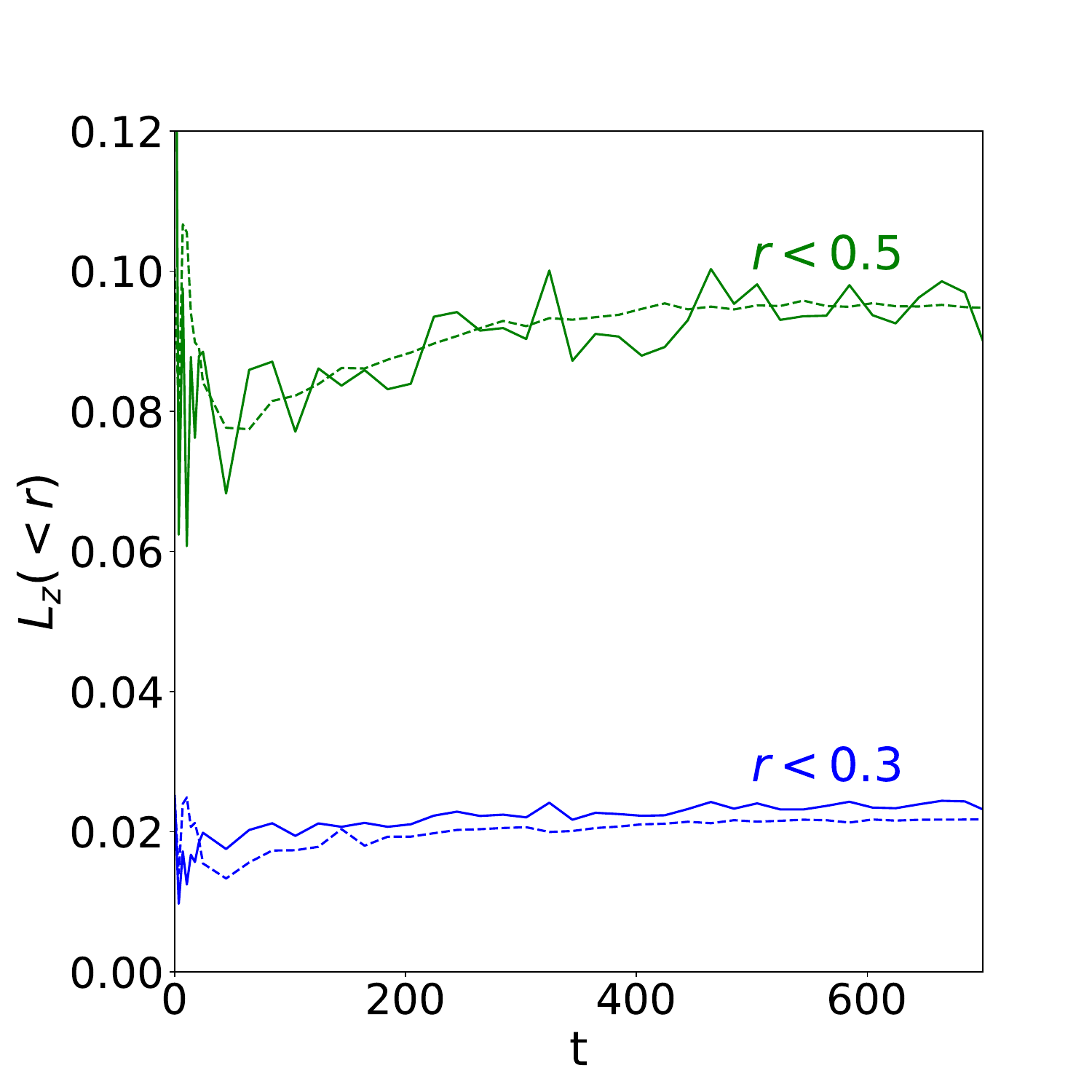}
\caption{
{\it Upper row:} for the case $[\epsilon=0.03, \alpha=1]$, rotation rate $\Omega(t)$ as
a function of time and angular momentum $L_z(<r,t)$ within the two
radii $r=0.3$ and $r=0.5$, as in Fig.~\ref{fig:Om-t-mu1-0p01}.
{\it Lower row:} same plots for the case $[\epsilon=0.005, \alpha=1]$.
}
\label{fig:Om-t-mu1}
\end{figure}

\begin{figure}
\centering
\includegraphics[height=5.5cm,width=0.41\textwidth]{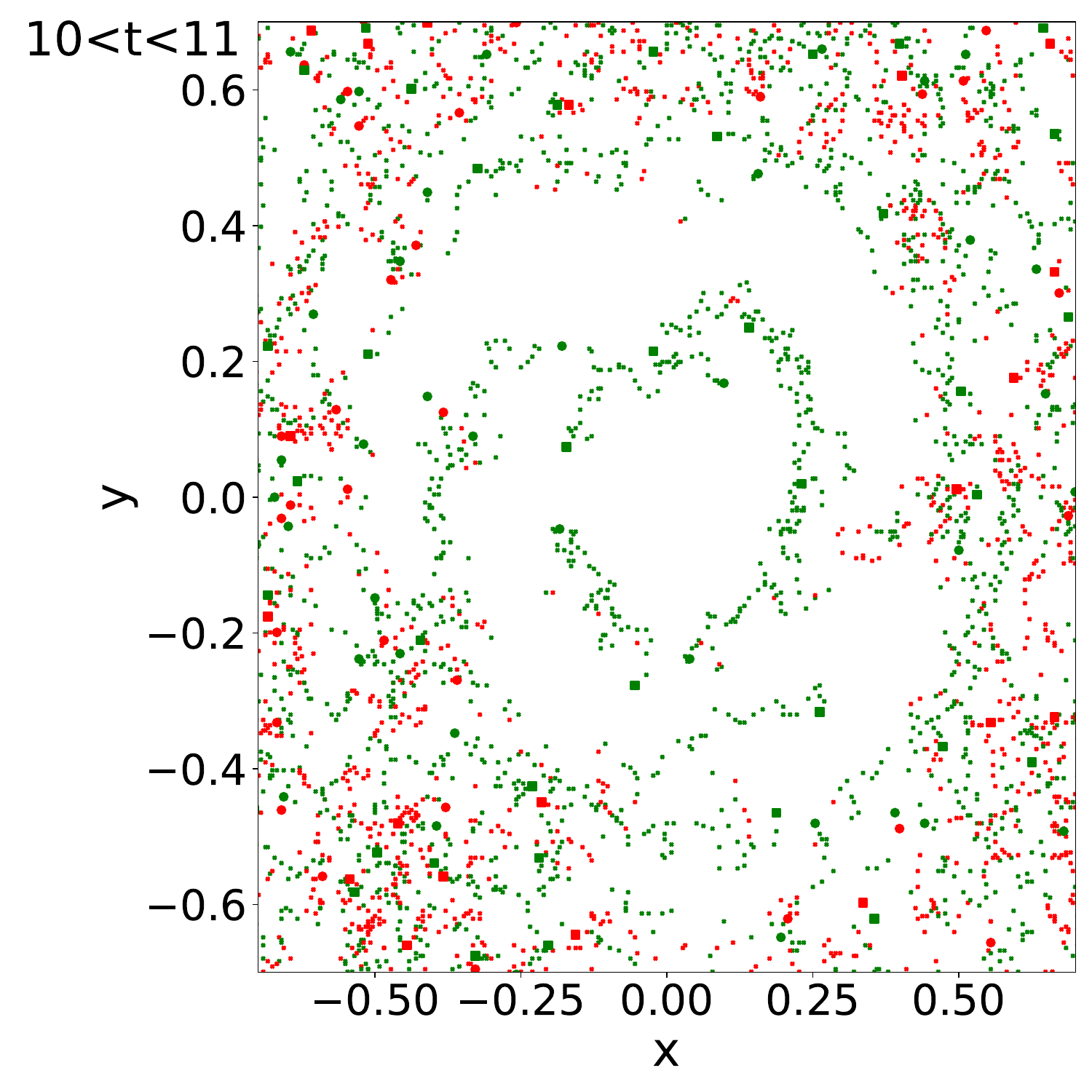}
\includegraphics[height=5.5cm,width=0.41\textwidth]{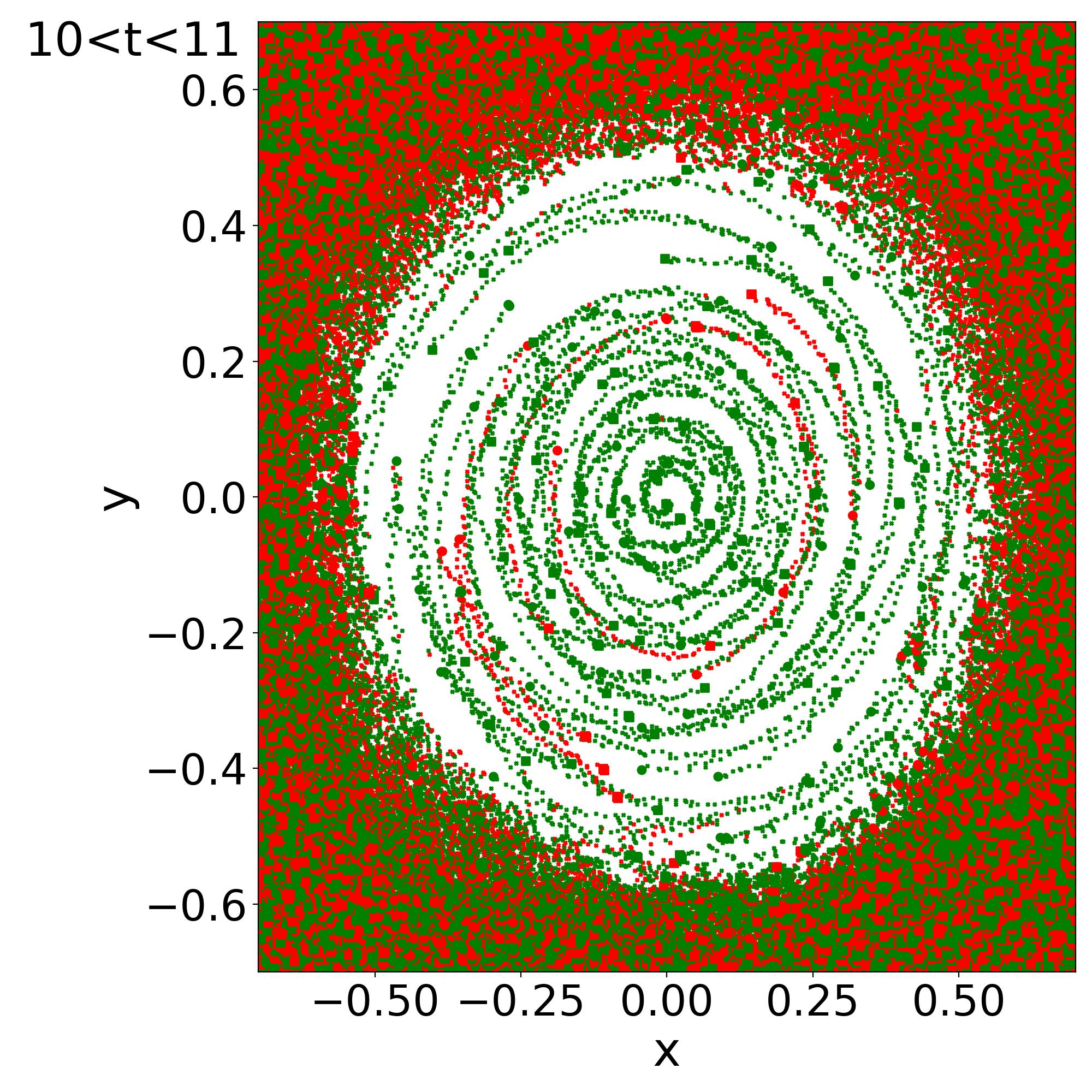}
\caption{
Superposition of many snapshots of the maps of vortices for the cases
$[\epsilon=0.03, \alpha=1]$ (upper panel) and
$[\epsilon=0.005, \alpha=1]$ (lower panel),
as in Fig.~\ref{fig:wind-mu1-0p01}.
}
\label{fig:wind-mu1}
\end{figure}

These differences are most clearly apparent in the 2D maps shown in Fig.~\ref{fig:2D-rho-mu1}.
The soliton profiles are similar and the main difference is the reduced/increased number of vortices for
$\epsilon=0.03$/$\epsilon=0.005$, in agreement with the scaling (\ref{eq:vortex-density}).
For $\epsilon=0.03$ there only remain 6 vortices inside the soliton, whereas for $\epsilon=0.005$
there remain about 80 vortices.
In agreement with the results of Sec.~\ref{sec:vortices}, for smaller $\epsilon$ the healing length
decreases faster than the distance between neighbouring vortices, so that the system remains in the dilute
regime.
We can also see on the density and phase maps how structures develop on smaller scales as
$\epsilon$ decreases, following the linear scaling of the de Broglie wavelength
(\ref{eq:de-Broglie-rescaled}).

The same behaviours can be seen in the velocity maps shown in Fig.~\ref{fig:2D-v-mu1}.
We can see the solid-body rotation in both cases and the disordered velocity field outside of the soliton,
with again smaller structures in the lower-$\epsilon$ case.
The solid-body rotation is somewhat less clear in the velocity map in the case $\epsilon=0.03$ because
the small number of vortices and the larger de Broglie wavelength mean that the continuum limit
in the Thomas-Fermi regime derived in Sec.~\ref{sec:continuum} receives significant corrections.

\begin{figure*}
\centering
\includegraphics[height=4.8cm,width=0.32\textwidth]{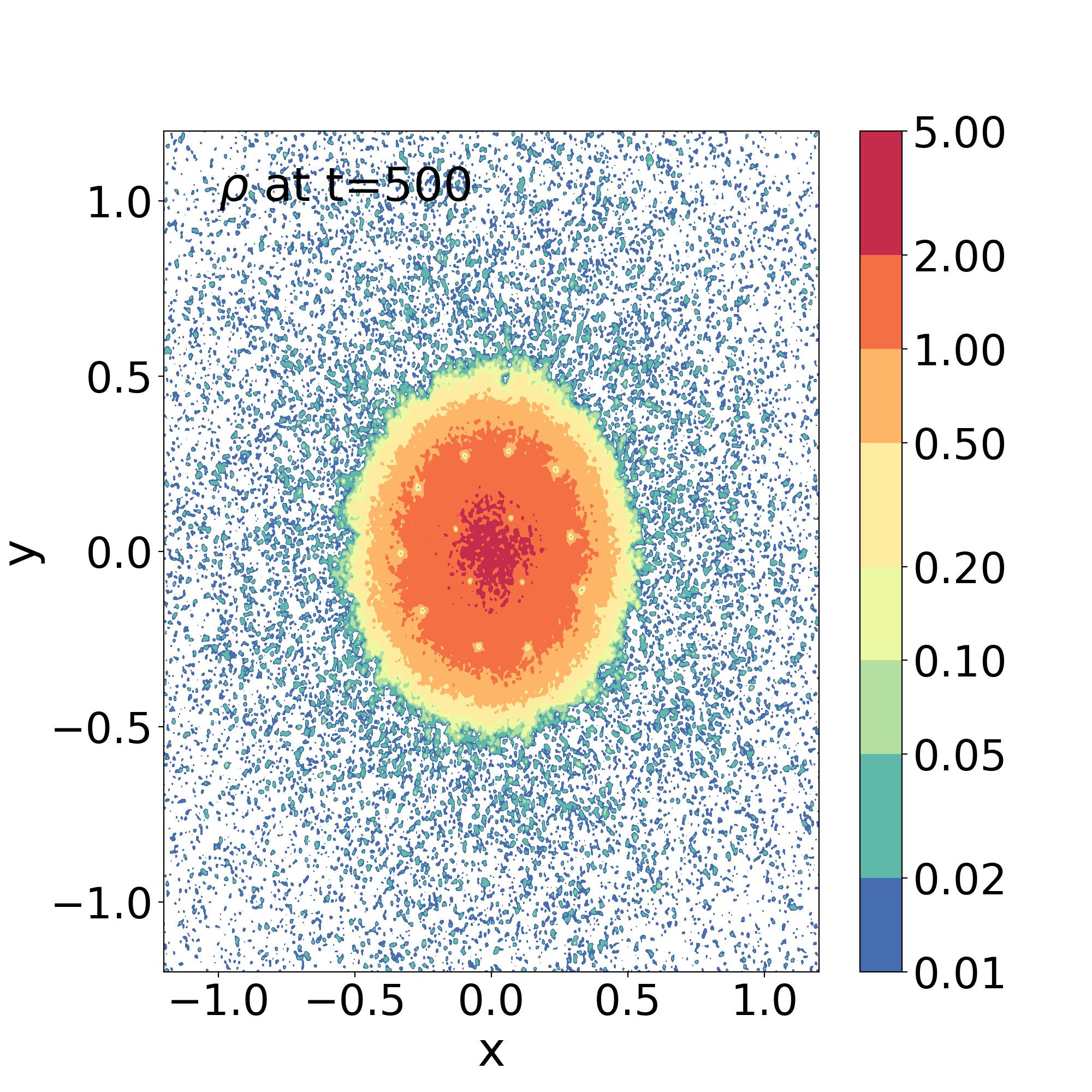}
\includegraphics[height=4.8cm,width=0.32\textwidth]{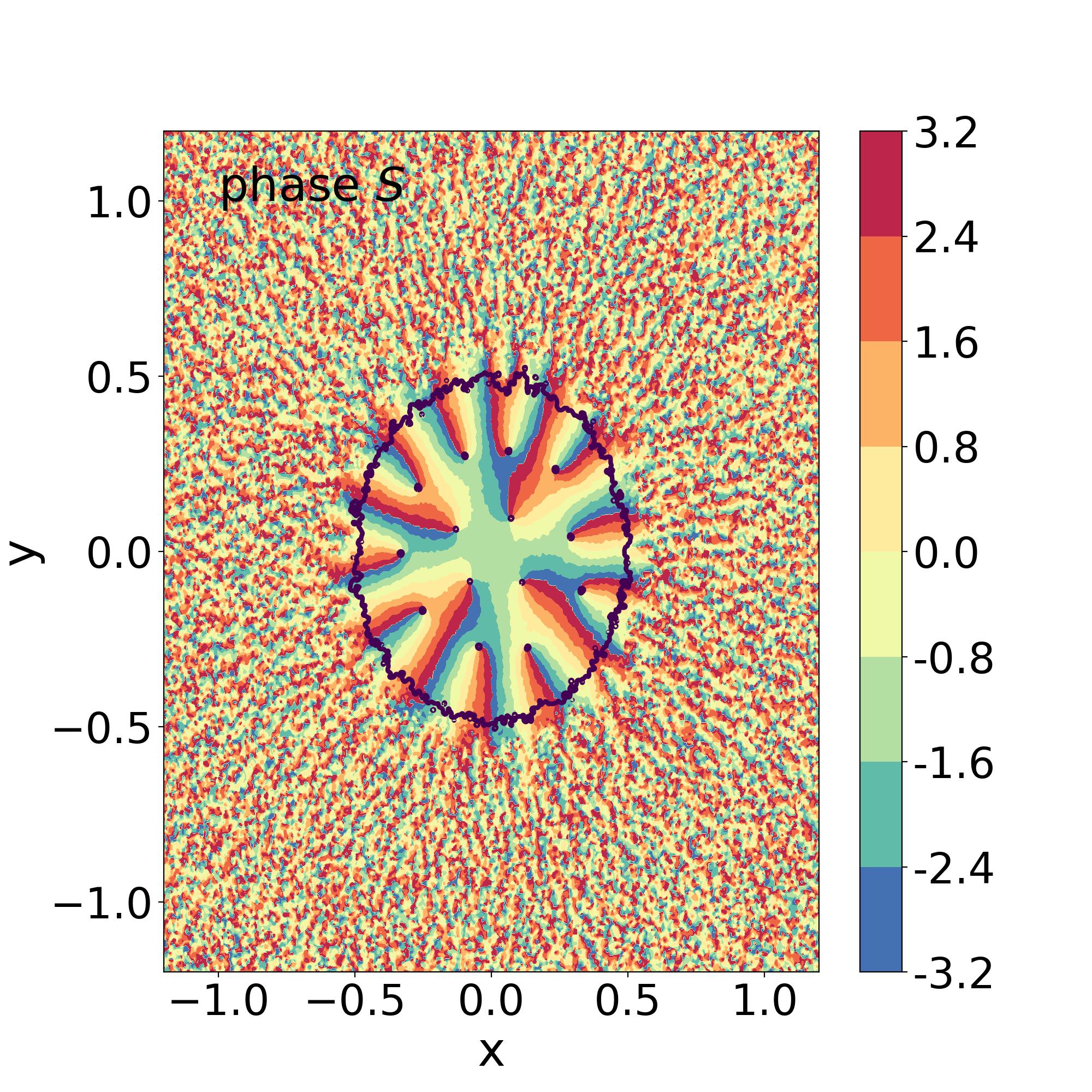}
\includegraphics[height=4.8cm,width=0.32\textwidth]{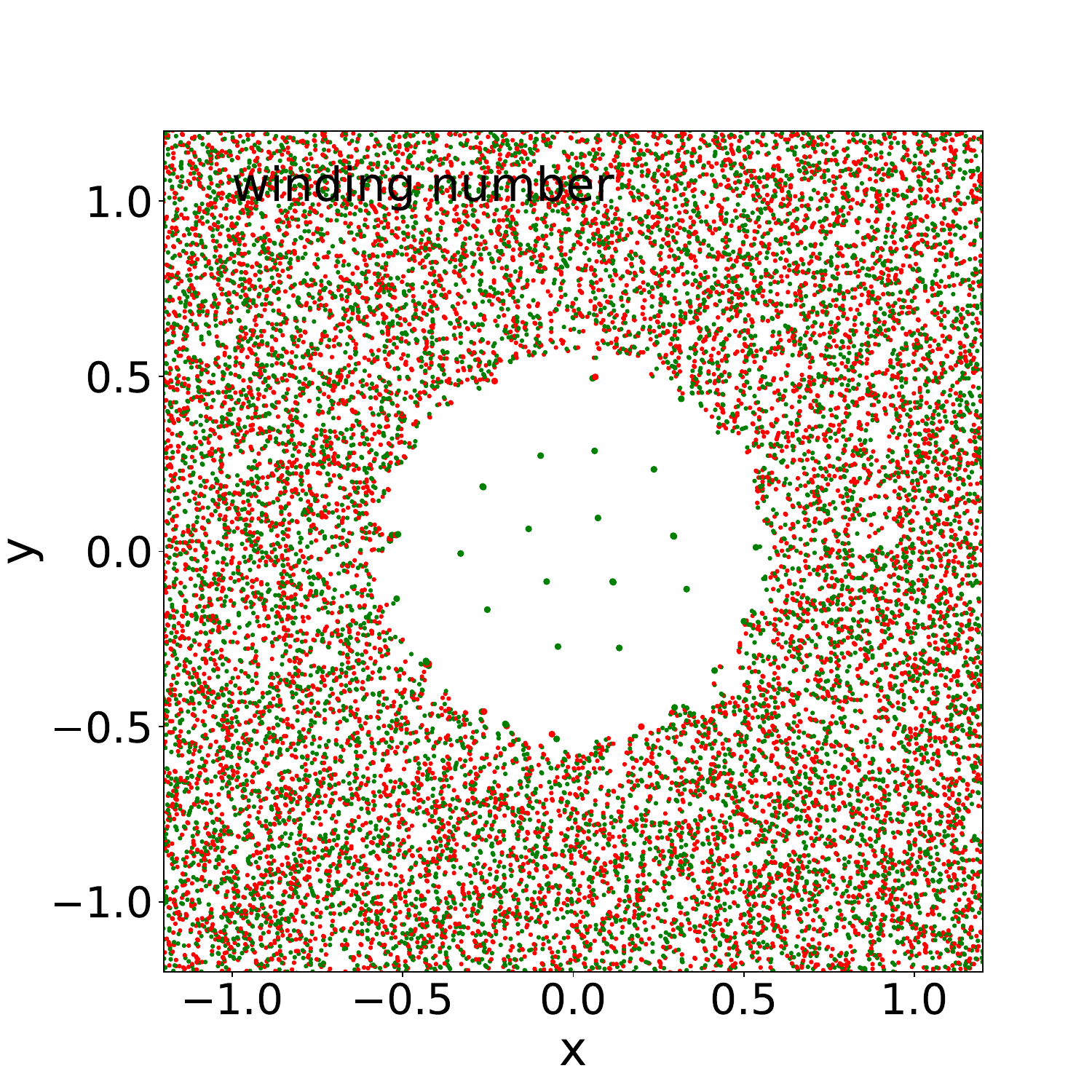}\\
\includegraphics[height=4.8cm,width=0.32\textwidth]{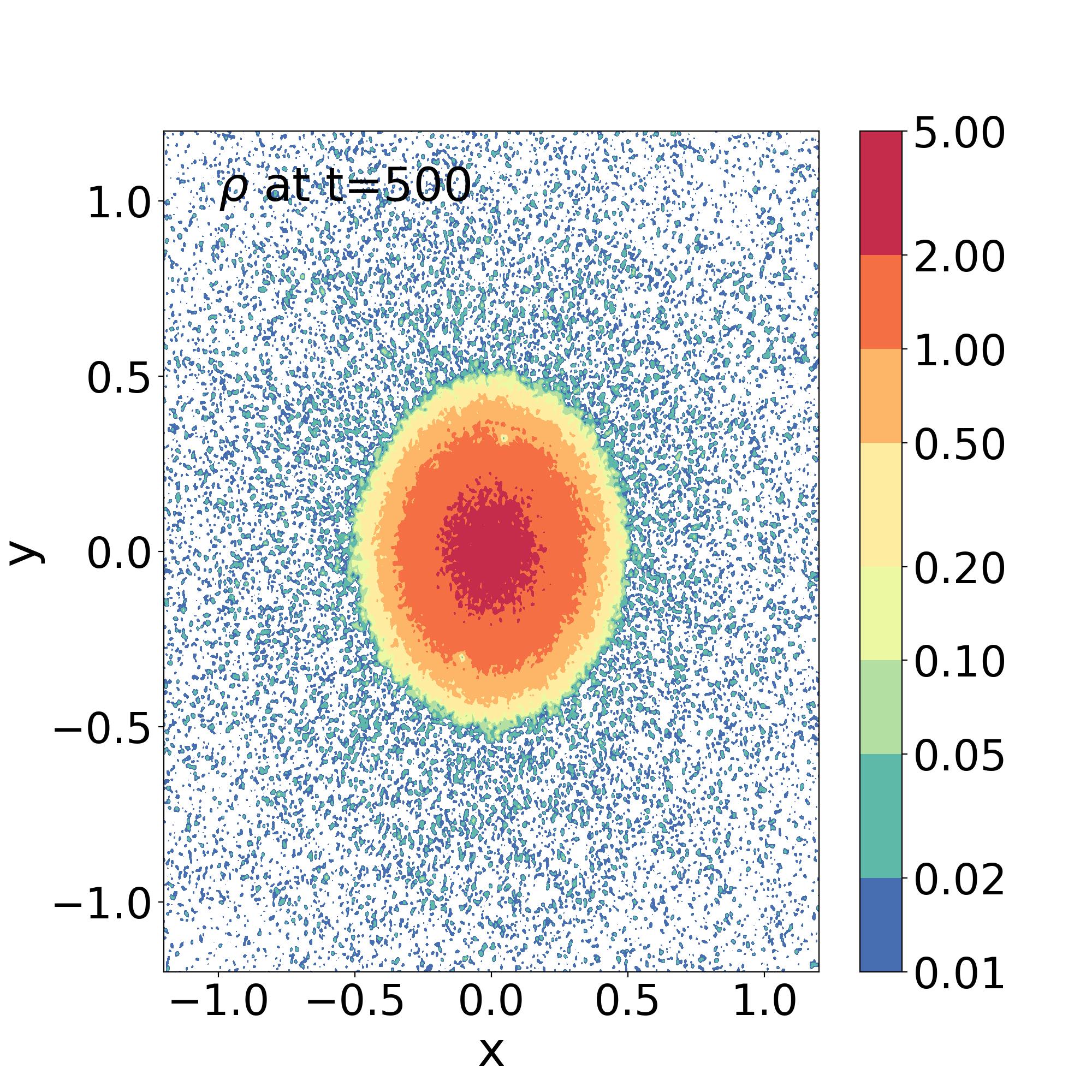}
\includegraphics[height=4.8cm,width=0.32\textwidth]{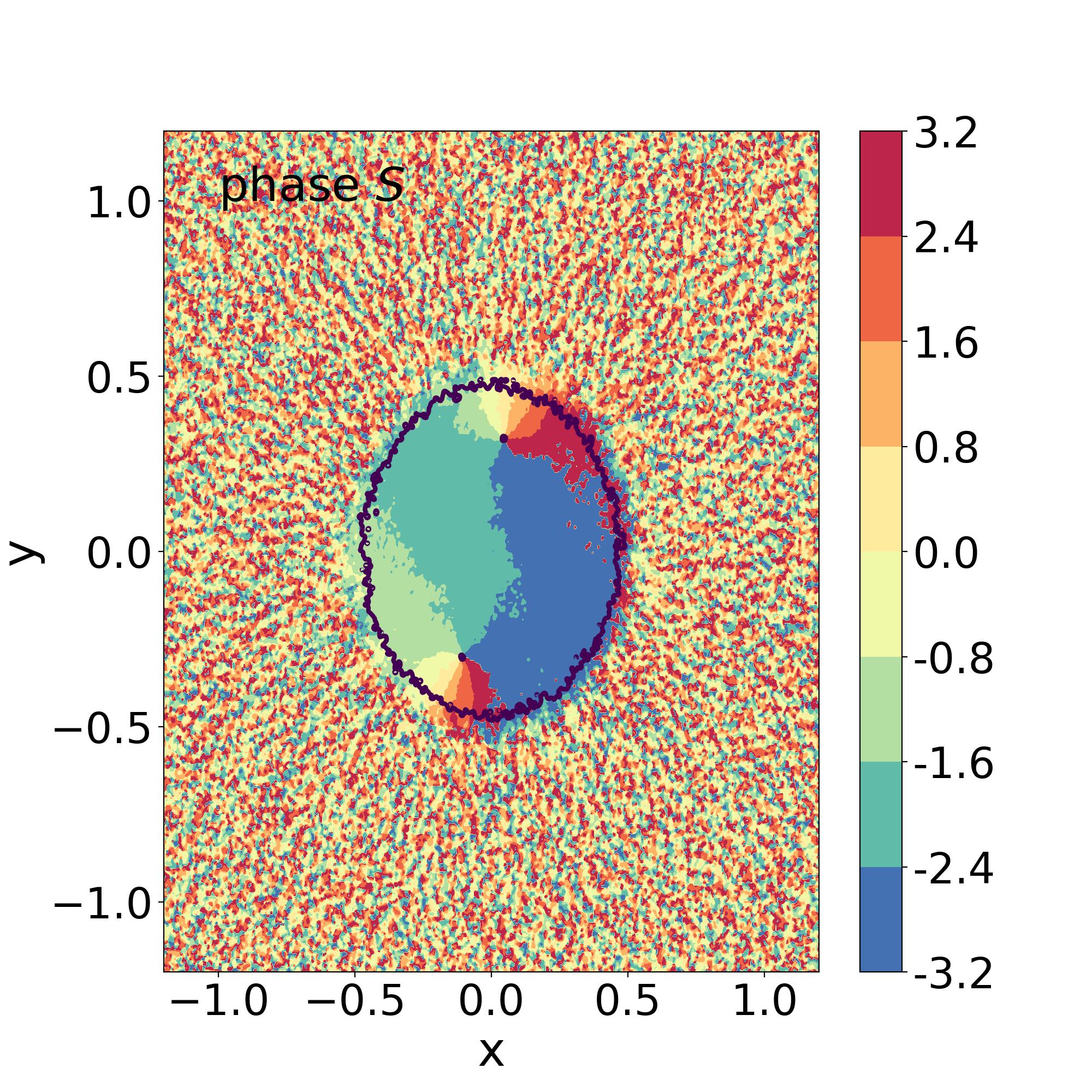}
\includegraphics[height=4.8cm,width=0.32\textwidth]{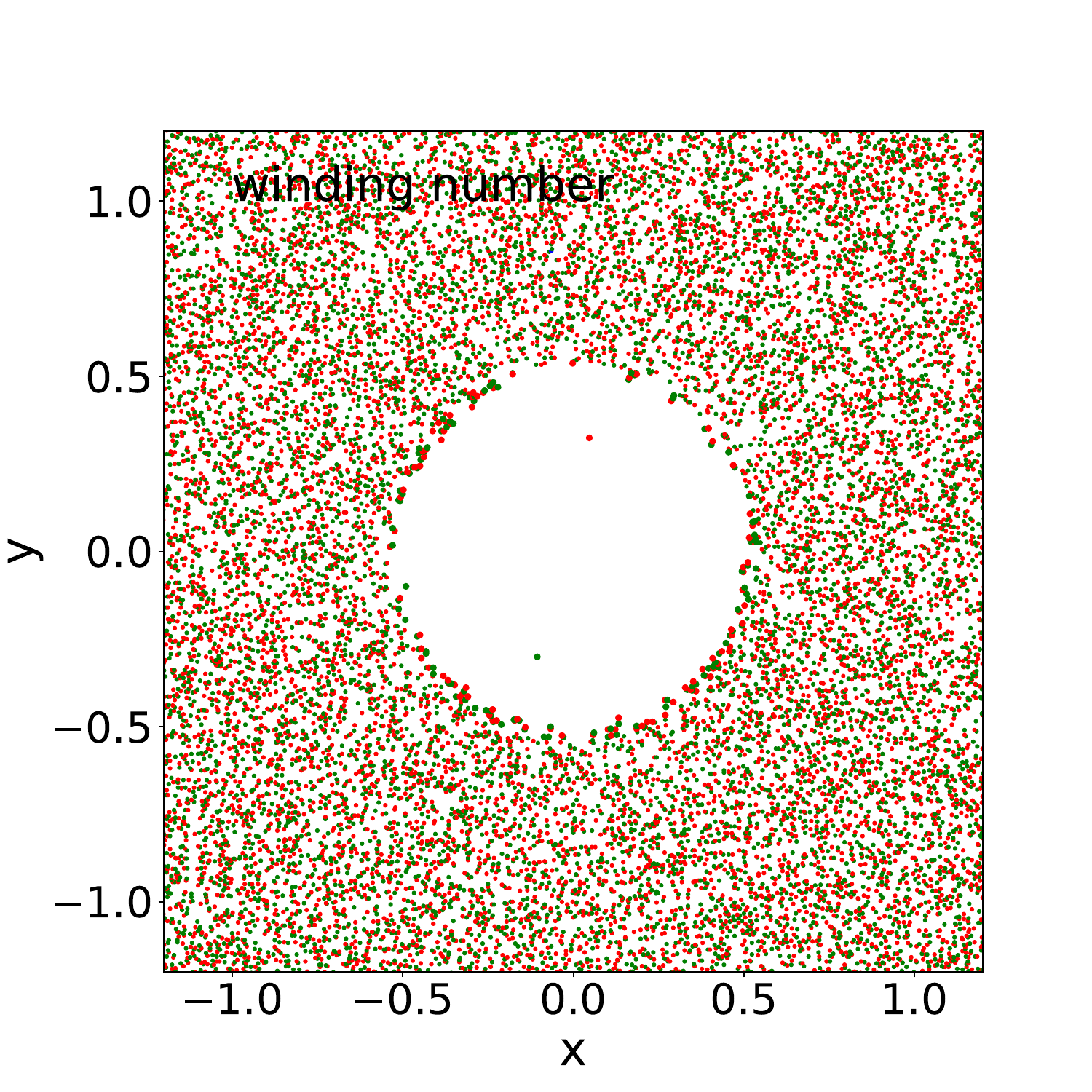}
\caption{
{\it Upper row:} for the case $[\epsilon=0.01, \alpha=0.5]$, maps of the 2D density field $\rho$,
of the phase $S$ of the wave function, and of the winding number $w$,
at time $t=500$, as in the upper row in Fig.~\ref{fig:2D-mu1-0p01}.
{\it Lower row:} same plots for the case $[\epsilon=0.01, \alpha=0]$.
}
\label{fig:2D-rho-0p01}
\end{figure*}

\begin{figure*}
\centering
\includegraphics[height=4.8cm,width=0.32\textwidth]{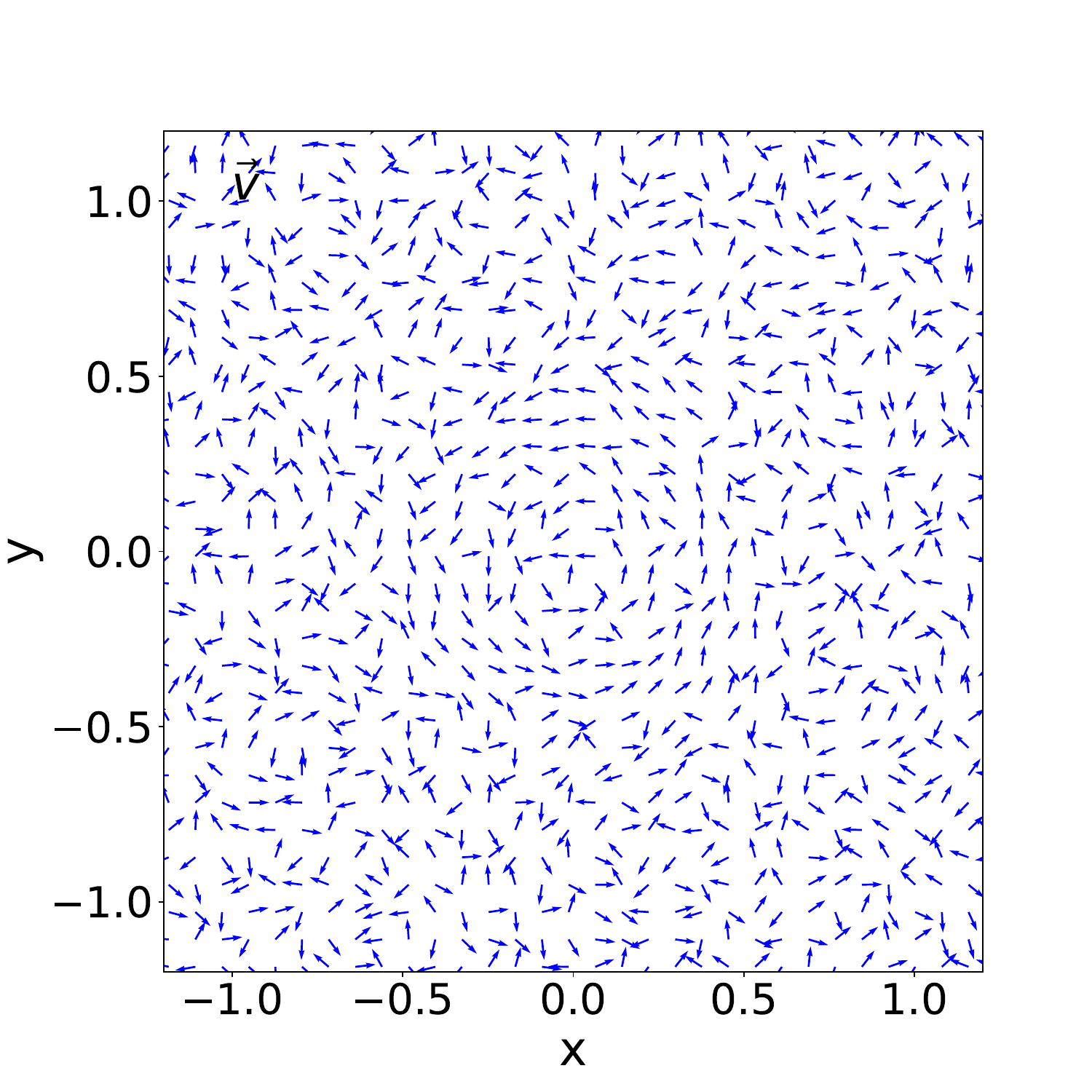}
\includegraphics[height=4.8cm,width=0.32\textwidth]{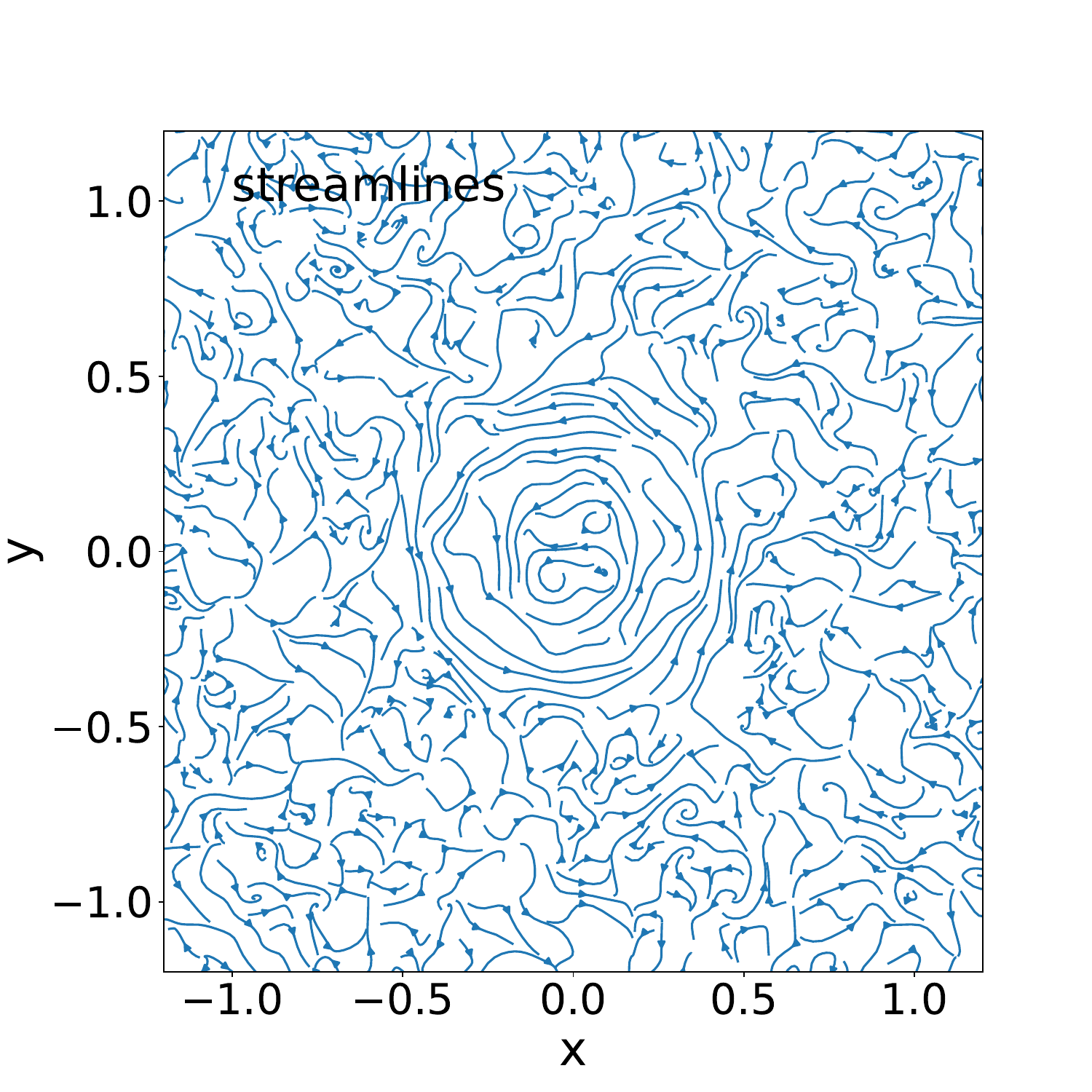}
\includegraphics[height=4.8cm,width=0.32\textwidth]{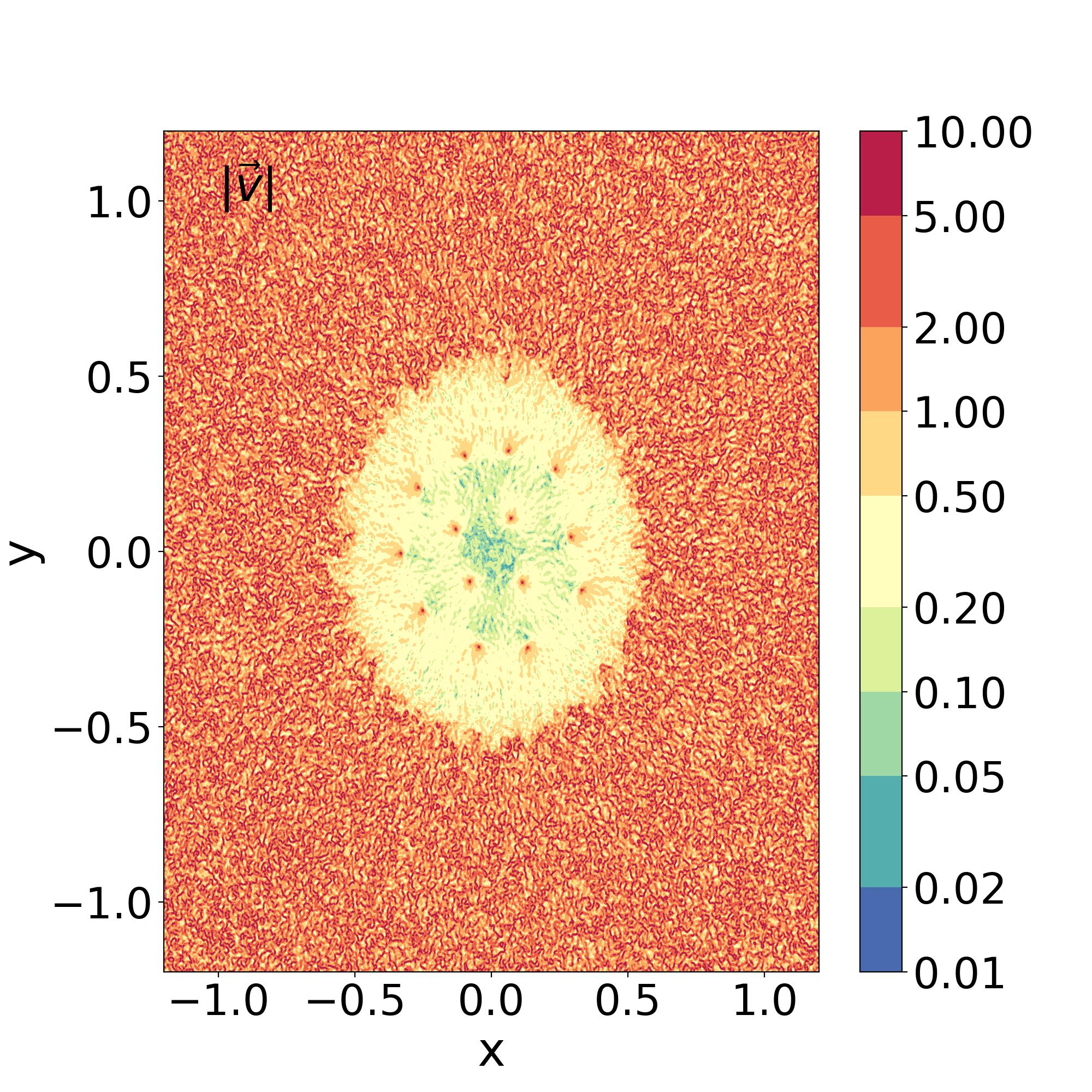}\\
\includegraphics[height=4.8cm,width=0.32\textwidth]{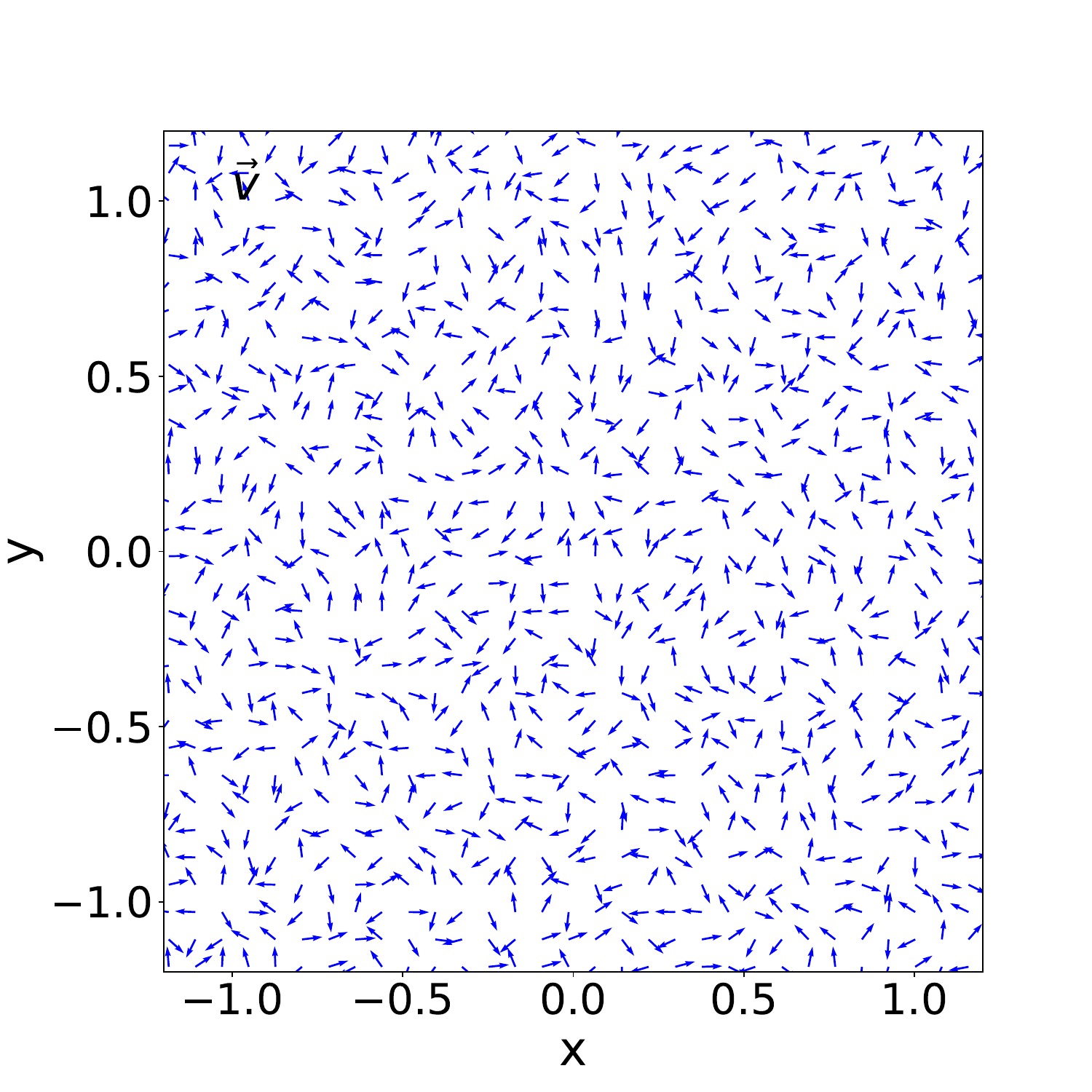}
\includegraphics[height=4.8cm,width=0.32\textwidth]{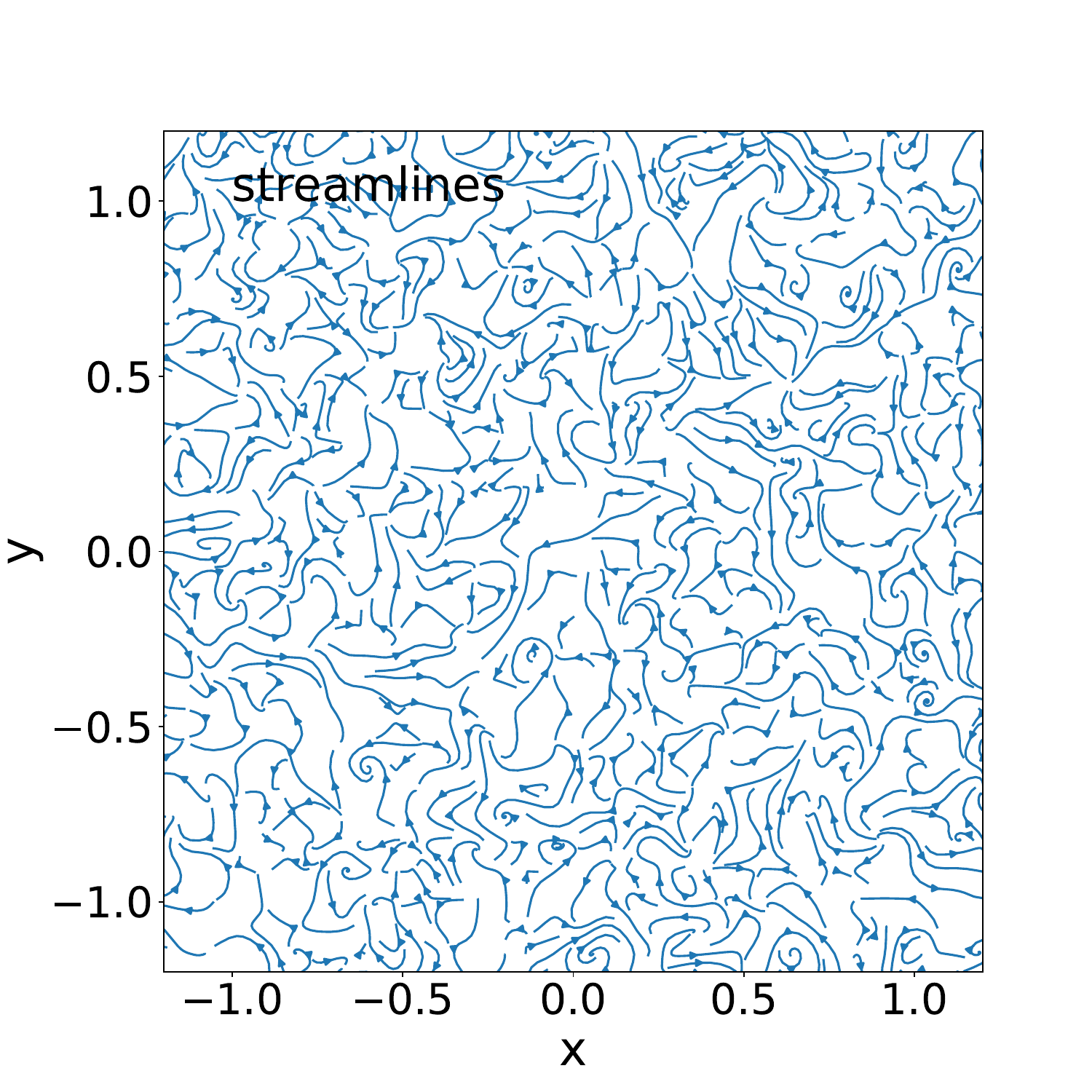}
\includegraphics[height=4.8cm,width=0.32\textwidth]{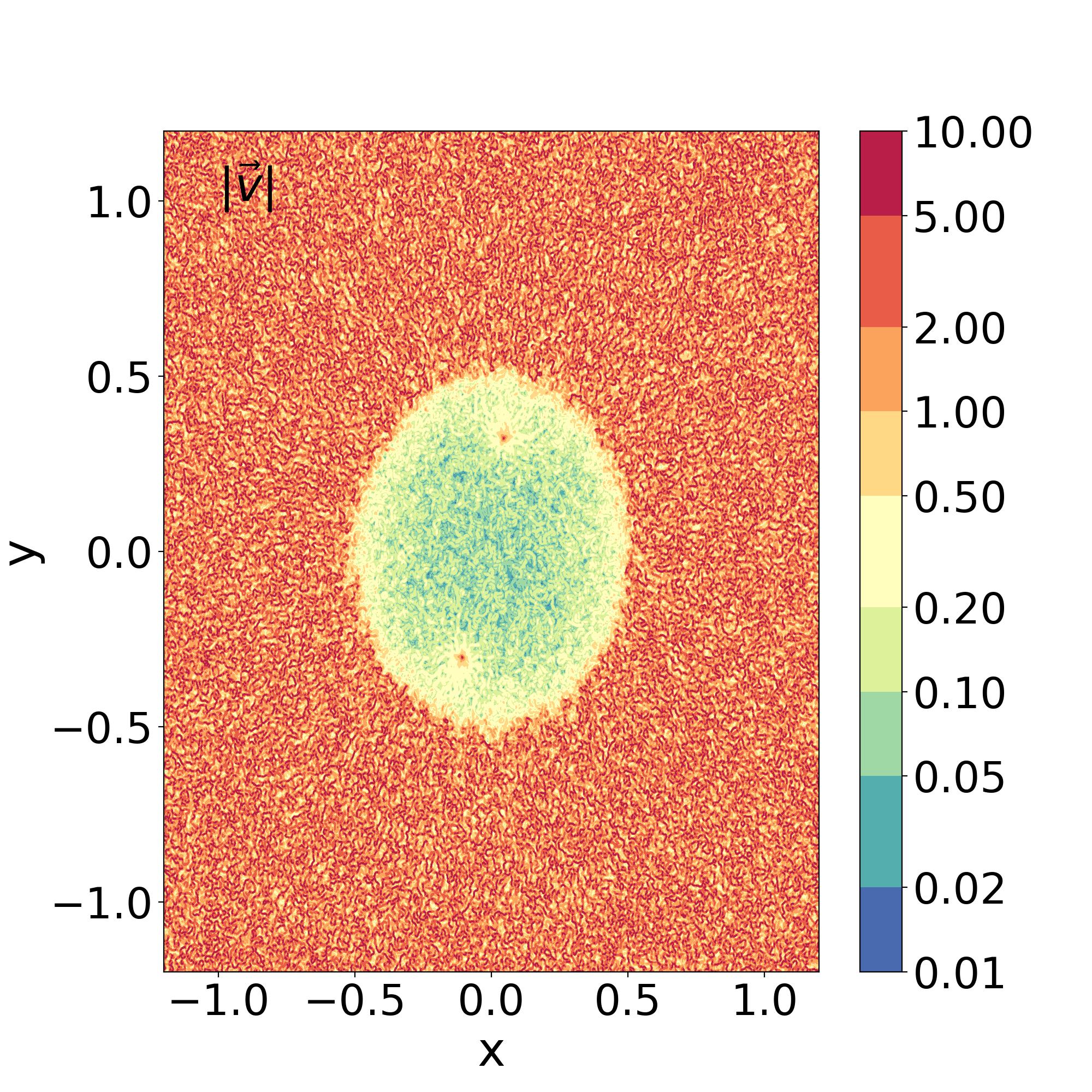}
\caption{
{\it Upper row:} for the case $[\epsilon=0.01, \alpha=0.5]$, maps of the 2D normalized velocity field
$\vec v/|\vec v|$, of the streamlines and of the velocity amplitude $|\vec v|$,
at time $t=500$, as in the lower row in Fig.~\ref{fig:2D-mu1-0p01}.
{\it Lower row:} same plots for the case $[\epsilon=0.01, \alpha=0]$.
}
\label{fig:2D-v-0p01}
\end{figure*}

\begin{figure}
\centering
\includegraphics[height=4.5cm,width=0.32\textwidth]{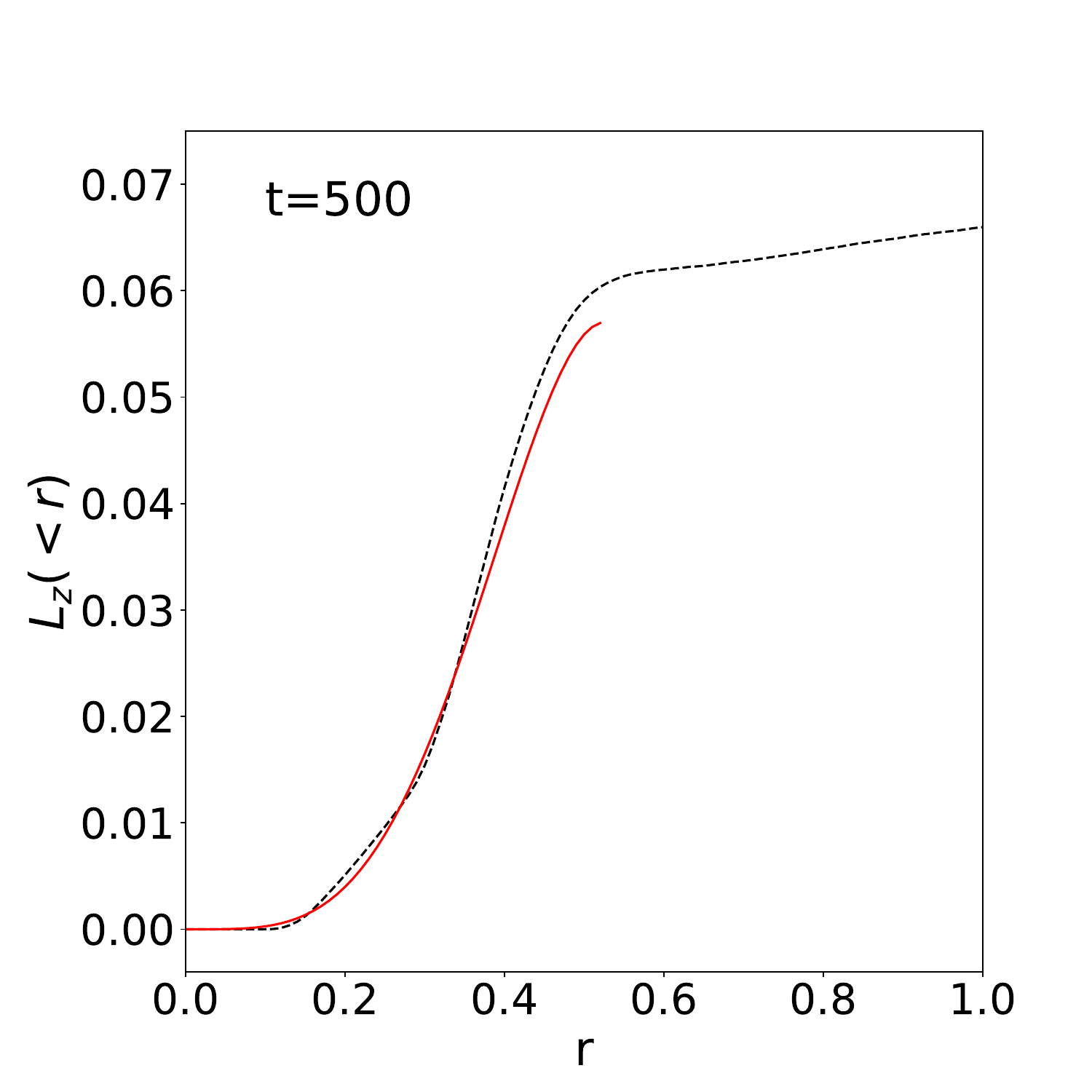}
\caption{
Angular momentum $L_z(<r)$ within radius $r$ for the case
$[\epsilon=0.01, \alpha=0.5]$.
}
\label{fig:Lz-r-0p01}
\end{figure}

\begin{figure}
\centering
\includegraphics[height=4.cm,width=0.235\textwidth]{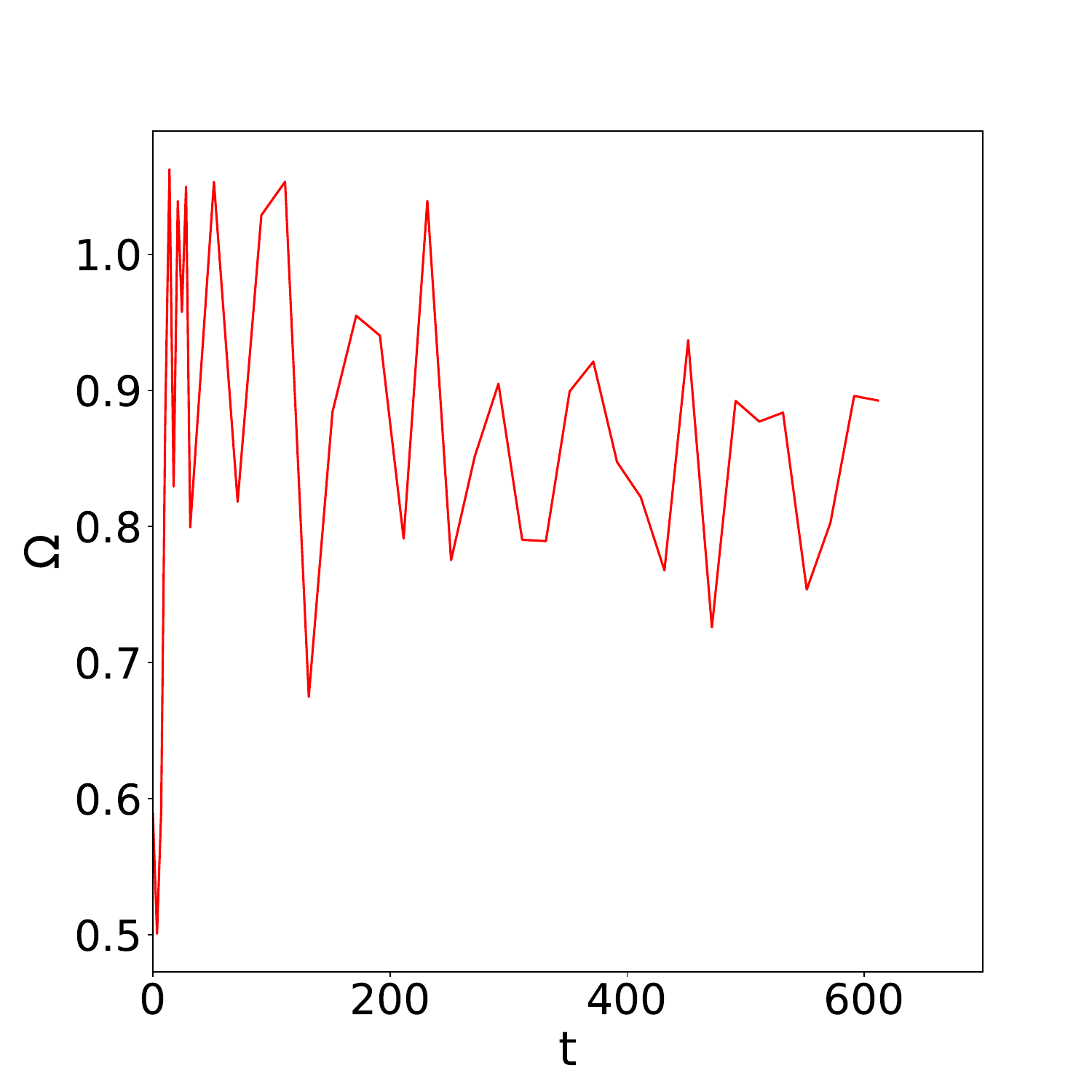}
\includegraphics[height=4.cm,width=0.235\textwidth]{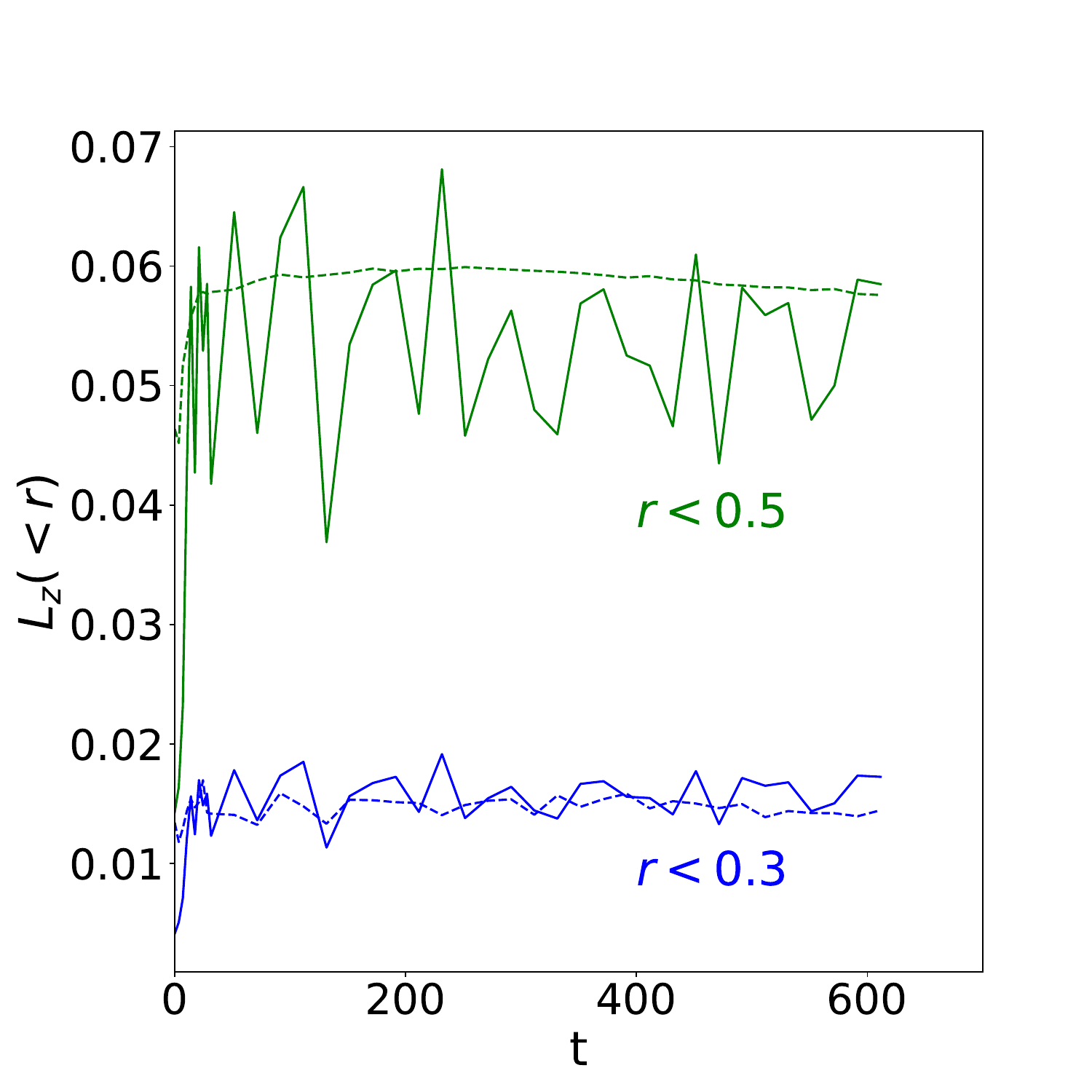}
\includegraphics[height=4.cm,width=0.235\textwidth]{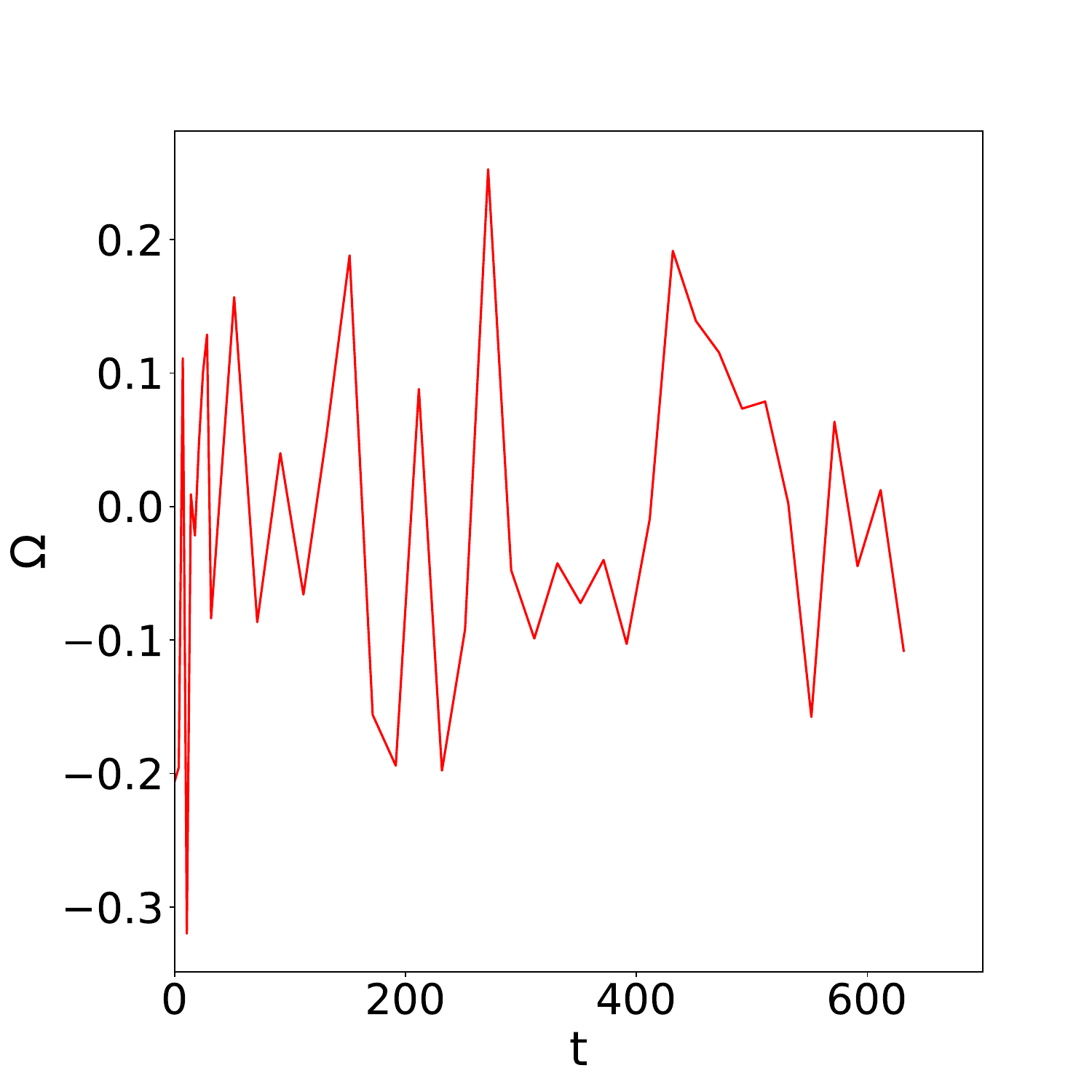}
\includegraphics[height=4.cm,width=0.235\textwidth]{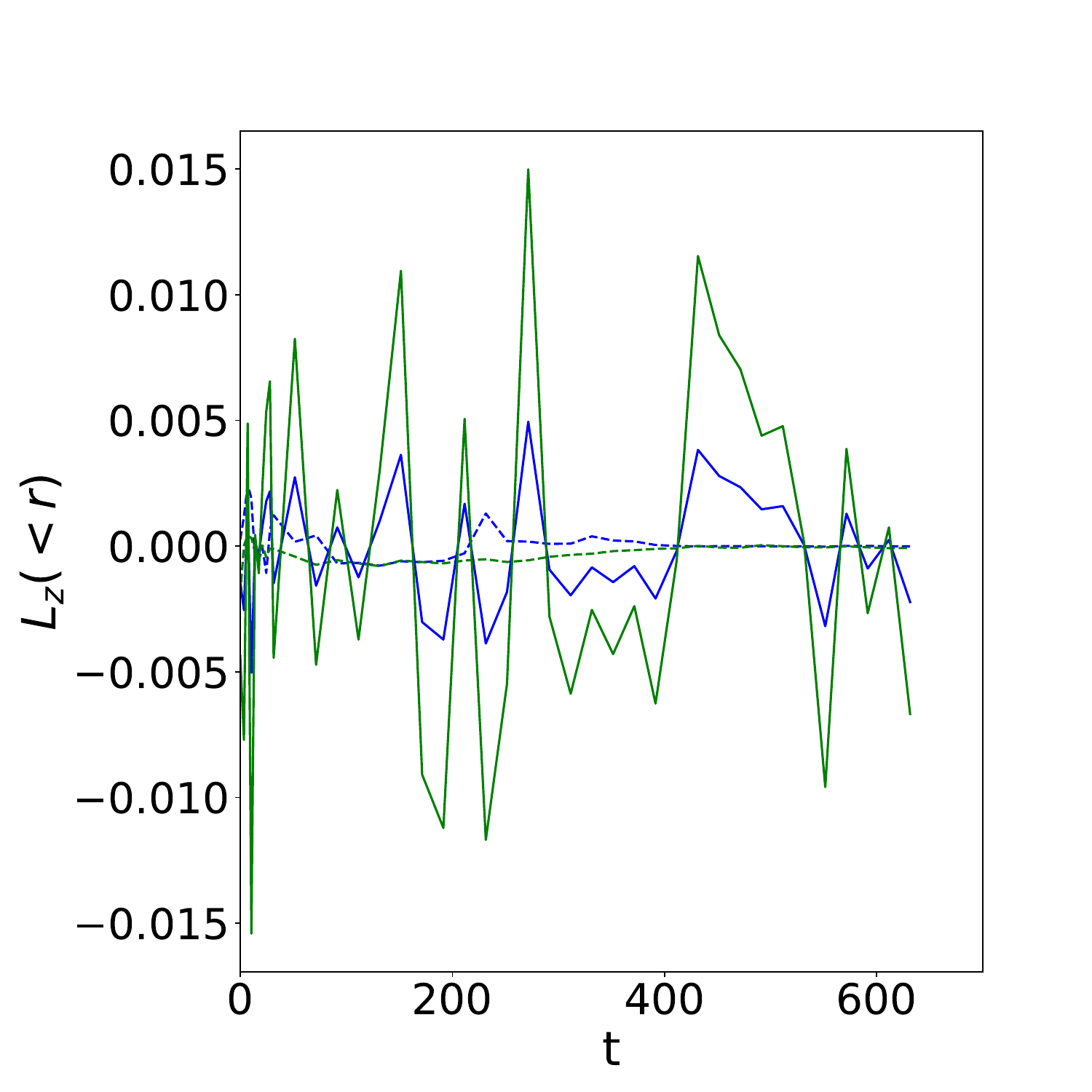}
\caption{
{\it Upper row:} for the case $[\epsilon=0.01, \alpha=0.5]$, rotation rate $\Omega(t)$ as
a function of time and angular momentum $L_z(<r,t)$ within the two
radii $r=0.3$ and $r=0.5$, as in Fig.~\ref{fig:Om-t-mu1-0p01}.
{\it Lower row:} same plots for the case $[\epsilon=0.01, \alpha=0]$.
}
\label{fig:Om-t-0p01}
\end{figure}

\begin{figure}
\centering
\includegraphics[height=5.5cm,width=0.41\textwidth]{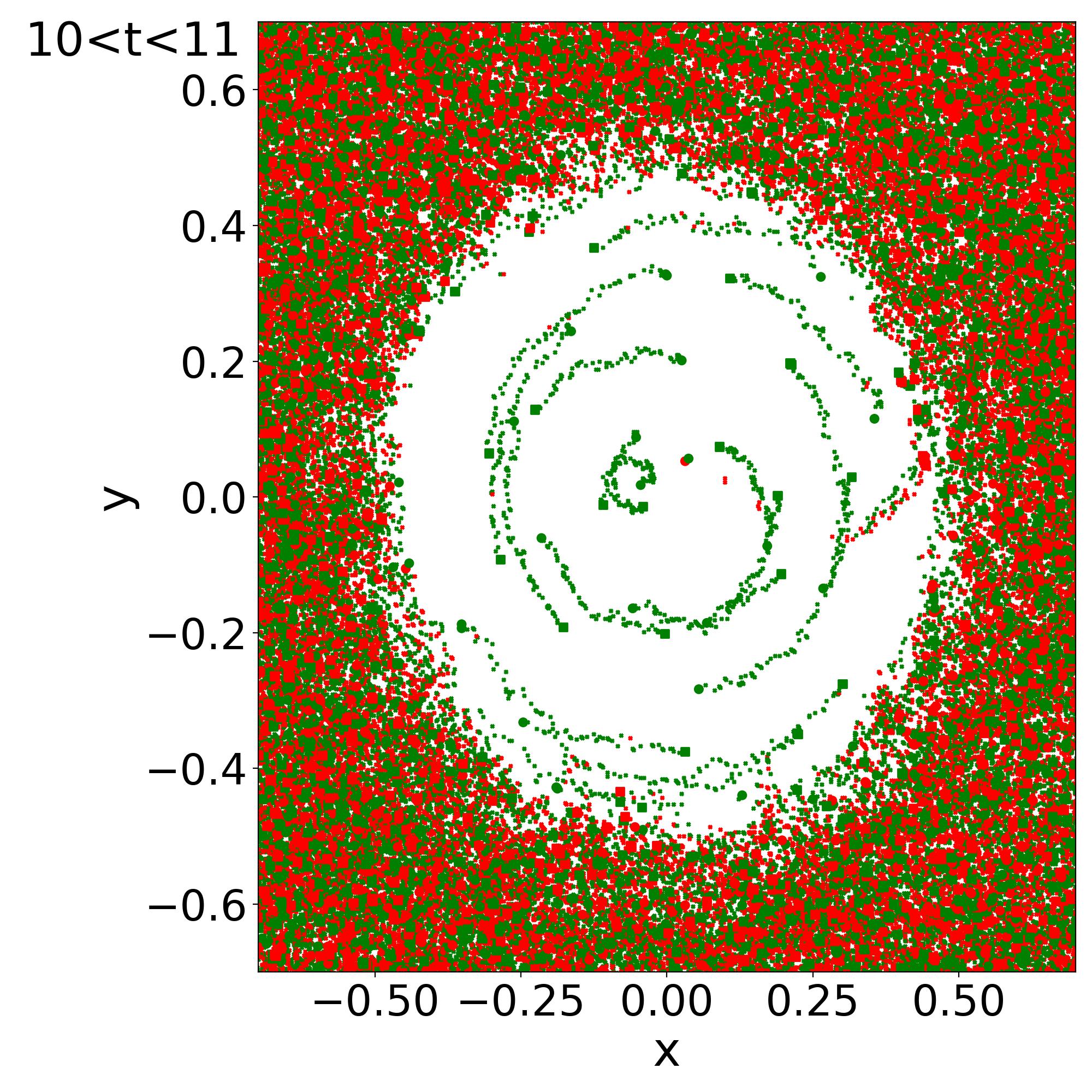}
\includegraphics[height=5.5cm,width=0.41\textwidth]{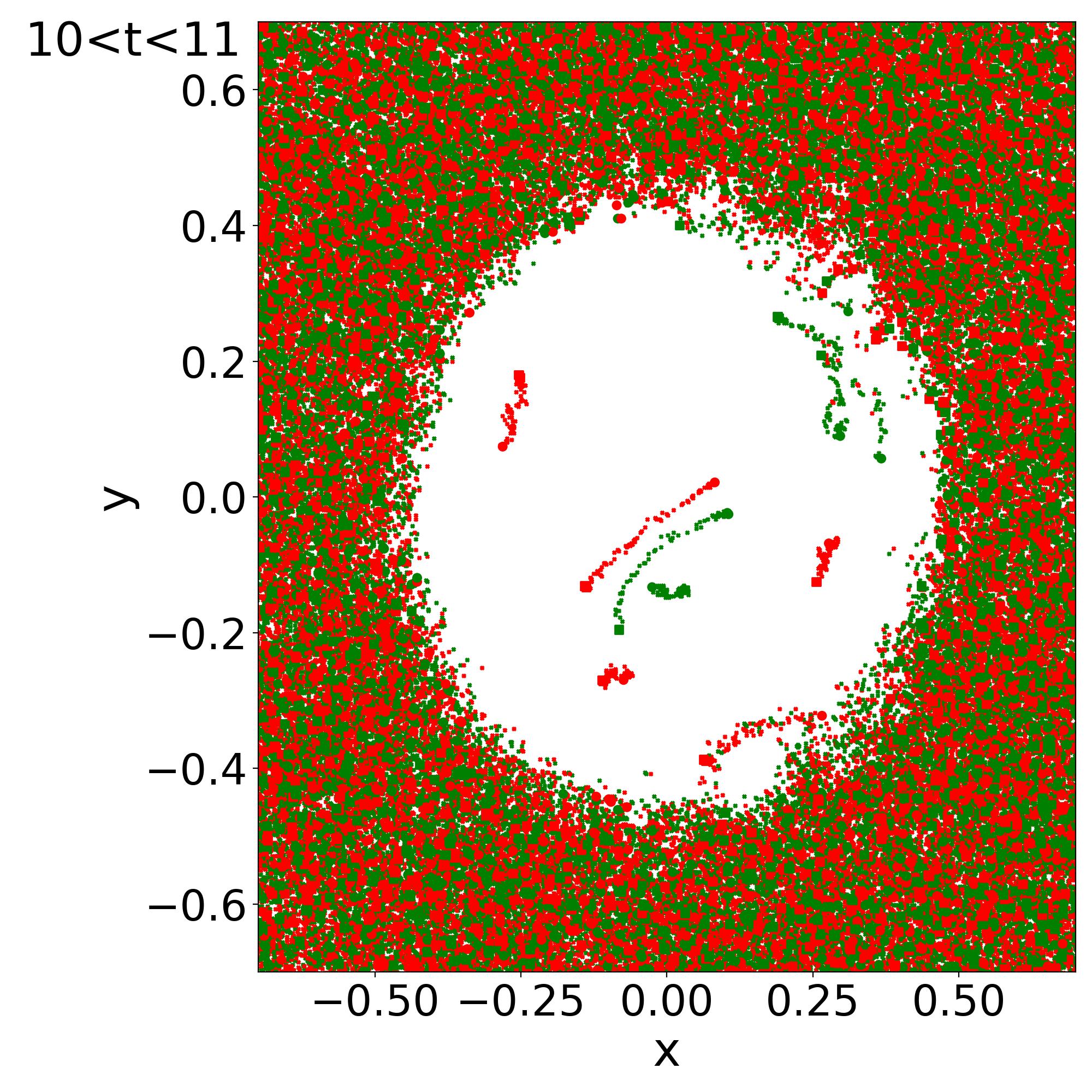}
\caption{
Superposition of many snapshots of the maps of vortices for the cases
$[\epsilon=0.01, \alpha=0.5]$ (upper panel) and
$[\epsilon=0.01, \alpha=0]$ (lower panel),
as in Fig.~\ref{fig:wind-mu1-0p01}.
}
\label{fig:wind-0p01}
\end{figure}

As for the reference case of Fig.~\ref{fig:Om-t-mu1-0p01}, the rotation rate and the angular momentum
inside the soliton, displayed in Fig.~\ref{fig:Om-t-mu1},  remain stable after its formation,
with fluctuations around a roughly constant value.
However, we note that for $\epsilon=0.03$ the rotation rate is somewhat smaller.
This may be due to the fact that with only 6 vortices left and a de Broglie wavelength that is not so small,
the continuum limit is not so well approximated and it is difficult for the system to keep a coherent
rotation in the central region.

This is also apparent in the trajectories of the vortices shown in Fig.~\ref{fig:wind-mu1}.
Whereas for $\epsilon=0.03$ the trajectories of the few vortices inside the soliton are more noisy
than in the reference case $\epsilon=0.01$ (upper panel in Fig.~\ref{fig:wind-mu1-0p01}),
for $\epsilon=0.005$ the trajectories are more regular and follow more closely circles with the same
angular velocity.

Even though $\epsilon \ll 1$, the system inside the soliton is far from systems described by the
Vlasov equation (i.e., the collisionless Boltzmann equation). This is because the self-interactions
play a dominant role and give rise to an effective pressure. Then, in the semi-classical limit
$\epsilon \to 0$ the system inside the soliton is well described by hydrodynamical equations
where, in contrast with a naive interpretation of the Madelung transform, the velocity field contains
a rotational component and nonzero vorticity. This behaviour, explained in Secs.~\ref{sec:vortices}
and \ref{sec:continuum}, is fully supported by our numerical results.
On the other hand, outside of the soliton, the self-interactions no longer play a dominant role
and gravity is balanced by the velocity dispersion. There, the dynamics are better described by
the Vlasov equation in the semi-classical limit, as for a system of collisionless particles.
These two different regimes are most clearly apparent in the 2D maps of the phase and of
the vortices.

Thus, the system shows a coexistence of two distinct phases, which would correspond to
two distinct classical systems, associated with either the Euler or the Vlasov equations.
Moreover, these two phases partially overlap in physical space, because the excited eigenmodes
associated with the outer halo also extend over the soliton (their wave functions extend over all space),
even though they only make a small fraction of the mass in the central region, which is dominated
by the hydrodynamical equilibrium associated with the soliton.
This shows that the Gross-Pitaevskii equation (\ref{eq:Schrod-eps}) can give rise to intricate behaviors
that cannot be fully captured by either the Euler or the Vlasov equations, as both frameworks
would be simultaneously needed to describe the system.

\subsection{Dependence on $\alpha$}

We now consider how the numerical results vary with $\alpha$,
the parameter that measures the initial angular momentum of the system
in Eq.(\ref{eq:Lz-init}).

We performed simulations for the cases $\alpha=0.5$ and $\alpha=0$.
The 1D profiles are similar to those shown in Figs.~\ref{fig:evol-mu1-0p01}
and \ref{fig:v-mu1-0p01}, but for $\alpha=0$ the soliton profile is the static prediction
(\ref{eq:rho-TF-0}).

We can see in the 2D maps shown in Fig.~\ref{fig:2D-rho-0p01} how the number of vortices
inside the soliton decreases with $\alpha$, as expected since both the angular momentum
and the number of vortices scale linearly with $\alpha$.
For the isotropic initial condition $\alpha=0$, there only remain two vortices of opposite signs
inside the soliton, which have not yet annihilated.
The outer halo is similar in both cases, as it is dominated by the large number of vortices of either
sign generated by the interferences between the many uncorrelated excited modes, rather than by
the angular momentum of the system.

We can also see in the velocity maps shown in Fig.~\ref{fig:2D-v-0p01}
the solid-body rotation inside the soliton for the case $\alpha=0.5$, whereas for the isotropic case
$\alpha=0$ the velocity field shows random fluctuations everywhere.
However, in both cases the magnitude of the velocity $|\vec v|$ is smaller inside the soliton,
except for the divergence at the vortices.
In the isotropic case associated with a static soliton, the velocity fluctuates around zero inside the soliton,
which leads to smaller values of $|\vec v|$ than for the rotating case.

As seen in Figs.~\ref{fig:Lz-r-0p01} and \ref{fig:Om-t-0p01}, the rotation rate $\Omega$ and the
angular momentum $L_z$ for the case $\alpha=0.5$ are roughly half of those in
Fig.~\ref{fig:Om-t-mu1-0p01} for $\alpha=1$.
This is because the initial angular momentum has been multiplied by a factor $1/2$ whereas
the soliton mass remains roughly the same.
This now corresponds to a soliton angular momentum that is twice smaller than the
upper bound (\ref{eq:Lz-max}).
In the isotropic case $\alpha=0$ the rotation rate $\Omega$ and the angular momentum $L_z$
fluctuate around zero.

We show the trajectories of the vortices in Fig.~\ref{fig:wind-0p01}. For the rotating case
$\alpha=0.5$ we recover roughly circular paths as in Fig.~\ref{fig:wind-mu1-0p01},
but with a smaller number of vortices and a smaller angular velocity because the rotation of the system
is smaller.
For the isotropic case, we find a few vortices of either spin sign and no collective rotation,
as vortices move along any direction, depending on the local fluctuations around zero of the
velocity field.

\section{Conclusion}
\label{sec:Conclusion}

In this paper we have studied the gravitational dynamics of ultralight dark matter halos,
in the case of non-negligible repulsive quartic self-interactions.
Considering the 2D case, which allows us to reach a greater numerical resolution
and simplifies the analysis, we have been able to investigate the
Thomas-Fermi regime, associated with a small de Broglie wavelength, for stochastic initial
conditions with a nonzero angular momentum.

We have shown that within a few dynamical times a rotating soliton forms at the center of
the system.
As in the zero angular momentum case, which leads to an axisymmetric static soliton
with vanishing velocity, the rotating soliton shows an axisymmetric density profile
with a flat core.
This means that the rotation is not due to a large orbital quantum number $\ell$
of the wave function, which would lead to a vanishing density at the center of the soliton.
Instead, the rotation is generated by a regular lattice of vortices, associated with singularities
of the phase and velocity fields (while the wave function remains regular and vanishes
at these points).
These singularities generate a nonzero vorticity $\vec\nabla\times\vec v$
from the velocity field $\vec v = \epsilon \vec\nabla S$, even though the latter
is defined as the gradient of the phase of the wave function.

We have found that in the Thomas-Fermi regime the system is  also in the dilute regime, as the
distance between vortices decreases more slowly than their width as the de Broglie
wavelength diminishes. This allows us to write a simple ansatz for the wave function
that includes many vortices within a smooth background. Then, the system
can still be mapped to hydrodynamical equations of motion, but the velocity field is
no longer restricted to be curl-free. The vortices simply follow the matter velocity field,
as in classical hydrodynamics of ideal fluids.

As in the static case, the rotating soliton corresponds to a stable minimum
of the energy functional, but now with the additional constraint of a fixed nonzero angular
momentum.
In the continuum limit, this soliton displays a solid-body rotation that adds a centrifugal term
to the equation of hydrostatic equilibrium. The soliton remains circular but its radial
density profile is deformed from the static case.
The rotation flattens and expands the radius of the soliton, by a finite factor
$R_{\max}/R_0 \lesssim 1.593$.
We have also shown that for a given central density these solitons can only support rotation
up to a maximum rotation rate $\Omega_{\max}$.
Then, all these configurations are dynamically stable, as they correspond to a minimum of the energy.
The solid-body rotation is associated with a uniform vorticity and a uniform distribution of
vortices, which for finite nonzero de Broglie wavelength leads to a regular lattice of vortices.

We have checked that all these analytical results, the deformed density profile, the
solid-body rotation, the uniform distribution of vortices,
agree with our numerical simulations.
As predicted, the number of vortices increases for higher angular momentum and
for smaller de Broglie wavelength, as each vortex carries a vorticity quantum $\pm2\pi\epsilon$.
We find that the angular momentum inside the soliton can be estimated from its mass after formation
and the angular momentum in the initial state carried by the same mass fraction (i.e., assuming
the radial ordering has not significantly changed).

We have analyzed the system of vortices in 2D and we expect that vortex rings would be their
equivalents in 3D \cite{Hui:2020hbq}.
Topologically, the nature of a network of vortex lines embedded in a dark matter halo would certainly
deserve further studies.
The existence of topological defects in the scalar dark matter distribution is intrinsically linked to the
zeros of the associated wave-function where the Madelung transform is ill-defined.
This is a feature of the nonrelativistic approximation considered in this paper.
Lifting this approximation, it would be worth studying how these networks of vortices
extend into the relativistic regime, which could open up a wealth of new phenomena
for the dynamics of dark matter halos. This is left for future work.

The possible existence of vortex rings could have interesting phenomenological consequences.
In realistic cosmological settings there would be dark matter and baryonic substructures
around the vortices, which would then rotate around these vortices.
As pointed out by \cite{Alexander:2021zhx}, this could be associated with the observed spin
of cosmic filaments.
On the other hand, the rotation of the dark matter soliton could provide a good model
to reproduce the rotation curves of galaxies \cite{Boehmer:2007um,Kain:2010rb}.
In addition, in a mildly relativistic regime baryonic matter accreting in a disk around the
vortices would precess due to the frame-dragging effect of the rotating filaments.
In turns, this could lead to X-ray synchrotron radiation which might be observable.
Of course, more studies would be needed to validate such a scenario.
For DM detection experiments, the lower density inside the vortices implies a depletion
of the decay rate of the scalars into photons when such a coupling is taken into account. This would
be another signature of the existence of dark matter vortices.

\appendix

\section{Static soliton profile}
\label{app:static-soliton}

In this appendix we derive the full profile of the static soliton in the limit $\epsilon \to 0$.
We distinguish three regions, the bulk of the soliton at $r \ll R_0$ described by the Thomas-Fermi regime,
the extended exponential tail at $r \gg R_0$ described by the WKB regime, and the intermediate
boundary layer at $r \simeq R_0$.
Asymptotic matching on both sides of the boundary layer provides the global solution and the
normalization of the outer exponential tail in terms of the central density $\rho_0$.

As we consider axisymmetric solutions of the Gross-Pitaevskii equation of the form
\be
\psi(\vec r,t) = e^{-i \mu t/\epsilon} \hat\psi(r) ,
\ee
the time-independent Schr\"odinger equation reads as
\be
\frac{\epsilon^2}{2} \left( \frac{d^2\hat\psi}{dr^2} + \frac{1}{r} \frac{d\hat\psi}{dr} \right) = 
( \Phi_N + \Phi_I - \mu ) \hat\psi ,
\label{eq:Schrodinger-radial}
\ee
as in Eq.(\ref{eq:hydrostatic-full}).
For an axisymmetric density distribution we also have
\be
\Phi_N(r) = 4\pi \ln r \int_0^r dr' r' \rho(r') + 4\pi \int_r^{\infty} dr' r' \rho(r') \ln r' ,
\label{eq:PhiN-axisymmetric}
\ee
so that the gravitational potential behaves at large distance as
\be
r \gg R_0 : \;\; \Phi_N(r) \simeq 2 M \ln r ,
\label{eq:PhiN-large-distance}
\ee
up to exponentially small corrections.

\subsection{Bulk of the soliton}
\label{app:soliton-bulk}

As we focus on the limit $\epsilon \to 0$, the bulk of the soliton, where the density is of order unity,
is described by the Thomas-Fermi regime where we can neglect the left-hand side in 
Eq.(\ref{eq:Schrodinger-radial}).
This gives
\be
\Phi_N + \Phi_I = \mu ,
\label{eq:TF-eps}
\ee
with the solution
\be
r \ge 0, \;\; R_0 - r \gg \epsilon^{2/3} : \;\;\; \hat\psi(r) = \sqrt{ \rho_0 J_0(z_0 r/R_0) } ,
\label{eq:TF-hat-psi}
\ee
as in Eqs.(\ref{eq:TF-static})-(\ref{eq:rho-TF-0}).
Here we wrote that the Thomas-Fermi approximation breaks down within a boundary layer
below $R_0$ of width of the order of $\epsilon^{2/3}$, as we shall see in App.~\ref{app:soliton-matching}
below.
The gravitational potential reads at leading order as
\be
r \! \geq \! 0, \; R_0 - r \! \gg \! \epsilon^{2/3} \! : \; \Phi_N(r) \! = \! 2 M \! \ln R_0 
- \lambda \rho_0 J_0(z_0 r/R_0) ,
\label{eq:PhiN-TF}
\ee
where we used Eq.(\ref{eq:PhiN-axisymmetric}) to normalize $\Phi_N$, and we obtain
\be
\mu = 2 M \ln R_0 , \;\;\; M = \frac{\lambda}{2} \rho_0 z_0 J_1(z_0) .
\ee

\subsection{Exponential tail}
\label{app:soliton-tail}

At large distance beyond $R_0$, the density becomes vanishingly small and the Thomas-Fermi
approximation no longer applies. Indeed, for $r \to \infty$ we have $\Phi_I \to 0$
while $\Phi_N$ keeps growing as in Eq.(\ref{eq:PhiN-large-distance}).
Making the change of variable $\hat\psi = u/\sqrt{r}$, the Schr\"odinger equation (\ref{eq:Schrodinger-radial})
becomes
\be
\frac{\epsilon^2}{2} \left( \frac{d^2u}{dr^2} + \frac{u}{4 r^2} \right) = ( \Phi_N + \Phi_I - \mu ) u .
\ee
Using the WKB approximation, $u = \exp[(S_0+\epsilon S_1 + \dots)/\epsilon]$,
we obtain
\be
r - R_0 \gg \epsilon^{2/3} \! : \; \hat\psi(r) = \frac{C}{\sqrt{r} ( \Phi_N \! - \! \mu )^{1/4} } 
e^{-\frac{1}{\epsilon} \int_{R_0}^r dr \sqrt{2 (\Phi_N \! - \! \mu)} } ,
\label{eq:WKB-C}
\ee
where $C$ is a constant to be determined.
Here we again wrote that the WKB approximation breaks down within a boundary
layer of width $\epsilon^{2/3}$ above $R_0$, as we shall see in App.~\ref{app:soliton-matching}
below, and that the density and $\Phi_I$ are exponentially
small beyond this boundary layer.  
Using Eq.(\ref{eq:PhiN-large-distance}) we obtain the large-distance tail
\be
r \gg R_0 : \;\;\; \hat\psi(r) \sim e^{-\frac{2}{\epsilon} \sqrt{M} r \sqrt{\ln r} } .
\label{eq:psi-decay}
\ee
Thus, the density decays slightly faster than a pure exponential because of the logarithmic
factor in the exponent.

\subsection{Boundary layer}
\label{app:soliton-boundary-layer}

Around $R_0$ we have a transition between the Thomas-Fermi and WKB regimes.
In this region the density shows a sharp bend while the gravitational potential remains smooth.
From Eq.(\ref{eq:PhiN-TF}) we write
\be
r \simeq R_0 : \;\; \Phi_N - \mu \simeq \frac{2M}{R_0} (r-R_0)  .
\label{eq:PhiN-R0-lin}
\ee
We can also neglect the first derivative as compared with the second derivative in the left-hand side
in Eq.(\ref{eq:Schrodinger-radial}) and we obtain
\be
\frac{\epsilon^2}{2} \frac{d^2\hat\psi}{dx^2} = \frac{2M}{R_0} x \hat\psi + \lambda \hat\psi^3 ,
\ee
where we introduced $x=r-R_0$.
Making the changes of variables
\be
x = (4M/R_0)^{-1/3} \epsilon^{2/3} \tau , \;\;\; 
\hat\psi = \lambda^{-1/2} (4M/R_0)^{1/3} \epsilon^{1/3} y ,
\ee
we obtain
\be
\frac{d^2y}{d\tau^2} = 2 y^3 + \tau y .
\label{eq:Painleve-2}
\ee
We recognize the second Painlev\'e equation \cite{Olver} and the unique solution that satisfies the
correct boundary conditions is the one at the transition between an oscillatory behavior and the
appearance of a pole on the negative real axis. It satisfies the asymptotic behaviors \cite{Olver}
\be
\tau \to \infty: \;\; P_{\rm II}(\tau) \simeq {\rm Ai}(\tau) , 
\tau \to -\infty : \;\; P_{\rm II}(\tau) \simeq \sqrt{-\tau/2} ,
\label{eq:P-II-asymp}
\ee
where ${\rm Ai}(\tau)$ is the Airy function.
Thus, we obtain in the intermediate regime between the Thomas-Fermi and WKB regions
\ba
&& 0 \leq  R_0 -r \ll 1 , \;\; 0 \leq r - R_0 \ll \epsilon^{2/5} : \nonumber \\
&& \hspace{-0.1cm} \hat\psi(r) \! = \! \lambda^{-1/2} (4 M \! / \! R_0)^{1/3} \epsilon^{1/3} 
P_{\rm II} \! \left[  (4 M \! / \! R_0)^{1/3} \epsilon^{-2/3} ( r \! - \! R_0) \right]
\!  . \nonumber \\
&&
\label{eq:P-II}
\ea
The widths of the intervals to the left and to the right of $R_0$ where Eq.(\ref{eq:P-II}) applies 
will be obtained in App.~\ref{app:soliton-matching} below,
from the asymptotic matching with the Thomas-Fermi 
expression (\ref{eq:TF-hat-psi}) and the WKB expression (\ref{eq:WKB-C}).
However, we can already see from Eq.(\ref{eq:P-II}) that the width of the boundary layer,
where the profile makes a transition from the Thomas-Fermi to the WKB regime, scales as
$\Delta r \sim \epsilon^{2/3}$.
As expected for a boundary layer, this goes to zero in the limit $\epsilon \to 0$.

\subsection{Asymptotic matchings}
\label{app:soliton-matching}

We first check the matching of the intermediate solution (\ref{eq:P-II}) with the Thomas-Fermi solution
(\ref{eq:TF-hat-psi}).
This Thomas-Fermi expression gives
\be
\epsilon^{2/3} \ll R_0 -r \ll 1 : \;\; \hat\psi(r) = \sqrt{\frac{2 M}{\lambda R_0}} \sqrt{R_0 -r} .
\label{eq:psi-bl-left}
\ee
Using the asymptotic behavior (\ref{eq:P-II-asymp}), we can check that this agrees with the boundary 
layer expression (\ref{eq:P-II}) over the overlapping domain $\epsilon^{2/3} \ll R_0 -r \ll 1$.
This asymptotic matching also provides the width $\epsilon^{2/3}$ of the boundary layer to the left
of $R_0$, where the Thomas-Fermi approximation breaks down, as well as the width 
$R_0 -r \ll 1$ where the intermediate solution (\ref{eq:P-II}) applies.
We already wrote these validity boundaries in (\ref{eq:TF-hat-psi}) and (\ref{eq:P-II}).

We now consider the matching to the right of $R_0$. From the asymptotic behavior of the Airy function,
\be
x \to \infty: \;\; {\rm Ai}(x) \simeq \frac{1}{2\sqrt{\pi}} x^{-1/4} e^{-2 x^{3/2}/3} 
\ee
and Eq.(\ref{eq:P-II-asymp}), the boundary-layer solution (\ref{eq:P-II}) gives
\ba
&& \epsilon^{2/3} \ll r - R_0 \ll \epsilon^{2/5} : \;\;\; \hat\psi = \sqrt{ \frac{\epsilon}{\lambda 2 \pi}} (M/R_0)^{1/4} 
\nonumber \\
&&\hspace{1cm}  \times (r-R_0)^{-1/4}  e^{-4 \sqrt{M/R_0} (r-R_0)^{3/2} /(3\epsilon)} . \hspace{1cm}
\label{eq:Ai-asymp}
\ea
The comparison with Eq.(\ref{eq:WKB-C}), where we use Eq.(\ref{eq:PhiN-R0-lin}) shows that 
these two approximation match over the range $\epsilon^{2/3} \ll r-R_0 \ll \epsilon^{2/5}$ with the normalization constant $C$ given by
\be
C = 2^{-1/4} \sqrt{\epsilon M/(\lambda\pi)} .
\label{eq:C-def}
\ee
The upper bound on the overlapping interval, $r-R_0 \ll \epsilon^{2/5}$, which we already wrote 
in (\ref{eq:P-II}) and (\ref{eq:Ai-asymp}), is set by the requirement that the subleading term 
of order $(r-R_0)^2$ in $(\Phi_N-\mu)$ in the exponent (\ref{eq:WKB-C}) 
gives a contribution that vanishes in the limit $\epsilon \to 0$.
The value (\ref{eq:C-def}) provides the normalization of the WKB expression
(\ref{eq:WKB-C}), which reads
\ba
&& r - R_0 \gg \epsilon^{2/3} : \;\;\; \hat\psi(r) = \frac{\sqrt{\epsilon M}}{\sqrt{\lambda \pi r} 
[ 2 (\Phi_N \! - \! \mu )]^{1/4} } \nonumber \\
&& \hspace{3cm} \times e^{-\frac{1}{\epsilon} \int_{R_0}^r dr \sqrt{2 (\Phi_N \! - \! \mu)} } ,
\label{eq:WKB-det}
\ea
and $\Phi_N(r) = 2 M \ln r$.
Thus, Eqs.(\ref{eq:TF-hat-psi}), (\ref{eq:P-II}) and (\ref{eq:WKB-det}) give the full profile of the static
soliton in the limit $\epsilon \to 0$.

\section{Vortex profile}
\label{app:vortex-profile}

Following the method presented in App.~\ref{app:static-soliton}, we compute the asymptotic profile
of a vortex of spin $\sigma$, going beyond the uniform-background treatment of
Sec.~\ref{sec:vortex-homogeneous-background} in the main text.
In contrast with the static soliton considered in App.~\ref{app:static-soliton}, to obtain the full profile
we shall need to restrict to the case of large spin,
\be
|\sigma| \gg 1 .
\label{eq:sigma-small}
\ee
For a vortex of spin $\sigma$, with a wave function of the form
\be
\psi_\sigma(\vec r,t) = e^{-i \mu t/\epsilon} \hat\psi_\sigma(r) e^{i \sigma \theta} , 
\label{eq:psi-sigma}
\ee
the time-independent Schr\"odinger equation reads
\be
\frac{\epsilon^2}{2} \left( \frac{d^2\hat\psi_\sigma}{dr^2} + \frac{1}{r} \frac{d\hat\psi_\sigma}{dr} 
- \frac{\sigma^2}{r^2} \hat\psi_\sigma \right) = ( \Phi_N + \Phi_I - \mu ) \hat\psi_\sigma .
\label{eq:Schrodinger-radial-sigma}
\ee
As compared with the static soliton case (\ref{eq:Schrodinger-radial}), we have the addition of the
orbital-barrier term $(\sigma^2/r^2) \hat\psi_\sigma$.
In the following we omit the subscript $\sigma$ to simplify notations.

\subsection{Bulk of the vortex}
\label{app:vortex-bulk}

As for the static soliton in App.~\ref{app:soliton-bulk}, in the limit $\epsilon \to 0$ the bulk of the vortex
is described by the Thomas-Fermi regime where we can neglect the spatial derivatives in 
Eq.(\ref{eq:Schrodinger-radial-sigma}).
However, because we can have $|\sigma | \gg 1$, we keep the orbital-barrier term in the left-hand side
and we generalize the Thomas-Fermi approximation (\ref{eq:TF-eps}) to
\be
\Phi_N + \Phi_I + \frac{\epsilon^2 \sigma^2}{2 r^2}= \mu .
\label{eq:TF-eps-sigma}
\ee
As seen from Eq.(\ref{eq:TF-eps-sigma}) and in agreement with Eq.(\ref{eq:rc-def}), the core radius
of the vortex is $r_c = \epsilon |\sigma|/\sqrt{2\lambda\rho_0}$ and we have
\be
r \gg r_c , \;\; R_0 - r \gg \epsilon^{2/3}  : \;\; \rho = \rho_0 J_0(z_0 r/R_0) ,
\label{eq:vortex-bulk}
\ee
as for the static soliton (\ref{eq:TF-hat-psi}).

Closer to the vortex radius and using Eq.(\ref{eq:PhiN-TF}), we can approximate at leading order 
in $\epsilon$ the gravitational potential by $\Phi_N \simeq \Phi_{N0} = \mu - \lambda \rho_0$, 
and we obtain the density falloff
\be
|\sigma|^{-2/3} r_c \ll r - r_c \ll R_0 : \;\;\; \rho = \rho_0 \left( 1 - \frac{r_c^2}{r^2} \right) .
\label{eq:rho-TF-rc}
\ee
This gives a vanishing density at the core radius $r_c$. As we shall see in App.~\ref{app:vortex-layer}
below, for $|\sigma| \gg 1$ this will be regularized by a boundary layer that connects this regime with 
the small-radius domain inside the vortex core.
Again, we already write the range of validity of these expressions, which will be justified from the
asymptotic matching described in App.~\ref{app:vortex-matchings} below.

At large radii $r \gtrsim R_0$, we recover the boundary layer (\ref{eq:P-II}) at $R_0$ and 
the exponential tail (\ref{eq:WKB-det}) at $r \gg R_0$, as for the static soliton.

\subsection{Vortex core}
\label{app:vortex-core}

Close to the center of the vortex the density vanishes as $\rho \propto r^{2|\sigma|}$ from
Eq.(\ref{eq:asymp-f}). More generally, the density is infinitesimally small below $r_c$ from the expression
(\ref{eq:rho-TF-rc}). Therefore, inside the vortex core we can neglect $\Phi_I$ in the right-hand side
in Eq.(\ref{eq:Schrodinger-radial-sigma}). Keeping also the approximation 
$\Phi_N \simeq \Phi_{N0} = \mu - \lambda \rho_0$, we obtain the linear Schr\"odinger equation
\be
\frac{d^2\hat\psi}{dr^2} + \frac{1}{r} \frac{d\hat\psi}{dr} + \left( \frac{\sigma^2}{r_c^2}
- \frac{\sigma^2}{r^2} \right) \hat\psi =  0 .
\label{eq:Schrodinger-radial-core}
\ee
The solution that is regular at the center reads
\be
r \geq 0, \;\; r_c - r \gg |\sigma|^{-2/3} r_c : \;\;\; \hat\psi =  C \, J_{|\sigma|}(|\sigma| r/r_c) ,
\label{eq:psi-J-|sigma|}
\ee
where $J_{|\sigma|}$ is the Bessel function of order $|\sigma|$ and $C$ is a normalization constant
to be determined.
This gives $\hat\psi \propto r^{|\sigma|}$ for $r\to 0$, in agreement with Eq.(\ref{eq:asymp-f}).

\subsection{Vortex boundary layer}
\label{app:vortex-layer}

At the boundary between the outer Thomas-Fermi regime (\ref{eq:rho-TF-rc}) and the inner
core regime (\ref{eq:psi-J-|sigma|}), the self-interaction $\Phi_I$, the orbital barrier $\sigma^2/r^2$
and the spatial derivatives are of the same order in the Schr\"odinger equation 
(\ref{eq:Schrodinger-radial-sigma}). As compared with the boundary layer at $R_0$ found for the
static soliton, described in App.~\ref{app:soliton-boundary-layer}, we can now approximate the
gravitational potential by a constant (its central value), but we have two new terms, the orbital barrier
and the first derivative $\frac{1}{r} \frac{d\hat\psi}{dr}$.
Close to $r_c$ we can again use a linear approximation in $r-r_c$ for the orbital term, but the first
derivative is not longer negligible in the general case, as $r \sim r_c \propto \epsilon$
and $\frac{1}{r} \frac{d\hat\psi}{dr} \sim \frac{d^2\hat\psi}{dr^2}$.
Nevertheless, in the limit $|\sigma| \gg 1$ a true boundary layer appears, which is much narrower
than the core radius $r_c \propto |\sigma|$. Then, we can neglect the first derivative
and approximate the Schr\"odinger equation (\ref{eq:Schrodinger-radial-sigma}) by
\be
\frac{\epsilon^2}{2} \frac{d^2\hat\psi}{dr^2} = \left[ - \frac{\epsilon^2 \sigma^2}{r_c^3} (r-r_c)
+ \lambda \hat\psi^2  \right] \hat\psi .
\label{eq:Schrodinger-sigma-bl}
\ee
Making the changes of variables
\be
r - r_c = - \frac{r_c}{2^{1/3} |\sigma|^{2/3}} \tau , \;\;\;
\hat\psi = \frac{2^{5/6} \sqrt{\rho_0}}{| \sigma |^{1/3}} y ,
\ee
we again obtain the second Painlev\'e equation (\ref{eq:Painleve-2}).
The solution that satisfies the correct boundary conditions is again determined by
(\ref{eq:P-II-asymp}) and we obtain the boundary-layer solution
\ba
&& 0 \leq r_c - r \ll |\sigma|^{-1/3} r_c , \;\; 0 \leq r - r_c \ll r_c : \nonumber \\
&& \hat\psi(r) = \frac{2^{5/6} \sqrt{\rho_0}}{| \sigma |^{1/3}} P_{\rm II} \left[ 
- 2^{1/3} | \sigma |^{2/3} \frac{r-r_c}{r_c} \right] .
\label{eq:vortex-bl}
\ea
In contrast with the boundary layer at $R_0$ in Eq.(\ref{eq:P-II}) and as announced earlier, 
we can see that the relative width of this transition region, $\Delta r/r$, does not vanish in the limit
$\epsilon \to 0$. Indeed, it decreases linearly with $\epsilon$ at the same rate as $r_c$.
However, in the limit of large spin (\ref{eq:sigma-small}) it becomes increasingly narrow as
$\Delta r/r \sim |\sigma|^{-2/3}$.
This is the regime where this boundary-layer analysis applies and we could neglect the
first derivative in Eq.(\ref{eq:Schrodinger-sigma-bl}).

\subsection{Asymptotic matchings}
\label{app:vortex-matchings}

We first check the outer matching of the boundary-layer solution (\ref{eq:vortex-bl}) with the 
generalized Thomas-Fermi solution (\ref{eq:rho-TF-rc}). The latter gives close to the core radius
\be
| \sigma |^{-2/3} r_c \ll r - r_c \ll r_c : \;\;\; \hat\psi = \sqrt{2\rho_0} \sqrt{\frac{r - r_c}{r_c}} .
\ee
Using Eq.(\ref{eq:P-II-asymp}), we can check that this agrees with the boundary-layer 
expression (\ref{eq:vortex-bl}) over the overlapping domain 
$|\sigma|^{-2/3} r_c \ll r-r_c \ll r_c$.
We can see that we require $|\sigma| \gg 1$ for this matching domain to exist.

We now consider the inner matching to the core solution (\ref{eq:psi-J-|sigma|}).
Using the asymptotic behaviors (\ref{eq:P-II-asymp}) and \cite{Olver}
\be
\nu\to\infty : \;\;\; J_\nu(\nu+a\nu^{1/3}) \sim 2^{1/3} \nu^{-1/3} {\rm Ai}(-2^{1/3} a) ,
\ee
we find that the inner solution (\ref{eq:psi-J-|sigma|}) matches with the boundary-layer solution
(\ref{eq:vortex-bl}), with the constant $C$ given by $C=\sqrt{2\rho_0}$, over the matching domain
$|\sigma|^{-2/3} r_c \ll r_c - r \ll |\sigma|^{-1/3} r_c$.
This fully determines the core solution (\ref{eq:psi-J-|sigma|}),
\be
r \geq 0, \;\; r_c \! - \! r \gg |\sigma|^{-2/3} r_c : \;\; \hat\psi \! = \sqrt{2\rho_0} \, J_{|\sigma|}(|\sigma| r/r_c) .
\label{eq:psi-J-|sigma|-norm}
\ee
At the center of the vortex this gives
\be
0 \leq r \ll \frac{r_c}{|\sigma|} : \;\;\; 
\hat \psi = \sqrt{\frac{\rho_0}{\pi |\sigma|}} \left( \frac{e}{2} \frac{r}{r_c} \right)^{|\sigma|} ,
\label{eq:vortex-center}
\ee
where we again used $|\sigma| \gg 1$.

For moderate values of the spin, $|\sigma| \sim 1$, the boundary layer described in 
App.~\ref{app:vortex-layer} no longer exists. The transition between the core and bulk regimes
takes place over a width $\Delta r \sim r_c$ rather than a narrow region 
$\Delta r \sim r_c/|\sigma|^{2/3}$ and we can no longer
neglect the first derivative in Eq.(\ref{eq:Schrodinger-sigma-bl}).
Nevertheless, we expect the normalization $C=\sqrt{2\rho_0}$ of the core solution
(\ref{eq:psi-J-|sigma|-norm}) to remain valid up to a factor of order unity.
Thus, for $|\sigma| \sim 1$ the bulk of the vortex remains described by Eqs.(\ref{eq:vortex-bulk})
and (\ref{eq:rho-TF-rc}), the boundary layer at $R_0$ and the outer exponential tail by Eqs.(\ref{eq:P-II})
and (\ref{eq:WKB-det}), the core by Eq.(\ref{eq:vortex-center}) but with a normalization that can be
modified by a factor of order unity. The remaining unknown is the exact shape of the transition at
$r \simeq r_c$. This requires a numerical computation of Eq.(\ref{eq:ODE-f-eta})
\cite{pitaevskii2003bose}. For $|\sigma|=1$ this gives $\hat \psi \simeq 0.6 \sqrt{\rho_0} r/r_c$
at the center, which is already not too far from Eq.(\ref{eq:vortex-center}) that would give 
$\hat\psi \simeq 0.77 \sqrt{\rho_0} r/r_c$.

\section{Energy of a vortex}
\label{app:vortex-energy}

\subsection{Energy difference}
\label{app:energy-difference}

For an axisymmetric profile with a vanishing winding number, such as the static soliton
(\ref{eq:hydrostatic-full}), the energy (\ref{eq:E-rho-v}) reads
\be
E_0[\rho] = \int d r \pi r \left[ \epsilon^2 \left( \frac{d\sqrt{\rho}}{dr} \right)^2 
+ \rho \Phi_N + \lambda \rho^2 \right] .
\label{eq:E0-def}
\ee
Then, the static soliton profile $\rho_{\rm sol}$ defined by the radial Schr\"odinger equation 
(\ref{eq:hydrostatic-full}) is also a minimum of the energy $E_0$ at fixed mass, that is
\be
\left. \delta^{(1)} ( E_0 - \mu M ) \right |_{\rho_{\rm sol}}= 0 ,
\label{eq:minimum-E-0}
\ee
over linear variations $\delta\rho$ of the density.

For a vortex of winding number $\sigma$ and wave function of the form (\ref{eq:psi-sigma}),
the energy (\ref{eq:E-rho-v}) reads
\ba
&& \hspace{-0.5cm} E_\sigma[\rho_\sigma] \! = \!\! \int \!\! d r \pi r \left[ \epsilon^2 
\left( \frac{d\sqrt{\rho_\sigma}}{dr} \right)^2 + \frac{\epsilon^2 \sigma^2 \rho_\sigma}{r^2} 
+ \rho_\sigma \Phi_{N\sigma} + \lambda \rho_\sigma^2 \right]   \nonumber \\
&& = E_0[\rho_\sigma] + {\cal E}_\sigma[\rho_\sigma] ,
\label{eq:E-sigma-def}
\ea
with
\be
{\cal E}_\sigma[\rho_\sigma] = \epsilon^2 \sigma^2 \pi \int \frac{d r}{r} \rho_\sigma .
\label{eq:Delta-E-sigma-def}
\ee

We wish to estimate the energy difference due to the vortex,
\be
\Delta E_\sigma = E_\sigma[\rho_\sigma] - E_0[\rho_{\rm sol}] 
= E_0[\rho_\sigma] - E_0[\rho_{\rm sol}] + {\cal E}_\sigma[\rho_\sigma] ,
\label{eq:Delta-E-sigma-split}
\ee
in the asymptotic limit $\epsilon \to 0$.
We shall find in Eq.(\ref{eq:E0-eps2}) below that the difference of the two $E_0$ terms scales
as $\epsilon^2$, whereas the new angular contribution ${\cal E}_\sigma$ scales as 
$\epsilon^2 \ln(1/\epsilon)$ from Eq.(\ref{eq:calE_sigma-eps}).
Therefore, the latter dominates, in agreement with the simpler analysis of
Eq.(\ref{eq:Delta-E_sigma-def}) in the main text.
Thus, we first consider the scaling over $\epsilon$ of the various terms in
Eq.(\ref{eq:Delta-E-sigma-split}), keeping $M$ and $R_0$ fixed of order unity.
The spin $\sigma$ can be large but does not grow with $1/\epsilon$.

\subsection{Scaling of the $E_0$ terms}
\label{app:scaling-E_0}

We first focus on the $E_0$ terms,
\be
\Delta E_0 = E_0[\rho_\sigma] - E_0[\rho_{\rm sol}] .
\ee
We have seen in Eq.(\ref{eq:rc-def}) that the core radius of the vortex is
$r_c = |\sigma|\xi \propto \epsilon$. Therefore, choosing a fixed large factor $\Lambda \gg 1$ 
(e.g., $\Lambda=10$ or $100$) that does not depend on $\epsilon$, we have
\be
\Lambda \gg 1, \;\; r \geq \Lambda r_c : \;\;\; 
\left | \frac{\rho_\sigma-\rho_{\rm sol}}{\rho_{\rm sol}} \right | \ll 1 .
\label{eq:Lambda-def}
\ee
Then, we define the truncated vortex profile $\tilde\rho_\sigma$ by
\be
r \geq \Lambda r_c \! : \; \tilde\rho_\sigma(r) = \rho_\sigma(r) , \;\;\;
r \leq \Lambda r_c \! : \; \tilde\rho_\sigma(r) = \rho_\sigma(\Lambda r_c) .
\label{eq:rho-sigma-tilde}
\ee
As the deviation $(\tilde\rho_\sigma - \rho_{\rm sol})/\rho_{\rm sol}$ is uniformly small,
we can use the vanishing of the first variation (\ref{eq:minimum-E-0}) to write at leading order
\be
E_0[\tilde\rho_\sigma] - E_0[\rho_{\rm sol}] \simeq \mu ( \tilde M_\sigma - M_{\rm sol} ) 
= \mu ( \tilde M_\sigma - M_\sigma ) .
\label{eq:E0-tilde-rho-rho-sol}
\ee
Here we used the fact that we look for the energy difference $\Delta E_\sigma$ due to the
vortex at fixed total mass, hence $M_\sigma = M_{\rm sol}$.
On the other hand, from Eqs.(\ref{eq:rho-sigma-tilde}) and (\ref{eq:asymp-f}) we obtain
\be
\tilde M_\sigma - M_\sigma \simeq 2 \pi \rho_0 r_c^2 \ln\Lambda \propto \epsilon^2 
\ee
and
\be
E_0[\tilde\rho_\sigma] - E_0[\rho_{\rm sol}] \simeq \mu 2 \pi \rho_0 r_c^2 \ln\Lambda \propto \epsilon^2 .
\ee

We now need to estimate the difference $(E_0[\rho_\sigma] - E_0[\tilde\rho_\sigma])$.
Let us first consider the outer region $r \geq \Lambda r_c$, where 
$\tilde\rho_\sigma = \rho_\sigma$.
We have
\be
\left( E_0[\rho_\sigma] - E_0[\tilde\rho_\sigma] \right)_{r \geq \Lambda r_c} = 
\int_{\Lambda r_c}^\infty dr \, \pi r \rho_\sigma ( \Phi_{N\sigma} - \tilde\Phi_{N\sigma} ) .
\ee
From Eq.(\ref{eq:PhiN-axisymmetric}) we have
\be
r \geq \Lambda r_c : \;\; \Phi_{N\sigma}(r) - \tilde\Phi_{N\sigma}(r) = 2 \ln r 
\left( M_\sigma - \tilde M_\sigma \right) ,
\label{eq:tilde-PhiN}
\ee
which gives
\be
\left( E_0[\rho_\sigma] - E_0[\tilde\rho_\sigma] \right)_{r \geq \Lambda r_c} \simeq
- 2 \pi^2 \rho_0^2 R_0^2 r_c^2 \ln R_0 \ln \Lambda \propto \epsilon^2 .
\label{eq:E0-r-large}
\ee
For the inner region, $r \leq \Lambda r_c$, we also obtain
\be
\left( E_0[\rho_\sigma] - E_0[\tilde\rho_\sigma] \right)_{r \leq \Lambda r_c} \sim 
\lambda \rho_0^2 r_c^2 \ln \Lambda \propto \epsilon^2  .
\label{eq:E0-r-small}
\ee
Collecting Eqs.(\ref{eq:E0-tilde-rho-rho-sol}), (\ref{eq:E0-r-large}) and (\ref{eq:E0-r-small}),
we obtain
\be
\Delta E_0 \propto  \epsilon^2 .
\label{eq:E0-eps2}
\ee
This result takes into account the finite mass of the system and the large-distance
exponential tail (\ref{eq:psi-decay}), which is beyond the reach of the Thomas-Fermi approximation.
Thanks to the leading-order variation (\ref{eq:E0-tilde-rho-rho-sol}), we do not need to
explicitly follow the precise profile of the system at large distance. This is because the exact soliton
is a minimum of the energy (\ref{eq:E0-def}) at fixed mass (i.e., without the need to use the Thomas-Fermi
approximation).

\subsection{Scaling of the ${\cal E}_{\sigma}$ term}
\label{app:scaling-cal-E_sigma}

Finally, we estimate the angular kinetic contribution (\ref{eq:Delta-E-sigma-def}).
From the soliton density profile obtained in App.~\ref{app:static-soliton}, with its sharp cutoff beyond
radius $R_0$, and the radius $r_c$ of the vortex core, we obtain
\be
{\cal E}_\sigma[\rho_\sigma] \simeq \epsilon^2 \sigma^2 \pi \rho_0 \ln(R_0/r_c) 
\propto \epsilon^2 \ln(1/\epsilon) . 
\label{eq:calE_sigma-eps}
\ee
As in the simple analysis of Eq.(\ref{eq:Delta-E_sigma-def}) in the main text and as in
standard textbooks \cite{Pethick_Smith_2008}, we recover that the angular kinetic contribution 
${\cal E}_\sigma$ dominates over the radial contributions $\Delta E_0$
by a logarithmic factor.
Therefore, we obtain for the energy of the vortex the expression (\ref{eq:Delta-E-L}).

\acknowledgments

This work was granted access to the CCRT High-Performance Computing (HPC) facility under the Grant CCRT2024-valag awarded by the Fundamental Research Division (DRF) of CEA.

\bibliography{ref}

\end{document}